\documentclass[a4paper,11pt]{article}
\usepackage{jheppub} % for details on the use of the package, please see the JINST-author-manual
\usepackage{tikz,ifthen}

\newcommand{\legbox}[1]{%
  \raisebox{-0.42\height}{%
  \begin{tikzpicture}[baseline,scale=0.30,line width=0.6pt]
    \draw[gray!60] (-1,-1) rectangle (1,1);
    \foreach \s in {#1}{%
      \ifthenelse{\equal{\s}{r}}{\draw[line width=1.6pt] (0,0)--(1,0);}{}%
      \ifthenelse{\equal{\s}{u}}{\draw[line width=1.6pt] (0,0)--(0,1);}{}%
      \ifthenelse{\equal{\s}{l}}{\draw[line width=1.6pt] (0,0)--(-1,0);}{}%
      \ifthenelse{\equal{\s}{d}}{\draw[line width=1.6pt] (0,0)--(0,-1);}{}%
    }
  \end{tikzpicture}}%
}
\newcommand{\plqbox}{\vcenter{\hbox{%
  \begin{tikzpicture}[scale=0.30,line width=0.6pt]
    \draw[gray!60] (-1,-1) rectangle (3,3);
    \draw[gray!60] (1,-1)--(1,3);
    \draw[gray!60] (-1,1)--(3,1);
    \draw[line width=1.6pt] (0,0) rectangle (2,2);
  \end{tikzpicture}}}}

\newcommand{\rectboxH}{\vcenter{\hbox{%
  \begin{tikzpicture}[scale=0.30,line width=0.6pt]
    \draw[gray!60] (-1,-1) rectangle (5,3);
    \draw[gray!60] (1,-1)--(1,3);
    \draw[gray!60] (3,-1)--(3,3);
    \draw[gray!60] (-1,1)--(5,1);
    \draw[line width=1.6pt] (0,0) rectangle (4,2);
  \end{tikzpicture}}}}

\newcommand{\rectboxV}{\vcenter{\hbox{%
  \begin{tikzpicture}[scale=0.30,line width=0.6pt]
    \draw[gray!60] (-1,-1) rectangle (3,5);
    \draw[gray!60] (1,-1)--(1,5);
    \draw[gray!60] (-1,1)--(3,1);
    \draw[gray!60] (-1,3)--(3,3);
    \draw[line width=1.6pt] (0,0) rectangle (2,4);
  \end{tikzpicture}}}}

\newcommand{\diagbox}{\vcenter{\hbox{%
  \begin{tikzpicture}[scale=0.30,line width=0.6pt]
    \draw[gray!60] (-1,-1) rectangle (5,5);
    \draw[gray!60] (1,-1)--(1,5); \draw[gray!60] (3,-1)--(3,5);
    \draw[gray!60] (-1,1)--(5,1); \draw[gray!60] (-1,3)--(5,3);
    \draw[line width=1.6pt] (0,0) rectangle (2,2);
    \draw[line width=1.6pt] (2,2) rectangle (4,4);
  \end{tikzpicture}}}}
\newcommand\beq{ \begin{eqnarray} }
\newcommand\eeq{ \end{eqnarray} }

\usepackage{amsmath}
\usepackage{amssymb}
\usepackage{mathrsfs}
\usepackage{graphicx}
\usepackage{bbold}
\usepackage{braket}
\usepackage{bm,bbm}
\usepackage{physics}
\usepackage{tikz}

\preprint{YITP-26-157, RIKEN-iTHEMS-Report-26}
\arxivnumber{1234.56789} % if you have one

\title{\boldmath Assessing the expressive power of a gauge-invariant PEPS ansatz
across a topological phase transition}

\author[a,b]{Etsuko Itou}
\author[c,d]{and Tsuyoshi Okubo}

\affiliation[a]{Yukawa Institute for Theoretical Physics, Kyoto University, Kitashirakawa Oiwakecho, Sakyo-ku, Kyoto 606-8502, Japan }
\affiliation[b]{RIKEN Center for Interdisciplinary Theoretical and Mathematical Sciences (iTHEMS), RIKEN,
2-1 Hirosawa, Wako, Saitama 351-0198 Japan }  

\affiliation[c]{Department of Physics, Niigata University, 8050 Ikarashi 2-no-cho, Nishi-ku, Niigata 950-2181, Japan }
\affiliation[d]{Department of Physics, University of Tokyo, 7-3-1 Hongo, Bunkyo-ku, Tokyo 113-0033, Japan }

\emailAdd{itou@yukawa.kyoto-u.ac.jp}

\abstract{
We investigate the expressive power of the gauged Gaussian fermionic projected entangled pair state (GGFPEPS) ansatz, which is a manifestly gauge-invariant tensor network ansatz for $(2+1)$-dimensional lattice gauge theories.
In the GGFPEPS ansatz, the gauge field on each link is represented by coupling its configuration to the fermion number of virtual fermions placed at both ends of the link.
This ansatz has been studied by variational Monte Carlo (VMC) methods in previous work~\cite{Emonts:2022yom}.
However, the statistical error of VMC makes a precise determination of the variational parameters difficult, so that the expressive power of the ansatz could not be examined accurately.
In this work, we reformulate the computation as a Grassmann tensor network (TN), in which the whole network is contracted deterministically, and introduce the higher-order tensor renormalization group (HOTRG), which allows us to perform calculations in large volumes up to $L=34$.
We also evaluate the infinite volume system using the corner transfer matrix renormalization group (CTMRG).
We find that the ansatz with only eight complex variational parameters accurately describes the pure $\mathbb{Z}_2$ gauge theory including the region of the topological phase transition, not only in the energy but also in the topological entanglement entropy.
The critical coupling obtained from the peaks of their susceptibilities agrees quantitatively with the one predicted by continuous-time quantum Monte Carlo simulations~\cite{wu2012phase}.
Furthermore, near the transition point, we show that the ansatz represents a state outside the plaquette product family, which is a one-parameter family of states assumed in the construction of the ansatz as a simple interpolation between the strong and weak coupling limits.
}

\begin{document}
\maketitle
%\flushbottom
\raggedbottom
%%%%%%%%%%%%%%%%%%%%%%%%%%%%%%%%%%%%%%%%%%%%%%%%%%%%%%%%%%
\section{Introduction}
%%%%%%%%%%%%%%%%%%%%%%%%%%%%%%%%%%%%%%%%%%%%%%%%%%%%%%%%%%
Studies of lattice gauge theories based on the Hamiltonian formalism have attracted growing attention in recent years~\cite{Banuls:2019rao, Banuls:2019bmf, Funcke:2023jbq, Bauer:2022hpo, DiMeglio:2023nsa, Dalmonte:2016alw, Magnifico:2024eiy}.
Quantum computing (QC) and tensor network (TN) methods are considered to be suitable for numerical calculations of gauge theories in the Hamiltonian formalism.
They give a sign-problem-free calculation framework instead of the conventional lattice Monte Carlo (MC) method based on the Euclidean Lagrangian formalism.
In these QC and TN methods, numerical calculations are performed in a finite-dimensional Hilbert space with discrete degrees of freedom such as spin variables.
Indeed, in $(1+1)$-dimensional gauge theories, even for a theory with a continuous gauge symmetry such as the Schwinger model, the gauge field has no dynamical degrees of freedom, so that the Hamiltonian can be written only in terms of the fermionic degrees of freedom by fixing the gauge and solving the Gauss law.
As a result, studies of the finite $\theta$ region and the real-time evolution, where the conventional method suffers from the sign problem, have been successful in recent years~\cite{Honda:2021aum, Honda:2021ovk, Mildenberger:2022jqr, Dempsey:2022nys, Okuda:2022hsq, Mueller:2022xbg, Honda:2022edn, Nagano:2023uaq, Itou:2023img, Lee:2023urk, Angelides:2023noe, Farrell:2024fit, Kaikov:2024acw, Itou:2024psm, Bazavov:2024spg, ArguelloCruz:2024xzi, Fujii:2024reh, Kanno:2024elz, Chen:2026sul, Lamm:2026cfq}.

In $(d+1)$ dimensions with $d>1$, on the other hand, the gauge field has dynamical degrees of freedom, and for continuous gauge groups its local Hilbert space on each link is infinite-dimensional.
Numerical calculations then require truncation to a finite-dimensional space, which often conflicts with the structure of the gauge theory, as we illustrate below.
The Kogut--Susskind Hamiltonian describing $SU(N)$ gauge theory in $(2+1)$ dimensions is given by
\beq
H = \frac{g^2}{2a} \sum_{\vb*{x}} \sum_{a=1}^{N^2-1}[( E_{\vb*{x},x}^a)^2 + (E_{\vb*{x},y}^a)^2] + \frac{1}{g^2a} \sum_p [N- \frac{1}{2}(U_p + U_p^\dag)].\label{eq:KS-Hamiltonian}
\eeq
Here, $E_{\vb*{x},x}^a$ and $E_{\vb*{x},y}^a$ denote the electric-field operators on the links in the $x$ and $y$ directions from the site $\vb*{x}$, and $U_p$ is the plaquette operator, $U_p= \tr \prod_{l \in p}U_l$.
Since $[E,U_p] \ne 0$, there are two representative choices of basis for defining the Hilbert space.
One is the eigenstates $| E_n \rangle$ of $E$ (strong coupling basis), and the other is the eigenstates $|U_q \rangle$ of the plaquette operator constructed from the eigenstates of each link variable $U_l$ (weak coupling basis).

For Abelian gauge groups such as $U(1)$ and $\mathbb{Z}_N$, these two bases are related by a Fourier transform.\footnote{For non-Abelian gauge groups, they are related by a character expansion instead.}
In the case of the compact $U(1)$ gauge theory, as an example,
if we write the link variable as $U=e^{iA}$ with $A \in [0,2\pi)$, then we have
$
|A\rangle = \frac{1}{\sqrt{2\pi}}\sum_{n\in\mathbb{Z}} e^{-inA} |E_n \rangle.
$
Since $U$ is periodic, $E$ is quantized and takes all integer values $n\in\mathbb{Z}$, so that the sum is infinite.

The truncation to a finite-dimensional space, however, breaks this Fourier transform.
If we truncate the electric basis to $|n|\le n_{\max}$, the Gauss law is kept exactly, since the Gauss law operator is diagonal in this basis.
However, the raising operation $U|E_n\rangle=|E_{n+1}\rangle$ stops at $n=n_{\max}$, so that $U$ is no longer unitary, $U^\dag U\neq1$.
If we instead discretize $A$ as $A=2\pi q/N$, $U$ remains unitary, but the eigenvalues of $E$ are identified modulo $N$.
Then, the gauge group is reduced from $U(1)$ to $\mathbb{Z}_N$.
In either case, it is difficult to cover the entire range of the coupling constant with a single truncation, and physical observables can depend on the truncation~\cite{Zohar:2014qma, Haase:2020kaj, Paulson:2020zjd}.
Various truncation schemes have been proposed to keep the algebraic structure of the gauge theory as much as possible~\cite{Chandrasekharan:1996ih, Brower:1997ha, Zache:2021ggw, Berenstein:2022wlm, Raychowdhury:2019iki, Kadam:2022ipf, Zache:2023dko, Hayata:2023bgh, Kaplan:2018vnj, Bauer:2021gek, Ciavarella:2025bsg, Bergner:2024qjl, Halimeh:2024bth}, but it remains a difficult problem.

For $\mathbb{Z}_N$ gauge theory, the degrees of freedom on each link are finite, and no truncation is needed in principle.
In practice, however, since they are present on each spatial link, the size of the Hilbert space grows exponentially with the volume.
Then, the sum over the gauge-field configurations is too large to handle directly.

In this situation, one possible approach is to assume a certain ansatz and construct the ground state variationally.
Among such approaches, the gauged Gaussian fermionic projected entangled pair state (GGFPEPS) ansatz was proposed in Refs.~\cite{Zohar:2015eda, Zohar:2017yxl}.
It represents the link variables, which carry the gauge-field degrees of freedom, by virtual fermions placed at the two ends of each link.
This ansatz is manifestly invariant under the action of the Gauss law operator.
It is also well suited to the TN methods.
Furthermore, the framework can apply to arbitrary gauge groups including continuous and non-Abelian ones, and also to dynamical fermions~\cite{Zohar:2015eda, Zohar:2017yxl, Kelman:2024gjt}.
Indeed, it has been applied to the pure $\mathbb{Z}_2$ gauge theory~\cite{Emonts:2022yom, Kelman:2026goj}, to the $\mathbb{Z}_3$ gauge theory~\cite{Emonts:2020drm}, and to the $\mathbb{Z}_2$ gauge theory with dynamical fermions~\cite{Kelman:2024exo}.
Moreover, the gauge field is treated in the group element basis.
It does not require the truncation of the electric field discussed above.

On the other hand, in the previous applications to the pure $\mathbb{Z}_2$ gauge theory in \mbox{$(2+1)$} dimensions~\cite{Emonts:2022yom, Kelman:2026goj}, the variational parameters were optimized by the variational MC (VMC) method.
With the statistical error of MC, however, it is difficult to determine the variational parameters precisely.
The usefulness of the ansatz thus cannot be examined accurately.
Indeed, in Ref.~\cite{Emonts:2022yom}, the minimal ansatz with $F=1$ fails to reproduce the exact diagonalization (ED) results around the critical coupling.
Here $F$ denotes the number of virtual fermionic modes at each end of a link.
The optimized ansatz with $F=2$, which has a larger variational space, was then introduced.
However, whether the $F=2$ ansatz is sufficient in the critical region has not been settled, because of the statistical error and the small volumes.
This $\mathbb{Z}_2$ gauge theory has the toric code ground state in the weak coupling limit and the confinement phase in the strong coupling regime.
The two phases are separated by a second-order topological phase transition.
Since the theory is dual to the Ising model~\cite{Wegner:1971app}, the critical point belongs to the 3D Ising universality class.
The aim of this paper is to settle whether the ansatz has enough expressive power around such a critical point.

In this paper, we use the Grassmann TN~\cite{Gu:2010yh, Akiyama:2020sfo}, which extends the TN methods for bosonic systems to fermionic systems, to examine the usefulness of the ansatz without statistical errors.
The Grassmann TN contracts the ansatz state exactly.
Thus, the optimization of the variational parameters is free from the statistical error of the MC sampling, whose convergence is not easy to control~\cite{Gomelski:2025enu}.
This allows us to test the ansatz systematically.
The systematic error of this method comes from the truncation of the bond dimension to $D_{\rm cut}$, which is under control.
To perform large volume calculations, we introduce the higher-order tensor renormalization group (HOTRG)~\cite{Xie:2012mjn} and carry out calculations up to $L=34$, where $L$ is the linear size of the lattice.
The infinite volume limit is evaluated by the corner transfer matrix renormalization group (CTMRG).
We also calculate the topological entanglement entropy (TEE) and the winding 't Hooft loop, which were not calculated in the previous works.
From the susceptibilities of the energy density and the TEE, we estimate the critical coupling of the topological phase transition and compare it quantitatively with the continuous-time quantum Monte Carlo (QMC) result~\cite{wu2012phase}.\footnote{The continuous-time QMC method is a numerically exact algorithm that satisfies detailed balance and ensures ergodic sampling.}
Furthermore, we compare the optimized state with the one-parameter plaquette product family introduced in the construction of the $F=2$ ansatz.
We show that, around the critical point, the state cannot be written within this one-parameter family and a larger variational space is required.
Even there, the $F=2$ ansatz can represent the state.

This paper is organized as follows.
Section~\ref{sec:ZN-LGT} reviews the $\mathbb{Z}_2$ gauge theory in the Hamiltonian formalism, its ground states in the two limits, and the strong and weak coupling expansions.
The GGFPEPS ansatz is introduced in Sec.~\ref{sec:GGFPEPS}, where we also prove its gauge invariance and discuss the $F=2$ ansatz in relation to the plaquette product family.
How to evaluate the physical observables, including the TEE, is explained in Sec.~\ref{sec:def-obs}.
Section~\ref{sec:calc-methods} describes the Grassmann TN and the HOTRG used in this work.
The numerical results are presented in Sec.~\ref{sec:results-finite-vol}, first for small volumes in comparison with ED, and then for large volumes up to $L=34$.
Section~\ref{sec:ctmrg} gives the results in the infinite volume limit by CTMRG.
Section~\ref{sec:summary} is devoted to the conclusions and outlook.
Appendix~\ref{sec:F1-ansatz} shows that the $F=1$ ansatz can also reproduce the toric code ground state, although it fails in the intermediate coupling region.
Appendix~\ref{sec:Dcut_CTMRG} examines the $D_{\rm cut}$ dependence in the CTMRG calculation.

%%%%%%%%%%%%%%%%%%%%%%%%%%%%%%%%%%%%%%%%%%%%%%%%%%%%%%%%%%
\section{$\mathbb{Z}_2$ gauge theory in (2+1) dimensions}\label{sec:ZN-LGT}
%%%%%%%%%%%%%%%%%%%%%%%%%%%%%%%%%%%%%%%%%%%%%%%%%%%%%%%%%%
\subsection{Lattice Hamiltonian}
%%%%%%%%%%%%%%%%%%%%%%%%%%%%%%%
We consider pure $\mathbb{Z}_N$ discrete gauge theory in $(2+1)$ dimensions.
This theory has degrees of freedom only on links, with an $N$-dimensional Hilbert space on each link $l = (\vb*{x},i)$.
Here, $\vb*{x}$ denotes a site in two-dimensional space, and $i$ denotes the link direction ($i=1,2$).
On each link, we introduce two unitary operators $P_l,Q_l$ satisfying
\begin{align*}
P_l^N = Q_l^N =1, \quad P_l^\dag P_l = Q^\dag_l Q_l =1, \quad
P_l Q_l P_l^\dag = e^{i\delta}Q_l, ~~ \delta = 2\pi/N.
\end{align*}
We denote the eigenstates of $P$ (p-basis) by $\{ |p \rangle \}_{p=0}^{N-1}$, with eigenvalues given by
\beq
P | p \rangle = e^{i p \delta } | p \rangle.
\eeq
Then, since $Q$ is also unitary, we find that with the cyclic identification $|N \rangle = | 0 \rangle$, $Q$ acts as a raising operator:
\beq
Q | p \rangle = | p+1 \rangle.
\eeq
Similarly, denoting the eigenstates of $Q$ (q-basis) by $\{ |q \rangle \}_{q=0}^{N-1}$ and writing their eigenvalues as
\beq
Q| q \rangle = e^{i q \delta } | q \rangle,
\eeq
$P$ acts as an operator that lowers the state cyclically:
\beq
P | q \rangle = | q-1 \rangle.
\eeq

\begin{figure}[htb]
\centering
\includegraphics[scale=0.13]{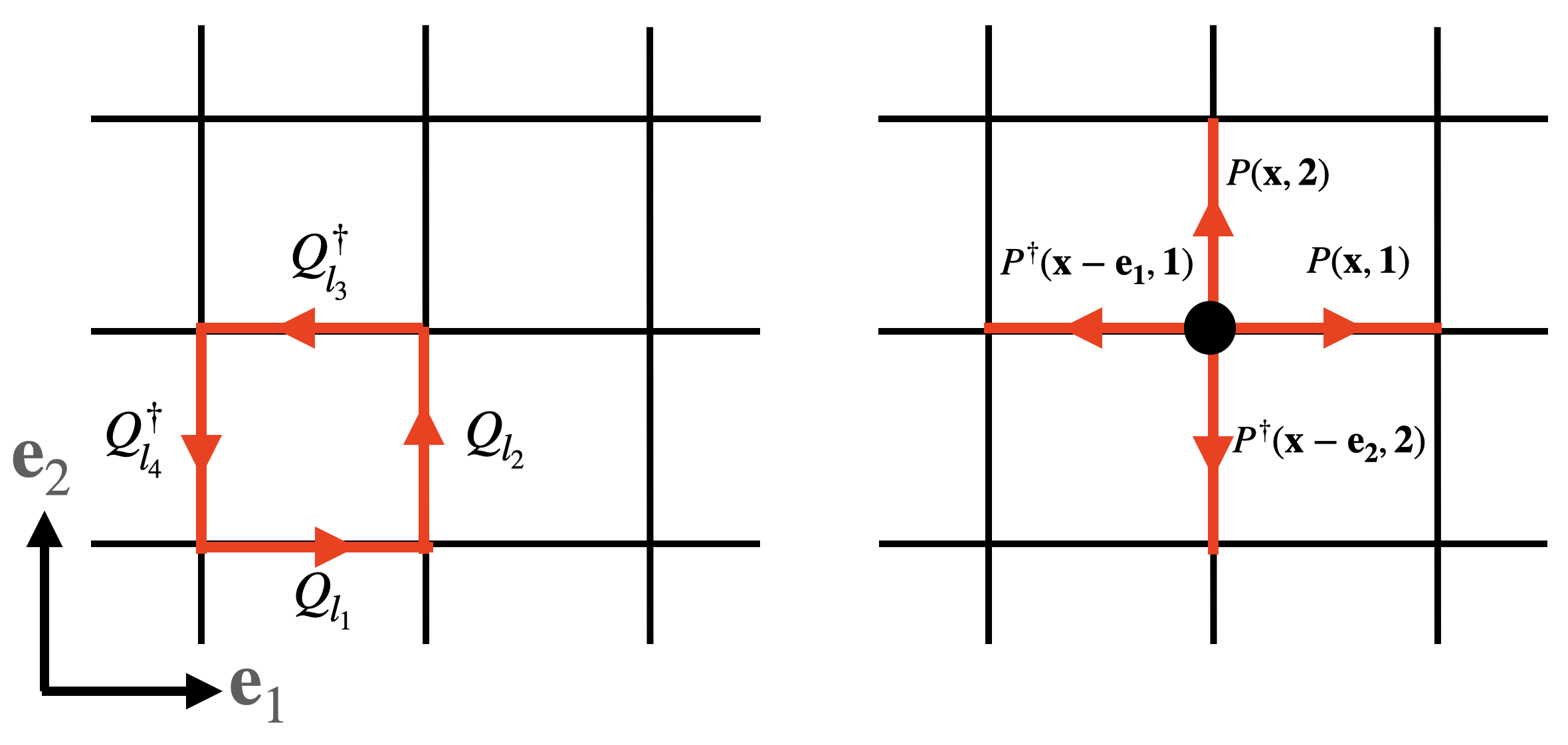}
\caption{(Left) Definition of the plaquette operator. (Right) The Gauss law operator $V(\vb*{x})$.}\label{fig:plaq}
\end{figure}

The Hamiltonian of pure $\mathbb{Z}_N$ gauge theory can be written in terms of $P,Q$ as
\begin{align*}
H = \frac{\lambda}{2} \sum_l \left[ 2 -(P_l + P_l^\dag) \right] +\frac{1}{2\lambda} \sum_{p} \left[2 - (Q_{l_1} Q_{l_2} Q^\dag_{l_3}Q^\dag_{l_4} + {\rm h.c.}) \right].
\end{align*}
Here, $\lambda$ denotes the coupling constant.
In the limit $N \rightarrow \infty$, this theory is naively expected to become $U(1)$ gauge theory. By this analogy, the first term is called the ``electric energy'', $E_e$, while the second is called the ``plaquette'' or ``magnetic energy'', $E_m$.
This plaquette term corresponds to an elementary closed loop on the lattice, and $l_1,l_2,l_3,l_4$ denote the four links surrounding the plaquette $p$.
The link orientations and the order of the product are defined as shown in Fig.~\ref{fig:plaq}.

%%%%%%%%%%%%%%%%%%%%%%%%%%%%%%
%%%%%%%%%%%%%%%%%%%%%%%%%%%%%%

Let us now consider $\mathbb{Z}_2$ gauge theory with $N=2$.
In the p-basis, we choose
\beq
P = P^\dag = \sigma_z, \quad Q = Q^\dag = \sigma_x. \label{eq:def-P-Q}
\eeq
The eigenstates can then be taken as those of $\sigma_z$ and $\sigma_x$, with eigenvalues $(-1)^p$ and $(-1)^q$, respectively.
Since $P$ and $Q$ are Hermitian, the link orientation does not matter for $N=2$, and the Hamiltonian becomes
\begin{align}
H = \lambda \sum_l \left[ 1 -P_l \right] +\frac{1}{\lambda} \sum_p \left[1 - (Q_{l_1} Q_{l_2} Q_{l_3}Q_{l_4}) \right]. \label{eq:Z2-Hamiltonian}
\end{align}
We adopt the convention that a link in the state $|p=0\rangle$ does not carry an electric field, while a link in the state $|p=1\rangle$ carries an electric field.

Furthermore, we define the Gauss law operator as 
\beq
V(\vb*{x}) = P(\vb*{x}, 1) P(\vb*{x}, 2) P^\dag(\vb*{x}-\vb*{e}_1, 1) P^\dag(\vb*{x}-\vb*{e}_2, 2), \label{eq:phys-gauge-trans}
\eeq
as shown in the right panel of Fig.~\ref{fig:plaq}.
It is a local unitary transformation and the Hamiltonian commutes with it.
Namely, $[H,V(\vb*{x})]=0$.
In addition, $[V(\vb*{x}),V(\vb*{y})]=0$ holds for arbitrary $\vb*{x},\vb*{y}$, since $V(\vb*{x})$ is a product of the operators $P$, which commute with each other.
Therefore, the full Hilbert space can be decomposed into a direct sum of sectors labeled by the eigenvalues of $V(\vb*{x})$.
Here we take the sector where the state satisfies
\beq
V(\vb*{x}) | \psi \rangle = |\psi \rangle, ~~~\forall \vb*{x}. \label{eq:Gauge-inv}
\eeq

This Hamiltonian~\eqref{eq:Z2-Hamiltonian} is known to be equivalent to the toric code model in an external magnetic field.
For example, in the notation of Ref.~\cite{wu2012phase}, the Hamiltonian is written as
\beq
H&=&H_{\rm TCM}-h_x \sum_b \sigma_b^x -h_z \sum_b \sigma^z_b, \label{eq:toric-code-H}
\eeq
where
\beq
H_{\rm TCM}&=&-J_s \sum_s \prod_{j \in s} \sigma_j^x -J_p \sum_p \prod_{j\in p}\sigma^z_j.
\eeq
By exchanging $\sigma^x \leftrightarrow \sigma^z$ on all links, it is equivalent to our $\mathbb{Z}_2$ gauge theory.
The parameters correspond to 
\beq
J_p=\frac{1}{\lambda},\ h_x=\lambda,\ h_z=0, \label{eq:toric-code-corresp}
\eeq
with the limit $J_s\to\infty$ imposing $\prod_{j\in s}\sigma_j^x=+1$ on every site. 
It corresponds to the Gauss law constraint $V(\vb*{x})|\psi\rangle=|\psi\rangle$ in our notation.

%%%%%%%%%%%%%%%%%%%%%%%%%%%%%%%%%%%%%%%%%%%%%%%%%%%%%%%%%%
\subsection{Ground states in the weak and strong coupling limits}\label{sec:Z2-ground-states}
%%%%%%%%%%%%%%%%%%%%%%%%%%%%%%%%%%%%%%%%%%%%%%%%%%%%%%%%%%
It is known that the $(2+1)$-dimensional $\mathbb{Z}_2$ gauge theory has the confinement phase in the strong coupling region and the topological (deconfinement) phase in the weak coupling region.
Here, the former is characterized by the area law of the spatial Wilson loop, $\mathcal{W}(C)\sim e^{-\sigma R_1R_2}$, for a closed loop $C$ of size $R_1\times R_2$, while the latter is characterized by the perimeter law, $\mathcal{W}(C)\sim e^{-a(R_1+R_2)}$~\cite{Wegner:1971app, Wilson:1974sk}.

The ground states in the two limits become simple in the two bases.
For $\lambda=\infty$, the electric term dominates.
The ground state is a product state in the p-basis (strong coupling basis),
\beq
|\psi (\lambda=\infty)\rangle = \otimes_l | p=0 \rangle_l \equiv |\psi_E \rangle.\label{eq:def-psi-E}
\eeq
Namely, all links are in the no-flux state, $|p=0 \rangle$.
On the other hand, for $\lambda=0$, the magnetic term dominates.
Its ground state can be given by the product state in the q-basis (weak coupling basis), $\otimes_l|q=0\rangle_l$.
However, the latter product state does not satisfy the Gauss law.
The physical ground state is obtained by projecting it onto the sector of the Gauss law.
In this paper, we would like to write the ground state at $\lambda=0$ using $|\psi_E\rangle$ as
\beq
| \psi (\lambda=0) \rangle = \mathcal{N} \prod_p (1+ Q_{l_1}Q_{l_2}Q_{l_3}Q_{l_4}) | \psi_E \rangle = | \psi_B\rangle.\label{eq:def-psi-B}
\eeq
Here, $\mathcal{N}$ is a normalization constant, which is given by $\mathcal{N}=2^{-(n_p+1)/2}$ under periodic boundary conditions.\footnote{In practical tensor network calculations, the wave function is normalized using $\langle \psi | \psi \rangle$, so this factor is irrelevant.}
Here $n_p=L\times L$ is the number of plaquettes.
This ground state is known as the toric code ground state.
It has a fourfold degeneracy on a two-dimensional torus.
Equation~\eqref{eq:def-psi-B} is the one among them which is continuously connected to $|\psi_E\rangle$.
The TEE, which will be discussed below, is given by $S_{\rm topo}=- \ln 2$, where $2$ is the total quantum dimension of the $\mathbb{Z}_2$ toric code.

These two phases are separated by a second-order topological phase transition.
For $\mathbb{Z}_2$, the critical point belongs to the 3D Ising universality class.
For compact $U(1)$ gauge theory obtained by taking $N \rightarrow \infty$, the topological transition point moves toward $\lambda \rightarrow 0$, where a continuum limit can be taken, and only the confinement phase remains~\cite{Polyakov:1976fu}.
Therefore, there is no continuum limit describing a theory with a gapless gauge boson for the $(2+1)$-dimensional compact $U(1)$ gauge theory.

According to the results of the QMC using an exact algorithm for the toric-code-type Hamiltonian~\eqref{eq:toric-code-H}, the critical point is located at $h_z=0,\ h_x^{(c)}=0.328474(3)$~\cite{wu2012phase}.
Translating this into our parameters using Eq.~\eqref{eq:toric-code-corresp}, we expect $h_x^{(c)}/J_p=\lambda_c^2=0.328474$, namely
\beq
\lambda_c &\simeq& 0.573. \label{eq:MC-lambda_c}
\eeq

%%%%%%%%%%%%%%%%%%%%%%%%%%%%%%%%%%%%%%%%%%%%%%%%%%%%%%%%%%
\subsection{Perturbative expansions}\label{sec:pert}
%%%%%%%%%%%%%%%%%%%%%%%%%%%%%%%%%%%%%%%%%%%%%%%%%%%%%%%%%%
We summarize the results of perturbative analyses of the energy and the winding 't Hooft loop in the strong and weak coupling regimes.

In general, we write the Hamiltonian as $H=H_0+V(\lambda)$ and consider the perturbative expansion in $V(\lambda)$.
Let $|\psi^{(0)}\rangle$ be the nondegenerate ground state of $H_0$, with energy $E^{(0)}$.
We expand the ground state and its energy as $|\psi\rangle=\sum_{n\ge0}|\psi^{(n)}\rangle$ and $E=\sum_{n\ge0}E^{(n)}$, where $|\psi^{(n)}\rangle$ and $E^{(n)}$ are the contributions with $n$ insertions of $V(\lambda)$, and we use the normalization $\langle\psi^{(0)}|\psi^{(n)}\rangle=\delta_{n0}$.

\subsubsection{Strong coupling expansion}\label{SCE}
To study the ground state in the strong coupling regime ($\lambda \gg 1$), we start from $|\psi^{(0)}\rangle =|\psi_E\rangle$ and write the Hamiltonian as
\beq
H_0 &=& \lambda \sum_l (1-P_l), \quad V(\lambda)=\frac{1}{\lambda} \sum_p (1-B_p)\\
B_p  &=& Q_{l_1} Q_{l_2} Q_{l_3}Q_{l_4}.
\eeq
Since the state $B_p|\psi_E\rangle$ has an energy $8\lambda$ above the one of $|\psi_E\rangle$, the ground state up to the order $n=1$ can be written as
\beq
| \psi^{(0)} \rangle + | \psi^{(1)} \rangle &=& \left( 1+\frac{1}{8\lambda^2} \sum_{p}B_p \right) | \psi_E \rangle,
\label{eq:plaq-prod-state}
\eeq
while the energy density per plaquette, which we denote by $E_{tot}$, up to the order $n=3$ is given by
\beq
E_{tot}^{(SC)} (\lambda) = \frac{1}{\lambda} (1-\frac{1}{8\lambda^2}) + \mathcal{O} ( 1/ \lambda^7 ). \label{eq:Etot-strong-coupling}
\eeq
Here, $E^{(3)}$ vanishes, so there is no $\mathcal{O}(1/\lambda^5)$ term.

Similarly, we consider the expectation value of the winding 't Hooft loop operator in the state including the first-order perturbative correction.
We define the winding 't Hooft loop operator as a product of the electric-field operators $P_l$ over the links on a closed loop winding around the torus.\footnote{Usually, the 't Hooft loop is defined on the dual lattice, namely, as a product of $P_l$ over the links crossed by a noncontractible loop on the dual lattice. Our definition differs from it in that $W$ is a product of $P_l$ over the links on a noncontractible loop of the original lattice. Its geometry is the same as that of the spatial Polyakov loop in lattice QCD, while the electric-field operator $P_l$ is used in place of the link variable.}
Here, we specifically define it as the product over links in the $x$ direction at $y_0=0$,
\beq
W&=&\prod_{x=0}^{L-1}P(x\vb*{e}_1+y_0\vb*{e}_2,1)
=\prod_{x=0}^{L-1}(\sigma_z)_{(x\vb*{e}_1+y_0\vb*{e}_2,1)}. \label{eq:def-W}
\eeq
Here, the second expression is written in the p-basis, Eq.~\eqref{eq:def-P-Q}, where $P=\sigma_z$.
In the actual numerical calculations, we use the q-basis, where $P$ is represented by $\sigma_x$, as explained in Sec.~\ref{sec:def-winding-loop}.
We can easily find that $\langle \psi_E | W | \psi_E \rangle =1$ while $\langle \psi_B | W | \psi_B \rangle =0$.
Thus, it is expected to be a good observable to distinguish the strong and weak coupling regimes.

To the leading order of the strong coupling expansion, we may replace the state in Eq.~\eqref{eq:plaq-prod-state} by the product form
\beq
\prod_p \left(1+\frac{1}{8\lambda^2}B_p\right)|\psi_E\rangle, \label{eq:plaq-prod-state-SC}
\eeq
which agrees with Eq.~\eqref{eq:plaq-prod-state} to first order in $1/\lambda^2$.
We call a state of this form a \emph{plaquette product state}.
The expectation value of $W$ in this state is given by
\beq
\langle W\rangle&=&\frac{\langle\psi_E|\prod_p \left(1+\frac{1}{8\lambda^2}B_p\right) W \prod_p \left(1+\frac{1}{8\lambda^2}B_p\right)|\psi_E\rangle}{\langle\psi_E|\prod_p \left(1+\frac{1}{8\lambda^2}B_p\right)^2|\psi_E\rangle},\nonumber\\
&=&\frac{(1-(1/8\lambda^2)^2)^{2L}(1+(1/8\lambda^2)^2)^{n_p-2L}}{(1+(1/8\lambda^2)^2)^{n_p}+(1/4\lambda^2)^{n_p}}
\xrightarrow{n_p\gg1}
\left[\frac{1-(1/8\lambda^2)^2}{1+(1/8\lambda^2)^2}\right]^{2L}.
\eeq

Now, we define
\beq
\alpha (\lambda) \equiv -\frac{1}{L}\ln \langle W \rangle \label{eq:def-alpha}
\eeq
to examine the $L$ dependence in this work.
At leading order (LO) in the strong coupling expansion, we obtain
\beq
\alpha^{(LO)} (\lambda)&=& - 2\ln \left[ \frac{1-(1/8\lambda^2)^2}{1+(1/8\lambda^2)^2} \right]. \label{eq:thooft-LO}
\eeq
Then, in the strong coupling limit ($\lambda \rightarrow \infty$), it approaches
\beq
\alpha^{(SC)} (\lambda)&=& \frac{1}{16 \lambda^4} + \mathcal{O} (\lambda^{-8}). \label{eq:alpha-sc}
\eeq

\subsubsection{Weak coupling expansion}\label{sec:WCE}
We also consider the perturbative expansion in the weak coupling regime.
We first split the Hamiltonian as
\beq
H_0=\frac1\lambda\sum_p(1-B_p),\qquad 
V(\lambda)=\lambda\sum_l(1-P_l).
\eeq
The ground state of $H_0$ is fourfold degenerate on the torus.
Among the four toric code ground states, we take $|\psi^{(0)}\rangle=|\psi_B\rangle$ as explained in Sec.~\ref{sec:Z2-ground-states}.
The other three are connected to $|\psi_B\rangle$ only at the $L$-th order in $V(\lambda)$, so that this degeneracy does not affect the expansion below.

Including the first-order correction to the ground state, we obtain
\beq
|\psi^{(0)}\rangle+|\psi^{(1)}\rangle &=&\left(1+\frac{\lambda^2}{4}\sum_l P_l\right)|\psi_B\rangle.\label{eq:leading-state-weak}
\eeq
Noting that the number of links is $n_l=2n_p$, the energy density per plaquette is given by
\beq
E^{(WC)}_{tot}(\lambda)=\frac{E^{(1)}+E^{(2)}}{n_p}=2\lambda-\frac{\lambda^3}{2}+\mathcal O(\lambda^7)
=2\lambda\Big(1-\frac{\lambda^2}{4}\Big)+\mathcal O(\lambda^7).
\label{eq:wc-energy-result}
\eeq
Here again, the $E^{(3)}$ term, namely the $\mathcal{O}(\lambda^5)$ contribution, vanishes as in the strong coupling expansion.

The expectation value of the winding 't Hooft loop vanishes in the state of Eq.~\eqref{eq:leading-state-weak}, since $W$ flips the $2L$ plaquettes adjacent to the winding-loop line, which cannot be compensated by the first-order correction for $L>2$.

%%%%%%%%%%%%%%%%%%%%%%%%%%%%%%%%%%%%%%%%%%%%%%%%%%%%%%%%%%
\section{Gauged Gaussian Fermionic PEPS ansatz}\label{sec:GGFPEPS}
%%%%%%%%%%%%%%%%%%%%%%%%%%%%%%%%%%%%%%%%%%%%%%%%%%%%%%%%%%
In this section, we review the GGFPEPS ansatz following Ref.~\cite{Emonts:2022yom}.
The notation and the construction of the ansatz are taken from that work unless otherwise stated.
The proof of the gauge invariance in Sec.~\ref{sec:gauge-inv} and the explanation of the construction of the $F=2$ ansatz in Sec.~\ref{sec:more-F2} are given in our own way.
%%%%%%%%%%%%%%%%%%%%%%%%%%%%%%%%%%%%%%%%%%%%%%%%%%%%%%%%%
\subsection{GGFPEPS ansatz}\label{sec:GGPEPS-basic}
%%%%%%%%%%%%%%%%%%%%%%%%%%%%%%%%%%%%%%%%%%%%%%%%%%%%%%%%%%
The basic idea of the GGFPEPS ansatz is to represent the link variables, which carry the gauge-field degrees of freedom, using virtual fermionic modes at each end of each link.
All site and link operators are then written in Gaussian form in terms of these virtual fermions.

The GGFPEPS ansatz is given by
\beq
|\psi \rangle = \langle \Omega | \prod_{\vb*{x},i} w^\dag (\vb*{x},i) \mathcal{U}_{\mathcal{G}} \prod_{\vb*{x}} A(\vb*{x}) | \Omega \rangle |\psi_E\rangle. \label{eq:Gauged-PEPS-ansatz}
\eeq
%%%%%%%%%%%%%%%%%%%%%%%%%%%%%%
\begin{figure}[htb]
\centering
\includegraphics[scale=0.5]{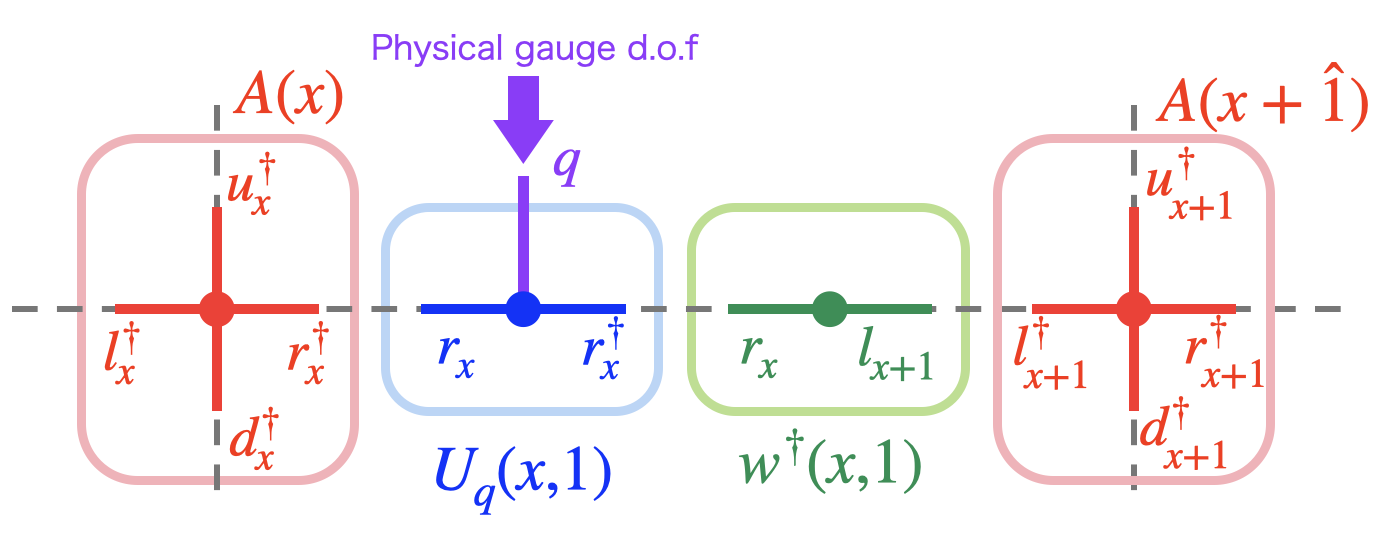}
\caption{GGFPEPS ansatz state.}\label{fig:GGPEPS-ansatz}
\end{figure}
%%%%%%%%%%%%%%%%%%%%%%%%%%%%%%
Here, $A(\vb*{x})$ is the \emph{site tensor} and $| \Omega \rangle$ is the Fock vacuum of the virtual fermions.
Introducing the virtual fermions at the two ends of each link corresponds to placing four kinds of virtual fermions at each site, associated with the four directions (right, up, left, and down) as shown in Fig.~\ref{fig:GGPEPS-ansatz}.
$\mathcal{U}_{\mathcal{G}}$ (denoted by $U_q$ in Fig.~\ref{fig:GGPEPS-ansatz}) is the \emph{link tensor}, which couples the gauge field on the link to the virtual fermions.
In addition to the virtual fermions, this tensor has a physical leg corresponding to the gauge-field configuration $q$ of the link.
The tensor $w$ is determined so as to connect the site tensor $A(\vb*{x})$ to the neighboring one, $A(\vb*{x}+\vb*{e}_i)$, while satisfying the Gauss law.
$| \psi_E \rangle$ is the no-flux state defined in Eq.~\eqref{eq:def-psi-E}.

In Eq.~\eqref{eq:Gauged-PEPS-ansatz}, 
the virtual fermions for the four directions, denoted by $\{ r, u, l, d\}$, are traced out, while the gauge-field degrees of freedom ($q$ in Fig.~\ref{fig:GGPEPS-ansatz}) remain untraced.
For pure gauge theory, both $|\psi_E \rangle$ and the factor $\langle \Omega | \prod_{\vb*{x},i} w^\dag (\vb*{x},i) \mathcal{U}_{\mathcal{G}} \prod_{\vb*{x}} A(\vb*{x}) | \Omega \rangle$ are invariant under the Gauss law operator $V(\vb*{x})$ in Eq.~\eqref{eq:phys-gauge-trans}, without explicitly solving the Gauss law, as will be explained in Sec.~\ref{sec:gauge-inv}.

Let us write down the explicit form of each operator in Eq.~\eqref{eq:Gauged-PEPS-ansatz}.
Each virtual fermion may have $F$ modes, which play the role of the bond in the tensor network.
We denote them generally by $\{ r_\alpha ^\dag (\vb*{x})\}_{\alpha=1}^F, \{ u_\alpha ^\dag (\vb*{x})\}_{\alpha=1}^F, \{ l_\alpha ^\dag (\vb*{x})\}_{\alpha=1}^F, \{ d_\alpha ^\dag (\vb*{x})\}_{\alpha=1}^F$, and collectively write these operators as $\{ a_\alpha ^\dag (\vb*{x})\}_{\alpha=1}^{4F}$.
Using $\{ a_\alpha ^\dag (\vb*{x})\}_{\alpha=1}^{4F}$, we define the site tensor, $A(\vb*{x})$, by
\beq
A(\vb*{x})= \exp \left( T_{\alpha \beta} a_\alpha^\dag a_{\beta}^\dag (\vb*{x}) \right).
\eeq
Here, $T_{\alpha \beta}$ can in principle be chosen independently at each site, but we impose translational invariance and take it to be site-independent.

For $F=1$, which is the minimal ansatz of Ref.~\cite{Emonts:2022yom}, $T_{\alpha \beta}$ is a $4 \times 4$ matrix.
Because of the anticommutation relations of the creation operators $a^\dag$, it can be written as an antisymmetric matrix with six complex parameters.
We further impose the condition corresponding to rotational symmetry on the square lattice, namely, that a $90$-degree rotation maps the state to itself up to a phase.
This is the $C_4$ invariance, $R^T T R = T$, where $R$ is the $90$-degree rotation matrix acting on the leg indices.
This leaves only two complex parameters.
The explicit form is given in Eq.~\eqref{eq:def-T-matrix-F1}.

In the case of the $F=2$ ansatz, namely, the optimized ansatz of Ref.~\cite{Emonts:2022yom}, $T$ is initially an $8 \times 8$ matrix, and antisymmetry allows $\binom{8}{2}=28$ complex parameters.
Imposing the $C_4$ invariance in the same way reduces them to eight complex parameters.
Their explicit form is given in Sec.~\ref{sec:more-F2}.

Next, we write down the explicit form of the link tensor $\mathcal{U}_{\mathcal{G}}$, which represents the coupling between the physical gauge-field degrees of freedom and the virtual fermions.
Using the eigenstates of $Q_l$, namely, in the q-basis, this operator can be expanded as
\beq
\mathcal{U}_{\mathcal{G}} (\vb*{x}, i) = \sum_q | q \rangle_{\vb*{x}, i} \langle q |_{\vb*{x},i} \otimes \mathcal{U}_q (\vb*{x},i). \label{eq:def-UG}
\eeq
Explicitly, we have
\beq
\mathcal{U}_q (\vb*{x},1) &=& \exp \left( i\pi q \sum_{\alpha=1}^{F} r_\alpha^\dag (\vb*{x}) r_\alpha (\vb*{x}) \right), \label{eq:def-Uq1}\\
\mathcal{U}_q (\vb*{x},2) &=& \exp \left( i\pi q \sum_{\alpha=1}^{F} u_\alpha^\dag (\vb*{x}) u_\alpha (\vb*{x}) \right).
\eeq
Thus, $\mathcal{U}_q$ has the usual form of a link variable $e^{iA}$ with $A=\pi q$ in the $x$ ($y$) direction, which couples to the number operator of $r_\alpha$ ($u_\alpha$).

Finally, $w(\vb*{x},i)$ is determined so as to connect the site tensor $A(\vb*{x})$ and the link tensor $\mathcal{U}_{\mathcal{G}}(\vb*{x},i)$ introduced above to the next site, while satisfying the Gauss law (flux conservation) and $C_4$ invariance.
In general, it can be written as 
\beq
w(\vb*{x},1)&=&\exp \left( W^1_{\alpha \beta} l^\dag_\alpha (\vb*{x}+\vb*{e}_1) r^\dag_{\beta} (\vb*{x}) \right), \\
w(\vb*{x},2)&=&\exp \left( W^2_{\alpha \beta} d^\dag_\alpha (\vb*{x}+\vb*{e}_2) u^\dag_{\beta} (\vb*{x}) \right).
\eeq
Here again, using translational invariance, the coefficient matrices $W^1$ and $W^2$ can be taken to be site-independent, and $C_4$ invariance relates $W^1$ to $W^2$.
Note that the Gauss law does not fix the form of $W^1$ and $W^2$ further, and several choices are allowed.
Our choice is given in Sec.~\ref{sec:more-F2}.

\begin{figure}[h]
\centering
\includegraphics[scale=0.45]{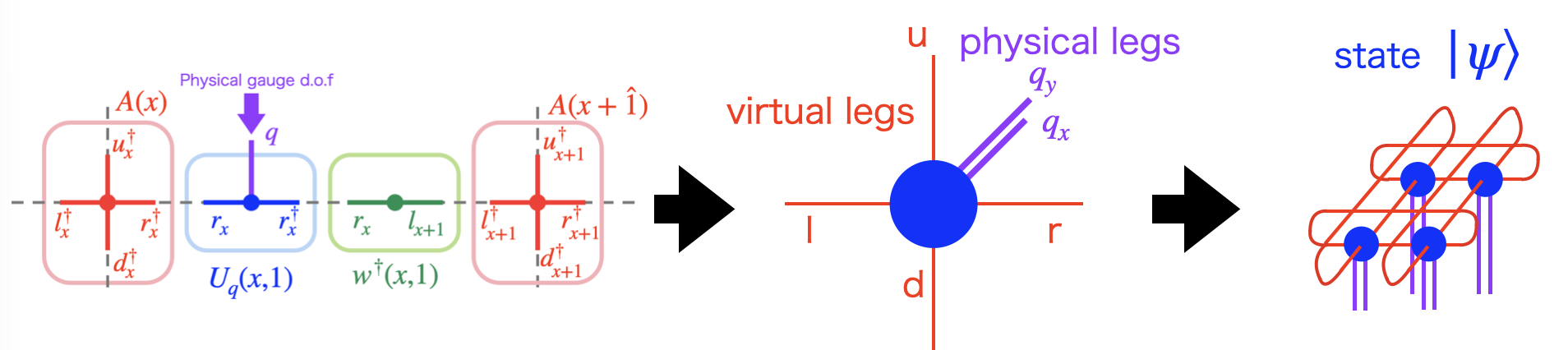}
\caption{Virtual and physical legs of the GGFPEPS ansatz.}\label{fig:state-prep}
\end{figure}

To obtain the GGFPEPS ansatz, $|\psi \rangle$, we first trace out the virtual fermions between two sites in each of the $x,y$ directions, as shown in Fig.~\ref{fig:state-prep}.
%%%%%%%%%%%%%%%%%%%%%%%%%%%%%%
%%%%%%%%%%%%%%%%%%%%%%%%%%%%%%
This leaves the physical legs $q_x,q_y$ at each site (more precisely, on the links), as shown in the middle panel of Fig.~\ref{fig:state-prep}.
By further tracing out all virtual fermions in the two-dimensional space with periodic boundary conditions, we obtain a state with two physical legs remaining at each site, as shown in the right panel of Fig.~\ref{fig:state-prep}.
This corresponds to Eq.~\eqref{eq:Gauged-PEPS-ansatz}.

Note that the legs remaining in the state $|\psi\rangle$ are physical legs, described in the q-basis as in Eq.~\eqref{eq:def-UG}.
Therefore, the operators for physical observables below must also be written as tensors in the q-basis.

%%%%%%%%%%%%%%%%%%%%%%%%%%%%%%%%%%%%%%%%%%%%%%%%%%%%%%%%%%
\subsection{Gauge invariance}\label{sec:gauge-inv}
%%%%%%%%%%%%%%%%%%%%%%%%%%%%%%%%%%%%%%%%%%%%%%%%%%%%%%%%%%
We now show that the state in Eq.~\eqref{eq:Gauged-PEPS-ansatz} is invariant under the Gauss law operator, namely, that it satisfies Eq.~\eqref{eq:Gauge-inv}.
We write Eq.~\eqref{eq:Gauged-PEPS-ansatz} as
\beq
|\psi \rangle &=& \langle \Omega | \prod_{\vb*{x},i} w^\dag (\vb*{x},i) \mathcal{U}_{\mathcal{G}} \prod_{\vb*{x}} A(\vb*{x}) | \Omega \rangle |\psi_E\rangle, \nonumber \\
&\equiv& \mathcal{O} |\psi_E\rangle. \label{eq:def-O}
\eeq
Since $|\psi_E\rangle = \otimes_l |p=0 \rangle$ is a product of eigenstates of $P_l$ with eigenvalue $+1$ on each link,
\beq
V(\vb*{x}) |\psi_E\rangle = |\psi_E\rangle
\eeq
holds trivially.
Thus, the question reduces to whether $V(\vb*{x})$ and $\mathcal{O}$ commute.
If $[V(\vb*{x}), \mathcal{O}]=0$, then
\beq
V(\vb*{x}) |\psi\rangle &=& V(\vb*{x}) \mathcal{O} | \psi_E \rangle, \nonumber\\
&=& \mathcal{O} V(\vb*{x}) | \psi_E \rangle = | \psi \rangle.
\eeq
Therefore, it is sufficient to show $V(\vb*{x}) \mathcal{O} V^\dag(\vb*{x})=\mathcal{O}$ to establish Eq.~\eqref{eq:Gauge-inv}.

The physical gauge-field degrees of freedom exist in $\mathcal{O}$ only through $\mathcal{U}_{\mathcal{G}}$, while the remaining factors are written in terms of virtual fermions.
In analogy with the Gauss law operator acting on the physical degrees of freedom, we define a local symmetry operator acting on the virtual fermions as
\beq
\mathcal{V}(\vb*{x}) a^\dag_\alpha(\vb*{x}) \mathcal{V}^\dag (\vb*{x})= - a^\dag_{\alpha} (\vb*{x}). \label{eq:virtual-V}
\eeq
Here, $\mathcal{V}(\vb*{x}) = \mathcal{V}^\dag (\vb*{x})$, and we take it to be a product of operators acting on the virtual fermions in each direction,
\beq
\mathcal{V}(\vb*{x}) = \mathcal{V}_r(\vb*{x}) \mathcal{V}_u(\vb*{x}) \mathcal{V}_l(\vb*{x}) \mathcal{V}_d(\vb*{x}).
\eeq
Each directional operator, such as $\mathcal V_r$, only changes the sign of the creation operators and leaves the vacuum invariant.
In the standard representation satisfying Eq.~\eqref{eq:virtual-V}, it can be written as $\mathcal V_r(\vb*{x})=\prod_\alpha\left(1-2 r^\dag_\alpha r_\alpha(\vb*{x})\right)=(-1)^{N_r(\vb*{x})}$ with $N_r(\vb*{x})=\sum_\alpha r^\dag_\alpha r_\alpha(\vb*{x})$, which measures the fermion parity.
Since the Fock vacuum $|\Omega \rangle$ has no virtual fermions, it is invariant,
\beq
\mathcal V_r(\vb*{x})|\Omega\rangle=|\Omega\rangle.
\eeq
The site tensor $A(\vb*{x})$ is written in Gaussian form in terms of virtual fermions and has even fermion parity.
It therefore transforms as
\beq
\mathcal V(\vb*{x})A(\vb*{x})\mathcal V^\dagger(\vb*{x})=A(\vb*{x}).\label{eq:A-trans}
\eeq

On the other hand, $\mathcal{U}_{\mathcal{G}}$ contains the physical degrees of freedom and transforms under the link operator $P(\vb*{x},i)$.
When $P(\vb*{x},i)$ acts on $|q\rangle_{\vb*{x},i}$ in Eq.~\eqref{eq:def-UG}, it gives $P|q\rangle = | q-1 \rangle$.
This can be regarded as a transformation of the virtual fermions introduced above.
Indeed, using $N_r(\vb*{x})$ defined above, we can write $\mathcal{U}_q(\vb*{x},1)$ in Eq.~\eqref{eq:def-Uq1} as
\beq
\mathcal{U}_q(\vb*{x},1) = \exp \left(i\pi q N_r(\vb*{x})\right)= \left(\mathcal{V}_r(\vb*{x})\right)^q,
\eeq
and obtain
\beq
P(\vb*{x},1)\mathcal{U}_{\mathcal{G}}(\vb*{x},1)P^\dagger(\vb*{x},1) &=& \sum_q | q-1 \rangle \langle q-1 | \otimes \left(\mathcal{V}_r(\vb*{x})\right)^q =\mathcal V_r(\vb*{x})\mathcal{U}_{\mathcal{G}}(\vb*{x},1),\nonumber\\
P(\vb*{x},2)\mathcal{U}_{\mathcal{G}}(\vb*{x},2)P^\dagger(\vb*{x},2) &=& \sum_q | q-1 \rangle \langle q-1 | \otimes \left(\mathcal{V}_u(\vb*{x})\right)^q =\mathcal V_u(\vb*{x})\mathcal{U}_{\mathcal{G}}(\vb*{x},2).\label{eq:U_G-trans}
\eeq
This is the essential step of the proof in this section. 
Namely, the transformation by $P$ on the physical leg can be rewritten as a transformation of the virtual fermions.\footnote{Equation~(22) of Ref.~\cite{Emonts:2022yom} reads $P(\vb*{x},1)\mathcal{U}_{\mathcal{G}}(\vb*{x},1)P^\dagger(\vb*{x},1)=\mathcal V_r(\vb*{x})\mathcal{U}_{\mathcal{G}}(\vb*{x},1)\mathcal V_r^\dagger(\vb*{x})$. Since $\mathcal V_r(\vb*{x})$ commutes with $\mathcal{U}_{\mathcal{G}}(\vb*{x},1)$, the right-hand side reduces to $\mathcal{U}_{\mathcal{G}}(\vb*{x},1)$ itself. We believe that this is a typographical error, and the gauge invariance of the ansatz is not affected.}

Since the $w$ tensor connects one virtual leg from each end of a link, the parity operation at one end can be transferred to the other end,
\beq
\mathcal V_r(\vb*{x}) w(\vb*{x},1) \mathcal V_r^\dagger(\vb*{x}) &=& \mathcal V_l(\vb*{x}+\vb*{e}_1) w(\vb*{x},1) \mathcal V_l^\dagger(\vb*{x}+\vb*{e}_1),\nonumber\\
\mathcal V_u(\vb*{x}) w(\vb*{x},2) \mathcal V_u^\dagger(\vb*{x}) &=& \mathcal V_d(\vb*{x}+\vb*{e}_2) w(\vb*{x},2) \mathcal V_d^\dagger(\vb*{x}+\vb*{e}_2).\label{eq:w-trans}
\eeq

Let us now consider $V(\vb*{x}) \mathcal{O}V^\dag (\vb*{x})$.
Since $V(\vb*{x})$ acts on the physical gauge field, it acts only on $\mathcal{U}_{\mathcal{G}}$ within $\mathcal{O}$,
\beq
V(\vb*{x}) \mathcal{O} V^\dagger(\vb*{x})
=\langle\Omega| \prod_{\vb*{x}',i} w^\dagger(\vb*{x}',i) \Big[V(\vb*{x})\mathcal{U}_{\mathcal{G}}V^\dagger(\vb*{x})\Big] \prod_{\vb*{x}'} A(\vb*{x}') |\Omega\rangle.
\eeq
Applying Eq.~\eqref{eq:U_G-trans} to the four $P$ operators in $V(\vb*{x})$, and then Eq.~\eqref{eq:w-trans} together with the invariance of $\langle\Omega|$ to move $\mathcal V_r(\vb*{x}-\vb*{e}_1)$ and $\mathcal V_u(\vb*{x}-\vb*{e}_2)$ to $\mathcal V_l(\vb*{x})$ and $\mathcal V_d(\vb*{x})$, we obtain
\beq
\langle\Omega| \prod_{\vb*{x}',i} w^\dagger(\vb*{x}',i) \Big[V(\vb*{x})\mathcal{U}_{\mathcal{G}}V^\dagger(\vb*{x})\Big]
=\langle\Omega|\prod_{\vb*{x}',i}w^\dagger(\vb*{x}',i) \mathcal V(\vb*{x}) \mathcal{U}_{\mathcal{G}}.
\eeq
Thus, the transformation can be rewritten as a local transformation of the virtual fermions at site $\vb*{x}$.

The operator $\mathcal{V}(\vb*{x})$ commutes with $\mathcal{U}_{\mathcal{G}}$, so that it finally acts on $\prod_{\vb*{x}'} A(\vb*{x}')$.
The invariance of the $A(\vb*{x})$ tensor, Eq.~\eqref{eq:A-trans}, and of the Fock vacuum gives
\beq
\mathcal{V}(\vb*{x})\prod_{\vb*{x}'} A(\vb*{x}') | \Omega \rangle = \prod_{\vb*{x}'} A(\vb*{x}') \mathcal{V}(\vb*{x})|\Omega \rangle = \prod_{\vb*{x}'} A(\vb*{x}') | \Omega \rangle.
\eeq
We conclude that $V(\vb*{x}) \mathcal{O} V^\dag (\vb*{x})=\mathcal{O}$, and thus the GGFPEPS ansatz satisfies Eq.~\eqref{eq:Gauge-inv}.

%%%%%%%%%%%%%%%%%%%%%%%%%%%%%%%%%%%%%%%%%%%%%%%%%%%%%%%%%%
\subsection{$F=2$ ansatz state}\label{sec:more-F2}
%%%%%%%%%%%%%%%%%%%%%%%%%%%%%%%%%%%%%%%%%%%%%%%%%%%%%%%%%%
In the $F=2$ ansatz, each link carries two virtual fermion modes.
There is a freedom in how the modes $\{r,u,l,d\}$ of the two kinds are connected by the $A$ and $w$ tensors.
In this subsection, we fix this freedom so that the one-parameter family of the plaquette product states introduced in Sec.~\ref{SCE} is realized.

The GGFPEPS ansatz must reproduce the known ground states at least in the weak and strong coupling limits.
The strong coupling ground state $| \psi_E \rangle$ is trivially reproduced for any $F$-mode ansatz by setting all variational parameters to $0$.
On the other hand, at $\lambda=0$, the ansatz must reproduce the toric code ground state $|\psi_B\rangle$ in Eq.~\eqref{eq:def-psi-B}, which is expanded as
\beq
|\psi_B\rangle=\mathcal{N} \prod_p(1+B_p)| \psi_E \rangle= \mathcal{N} \sum_S \prod_{p \in S} B_p |\psi_E \rangle.
\eeq
In this state, all subsets $S$ of plaquettes appear with equal weight $1$.
The key question is therefore whether the variational parameters can be chosen so that all the coefficients of $\prod_{p\in S}B_p$ in this expression are $1$.

The $F=1$ ansatz can reproduce the toric code ground state, as shown in Appendix~\ref{sec:F1-ansatz}, although Ref.~\cite{Emonts:2022yom} states that $F=1$ is not sufficient to represent the weak coupling ground state.
Around the critical coupling, however, it fails to reproduce the ED results even for small $L$, as reported in Ref.~\cite{Emonts:2022yom} and confirmed in Appendix~\ref{sec:F1-ansatz}.
This indicates that its variational parameter space is too small.
Therefore, we consider the $F=2$ ansatz.

To be specific, we consider the plaquette product family of Sec.~\ref{SCE}, which contains both the strong coupling ground state $|\psi_E\rangle$ and $|\psi_B\rangle$ above,
\beq
|\psi(t) \rangle &\propto& \prod_p (1+t B_p) | \psi_E \rangle.
\eeq
At the leading order of the strong coupling expansion, $t$ corresponds to $1/8\lambda^2$.
It becomes $|\psi_E \rangle$ at $t = 0$ and $|\psi_B \rangle$ at $t= 1$.

\begin{figure}[htb]
\centering
\includegraphics[scale=0.5]{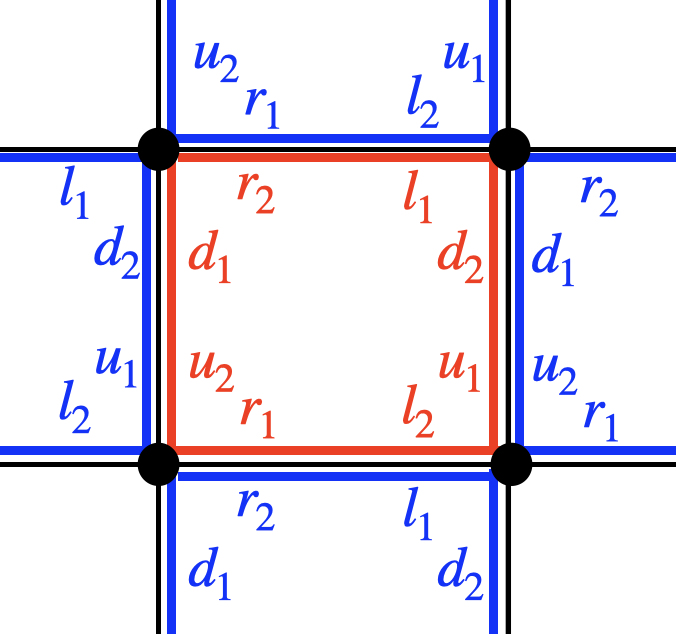}
\caption{Construction of a plaquette in the $F=2$ ansatz. The two modes on each link are shown in red and blue. The red modes form the loop around the central plaquette, and the blue modes belong to the neighboring plaquettes.}\label{fig:def-corner}
\end{figure}

An efficient way to realize the plaquette product family within the $F=2$ ansatz is to allow each plaquette to be excited independently, namely, to associate the two modes on each link with the excitations of the two plaquettes sharing that link.
To realize this, we define the $A$ tensor and the $w$ tensors connecting the links so that virtual fermions with different modes are connected at each corner of a plaquette, as shown in Fig.~\ref{fig:def-corner}.
%%%%%%%%%%%%%%%%%%%%%%%%%%%%%%
%%%%%%%%%%%%%%%%%%%%%%%%%%%%%%
One of the two modes on each link (red) is used for the loop around the central plaquette, while the other (blue) is used for the loops around the neighboring plaquettes.
The neighboring plaquettes can be excited independently of the central one.
This is the essential point of the $F=2$ ansatz.

Let us write down this construction explicitly.
The $C_4$-invariant form of the matrix $T_{\alpha\beta}$ in the $A(\vb*{x})$ tensor with eight complex parameters is given by
\beq
T=
\begin{pmatrix}
0 & -z_1 & -i y_1 & -iz_1 & ia & ib & ic & id \\
z_1 & 0 & -iz_1 & y_1 & -d & -a & -b & -c \\
iy_1 & iz_1 & 0 & z_1 & -ic & -id & -ia & -ib \\
iz_1 & -y_1 & -z_1 & 0 & b & c & d & a \\
-ia & d & ic & -b & 0 & -z_2 & -iy_2 & -iz_2 \\
-ib & a & id & -c & z_2 & 0 & -iz_2 & y_2 \\
-ic & b & ia & -d & iy_2 & iz_2 & 0 & z_2 \\
-id & c & ib & -a & iz_2 & -y_2 & -z_2 & 0
\end{pmatrix},\label{eq:def-T-matrix-F2}
\eeq
where the legs are ordered as $(r_1, u_1, l_1, d_1, r_2, u_2, l_2, d_2)$.
The parameters $y_f$ and $z_f$ with $f=1,2$ are those of the $F=1$ ansatz within each mode.
The remaining four parameters connect the two modes.
The same-leg type $a$ couples $(r_1,r_2)$, $(u_1,u_2)$, $(l_1,l_2)$, and $(d_1,d_2)$.
The cross-corner type $b$ couples $(r_1,u_2)$, $(u_1,l_2)$, $(l_1,d_2)$, and $(d_1,r_2)$.
The mixed straight-line type $c$ couples $(r_1,l_2)$, $(u_1,d_2)$, $(l_1,r_2)$, and $(d_1,u_2)$.
The reverse cross-corner type $d$ couples $(r_1,d_2)$, $(u_1,r_2)$, $(l_1,u_2)$, and $(d_1,l_2)$.
The way of connecting the modes in Fig.~\ref{fig:def-corner} corresponds to keeping only the cross-corner type parameter $b$ and setting all the other parameters to zero.

As for the matrices $W^1$ and $W^2$ in $w(\vb*{x},i)$, the two modes $f=1$ and $f=2$ are exchanged across the link in Fig.~\ref{fig:def-corner}, namely, $l_1^\dag$ is contracted with $r_2^\dag$ and $l_2^\dag$ with $r_1^\dag$.
This corresponds to
\beq
W^1 = \sigma_x, \quad W^2 = \eta^2 \sigma_x. \label{eq:def-W-matrix}
\eeq
Here, $\eta$ is required to satisfy $\eta^4=-1$, which still leaves an ambiguity in its choice.
In Ref.~\cite{Emonts:2022yom}, the choice $\eta=e^{i\pi/4}$ was adopted and we follow this.

Contracting the $A$ tensors at the four corners and the $w$ tensors on the four links around a single plaquette, we obtain
\beq
|\psi_\square\rangle = \left( 1 - 16 b^4 Q_{l_1} Q_{l_2} Q_{l_3} Q_{l_4} \right) |\psi_E\rangle,
\eeq
where the factor $-16b^4$ is the product of the four corner factors $2ib$, $-2b$, $-2ib$, and $2b$.
Since each plaquette can be excited independently in this construction, the full state is the plaquette product state $|\psi(t)\rangle$ with $t=-16b^4$.
The toric code ground state is obtained when 
\beq
b^4=-1/16, \quad \mbox{the others}=0. \label{eq:f-b-relation}
\eeq
Note that a subset $S$ and its complement give the same state because of $\prod_pB_p=1$ on the torus.
This redundancy appears in the singular value spectrum of HOTRG in Sec.~\ref{sec:sv-hotrg}.

There are other constructions of the toric code ground state with the same number of parameters as the $F=2$ ansatz.
Indeed, the realization of the toric code ground state for $F=1$ shown in Appendix~\ref{sec:F1-ansatz} differs from the present one, and taking two copies of the $F=1$ ansatz is one possibility.
This corresponds to $W^1=1$ and $W^2=\eta^2$, with which the two modes are never mixed.
In that case, however, the state is not of the plaquette product form, and hence it is not connected to the one-parameter family $|\psi(t)\rangle$.
Assuming the plaquette product form, on the other hand, allows us to extend the construction systematically to higher dimensions and obtain a sufficient condition on the size of $F$.
Indeed, in $(d+1)$ dimensions, each link belongs to $2(d-1)$ plaquettes, so a similar plaquette product state can be constructed with $F=2(d-1)$.
For $d=3$, $F=4$ is sufficient.

The discussion so far concerns the two limits $\lambda=0,\infty$.
The plaquette product family reproduces the ground states in both limits.
However, there is no guarantee that the true ground state in the intermediate region belongs to this family.
A larger variational space is probably needed there, as the failure of the $F=1$ ansatz in Appendix~\ref{sec:F1-ansatz} suggests.
We can gain indirect insight into this question by examining numerically whether parameters other than $b$ take nonzero values in Sec.~\ref{sec:results-finite-vol}.
In the next subsection, we analytically calculate $E_{tot}$ under the assumption of the plaquette product form to prepare a more direct test in terms of a physical quantity.

%%%%%%%%%%%%%%%%%%%%%%%%
\subsubsection{Variational energy within the plaquette product family}\label{sec:E-fam}
As a preparation for the comparison with the numerical results in Sec.~\ref{sec:comp-plaq-prod-state}, let us calculate the minimum energy attainable under the assumption of the plaquette product form.~\footnote{After posting the first version of this paper on arXiv, we became aware that an equivalent one-parameter variational ansatz and its energy minimization had already been studied in Ref.~\cite{Dusuel:2015sta}. }

We consider the thermodynamic limit $n_p\to\infty$, where the contribution of $\prod_pB_p=1$ on the torus is neglected since it is suppressed by $(2t/(1+t^2))^{n_p}$ for $t<1$.
The norm is then
\beq
\langle\psi(t)|\psi(t)\rangle = \langle\psi_E|\prod_p(1+tB_p)^2|\psi_E\rangle = (1+t^2)^{n_p},
\eeq
since $\langle\psi_E|B_p|\psi_E\rangle=0$.
For the magnetic term of the Hamiltonian, using $(1+tB_p)^2B_p=(1+t^2)B_p+2t$ ($B_p^2=1$), we obtain
\beq
\langle B_p\rangle = \frac{2t(1+t^2)^{n_p-1}}{(1+t^2)^{n_p}}=\frac{2t}{1+t^2}, \qquad
\langle 1-B_p\rangle = \frac{(1-t)^2}{1+t^2}.
\eeq
For the electric term, $P_l$ anticommutes with $B_p$ on the two plaquettes containing the same link $l$ and commutes with the others, so that
\beq
\langle P_l\rangle = \frac{(1-t^2)^2(1+t^2)^{n_p-2}}{(1+t^2)^{n_p}}=\frac{(1-t^2)^2}{(1+t^2)^2}, \qquad
\langle 1-P_l\rangle = \frac{4t^2}{(1+t^2)^2}.
\eeq

Since the number of links is $n_l=2n_p$, the energy density per plaquette is
\beq
E_{\rm fam}(t;\lambda) = \frac{8\lambda t^2}{(1+t^2)^2}+\frac{1}{\lambda}\frac{(1-t)^2}{1+t^2}. \label{eq:E-fam}
\eeq
Because of $\prod_pB_p=1$ on the torus, $|\psi(t)\rangle$ and $|\psi(1/t)\rangle$ are the same state, and $E_{\rm fam}$ is indeed invariant under $t\rightarrow1/t$.
To minimize it with respect to $t$, we introduce $s\equiv t/(1+t^2)$, which increases monotonically from $0$ to $1/2$ as $t$ goes from $0$ to $1$.
Using $(1-t)^2/(1+t^2)=1-2s$, Eq.~\eqref{eq:E-fam} becomes a quadratic function of $s$,
\beq
E_{\rm fam} = \frac{1}{\lambda}-\frac{2}{\lambda}s+8\lambda s^2,
\eeq
whose minimum is at $s=1/8\lambda^2$.
This is within the allowed range $s\le1/2$ only for $\lambda\ge1/2$.
We therefore obtain
\beq
\min_t E_{\rm fam}(t;\lambda)=
\begin{cases}
\dfrac{1}{\lambda}\left(1-\dfrac{1}{8\lambda^2}\right), & \lambda\ge\dfrac12, \quad 
t_{\rm opt}=4\lambda^2\left(1-\sqrt{1-\dfrac{1}{16\lambda^4}}\right),\\
2\lambda \quad (t_{\rm opt}=1,\ \text{toric code}), & \lambda<\dfrac12.
\end{cases}\nonumber \\
\label{eq:E-fam-min}
\eeq
For $\lambda\ge1/2$, the minimum coincides with the strong coupling expansion, Eq.~\eqref{eq:Etot-strong-coupling}, up to $\mathcal{O}(\lambda^{-7})$, and $t_{\rm opt}\to1/8\lambda^2$ as $\lambda\to\infty$, consistently with Eq.~\eqref{eq:plaq-prod-state-SC}.
For $\lambda<1/2$, on the other hand, the family cannot lower the energy below that of the toric code ground state, $2\lambda$, whereas the weak coupling expansion, Eq.~\eqref{eq:wc-energy-result}, gives $E_{tot}^{(WC)}(\lambda) \approx 2\lambda-\lambda^3/2$.
The winding 't Hooft loop within the family is also obtained from Eq.~\eqref{eq:thooft-LO} by replacing $1/8\lambda^2$ with $t_{\rm opt}$.
The corresponding $\alpha$ diverges as $t_{\rm opt}\rightarrow1$, i.e., $\langle W\rangle=0$ for $\lambda<1/2$.

%%%%%%%%%%%%%%%%%%%%%%%%%%%%%%%%%%%%%%%%%%%%%%%%%%%%%%%%
%%%%%%%%%%%%%%%%%%%%%%%%%%%%%%%%%%%%%%%%%%%%%%%%%%%%%%%%
\section{Physical observables}\label{sec:def-obs}
%%%%%%%%%%%%%%%%%%%%%%%%%%%%%%%%%%%%%%%%%%%%%%%%%%%%%%%%
%%%%%%%%%%%%%%%%%%%%%%%%%%%%%%%%%%%%%%%%%%%%%%%%%%%%%%%%
\subsection{Calculation of local operators $\langle \hat{O} \rangle$}\label{sec:local-O}
%%%%%%%%%%%%%%%%%%%%%%%%%%%%%%%%%%%%%%%%%%%%%%%%%%%%%%%%
We consider lattices whose extent in each direction is $L=2+2^k$ with $k=0,1,2,\cdots$, in addition to $L=2$, which is contracted exactly.
%%%%%%%%%%%%%%%%%%%%%%%%%%%%%%%%%%%%
\begin{figure}[h]
\centering
\includegraphics[scale=0.45]{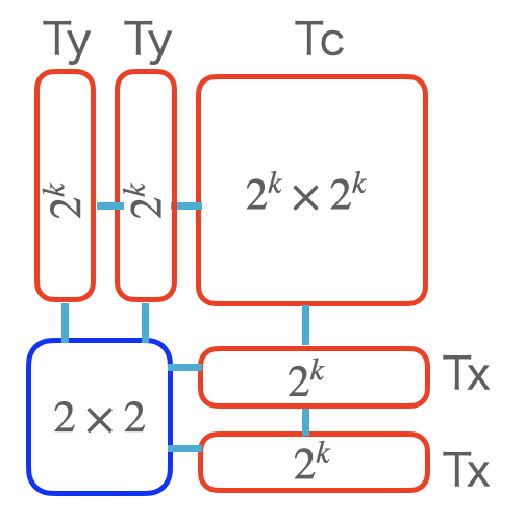}\quad \quad
\includegraphics[scale=0.5]{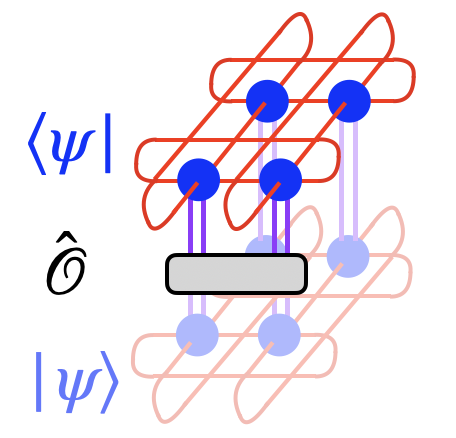}
\caption{Evaluation of a local observable $\hat{O}$.
(Left) Decomposition of the $L\times L=(2+2^k)\times(2+2^k)$ lattice. The $2\times2$ part (blue square) is treated as the local cell, and the remaining part (red) as the environment tensor.
(Right) To evaluate $\langle\hat{O}\rangle$, the local operator $\hat{O}$ is sandwiched between the GGFPEPS ansatz and its conjugate.}
\label{fig:eval-O-HOTRG}
\end{figure}
%%%%%%%%%%%%%%%%%%%%%%%%%%%%%%%%%%%%
As shown in the left panel of Fig.~\ref{fig:eval-O-HOTRG}, we decompose the $(2+2^k)\times(2+2^k)$ lattice into a $2\times2$ local cell (blue square) and the remaining region (red), and apply the HOTRG method~\cite{Xie:2012mjn} to the parts of size $2^k$ in the remaining region to reduce the computational cost.
The remaining region is treated as the environment tensor, in which both the virtual fermions and the physical legs ($q_x,q_y$) are contracted.
A local observable $\hat{O}$ is inserted in the $2\times2$ local cell, and its expectation value is evaluated by sandwiching $\hat{O}$ between the GGFPEPS ansatz and its conjugate and contracting the whole network, as shown in the right panel of Fig.~\ref{fig:eval-O-HOTRG}.

The HOTRG, which will be explained in Sec.~\ref{sec:hotrg}, is applied only to the environment region without operator insertions.
The tensors with operator insertions remain in the $2\times2$ local cell without being coarse-grained and are contracted with the environment tensor at the final step.
Since the physical legs of the state $|\psi\rangle$ are in the q-basis, the actual calculations use the q-basis operators $P=\sigma_x$ and $Q=\sigma_z$.

We define the energy density per plaquette as
\beq
E_{tot}=\left[ n_l \lambda (1-\langle P \rangle) + n_p \frac{1}{\lambda}(1-\langle Q_{l_1} Q_{l_2} Q_{l_3} Q_{l_4} \rangle) \right] / n_p.\label{eq:def-Etot}
\eeq
We also write the electric and magnetic parts separately as
\beq
E_e = n_l \lambda (1-\langle P \rangle)/n_p, \qquad
E_m = \frac{1}{\lambda}(1-\langle Q_{l_1} Q_{l_2} Q_{l_3} Q_{l_4} \rangle),
\eeq
so that $E_{tot}=E_e+E_m$.
In practice, we measure one link in the $x$ direction and one plaquette within the $2\times2$ local cell, and multiply the results by $n_l$ and $n_p$, respectively, assuming translational symmetry.
Only for the exact contraction at $L=2$, we measure one link in each of the $x$ and $y$ directions and take their average.

%%%%%%%%%%%%%%%%%%%%%%%%%%%%%%%%%%%%%%%%%%%%%%%%%%%%%%%%
\subsection{Wilson loop and winding 't Hooft loop operators}
\label{sec:def-winding-loop}
%%%%%%%%%%%%%%%%%%%%%%%%%%%%%%%%%%%%%%%%%%%%%%%%%%%%%%%%
The Wilson loop is defined as the product of the operators $Q_l$ along a contractible loop $C$ in two-dimensional space,
\beq
\mathcal{W}(C) \mathrel{\mathop:}= \prod_{l \in C} Q_l .
\eeq
Its expectation value is straightforward to evaluate in the q-basis, since
\beq
\mathcal{W}(C) | \{q_l\} \rangle = \prod_{l \in C} (-1)^{q_l} |\{q_l\} \rangle
\eeq
for a basis state $|\{q_l\}\rangle=\otimes_l|q_l\rangle$, so that it can be calculated from the eigenvalues $q_l$ on the links along $C$.
The evaluation of the Wilson loop for gauge-invariant PEPS has been studied in Ref.~\cite{Zohar:2021wqy}, and the corresponding formula for the GGFPEPS ansatz is given in Ref.~\cite{Emonts:2022yom}.

Here, we consider another quantity, namely, the winding 't Hooft loop operator, $W$, defined in Eq.~\eqref{eq:def-W} as a product of the operators $P_l$ along one spatial direction of the torus.
There, we took $P_l=\sigma_z$ in the p-basis.
In the actual calculations, however, we use the q-basis, in which $P_l$ is represented by $\sigma_x$.
We therefore calculate the expectation value by inserting $P_l$ on the links at $y_0=0$ in Fig.~\ref{fig:eval-O-HOTRG}, in both the local cell and the environment region.

In the HOTRG, the string of operators for the winding 't Hooft loop passes through the environment region as well.
We therefore coarse-grain the row tensors with operator insertions in the $x$ direction.
Of the two row tensors in Tx in Fig.~\ref{fig:eval-O-HOTRG}, the one at $y=y_0$ is replaced by this coarse-grained tensor.
For the coarse-graining, we use the same isometry $U$ as for the environment tensor.
This isometry is determined from the tensor without operator insertions by the HOSVD~\cite{Xie:2012mjn}, namely, the singular value decomposition (SVD) of the tensor reshaped into a matrix with ``leg $k$'' $\times$ ``all the remaining legs,'' which gives the isometry $U_k$ for leg $k$.

The use of the same isometry is a kind of approximation.
The transformation $U$ changes the basis so as to retain the states with large weights in the bond space of the environment tensor.
The idea is therefore to use the same basis for the tensors with operator insertions, since they are contracted with the same environment tensor.
Depending on the observable, however, the tensor with operator insertions may have large weights in the subspace discarded by $U$, which is determined from the tensor without operator insertions.
In fact, this happens for the winding 't Hooft loop in the weak coupling region, where $\langle W\rangle$ becomes exactly zero, as we will see in Sec.~\ref{sec:thooft-large}.
There, however, $\langle W\rangle$ is of $\mathcal{O}(\lambda^{2L})$ in perturbation theory and practically zero, so that the approximation does not affect the physical conclusion.

%%%%%%%%%%%%%%%%%%%%%%%%%%%%%%%%%%%%%%%%%%%%%%%%%%%%%%%%
\subsection{Topological entanglement entropy (TEE)}
%%%%%%%%%%%%%%%%%%%%%%%%%%%%%%%%%%%%%%%%%%%%%%%%%%%%%%%%
In $(2+1)$ dimensions, the entanglement entropy (EE) of a subsystem $A$ can be written as
\beq
S_A=\beta \ell_A/a+S_{\rm topo}.
\eeq
Here, the first term represents the so-called area law with the boundary length $\ell_A$, although there is some ambiguity in how to count the boundary.
The constant term $S_{\rm topo}$ is universal and independent of the details of the boundary.
Following Ref.~\cite{Kitaev:2005dm} (see also Ref.~\cite{levin2006detecting}), let us consider three subsystems $A$, $B$, and $C$ and take the combination,
\beq
S_{\rm topo}=S_A+S_B+S_C-S_{AB}-S_{BC}-S_{CA}+S_{ABC}.
\label{eq:def-tee}
\eeq
This quantity, called the \emph{topological entanglement entropy} (TEE), provides a useful probe of topological phase transitions.
The TEE of the toric code state has also been studied by quantum computation~\cite{Satzinger:2021eqy, Sun:2022sux}.

\begin{figure}[htb]
\centering
\includegraphics[scale=0.5]{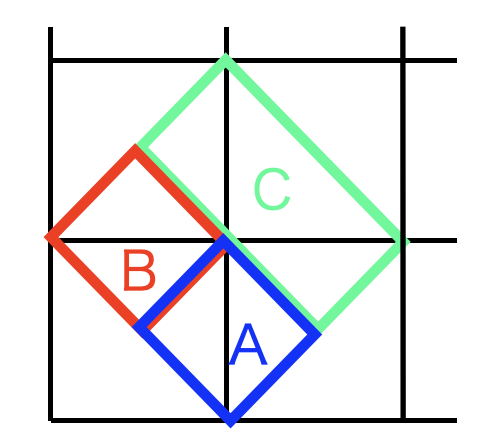}\quad\quad
\includegraphics[scale=0.5]{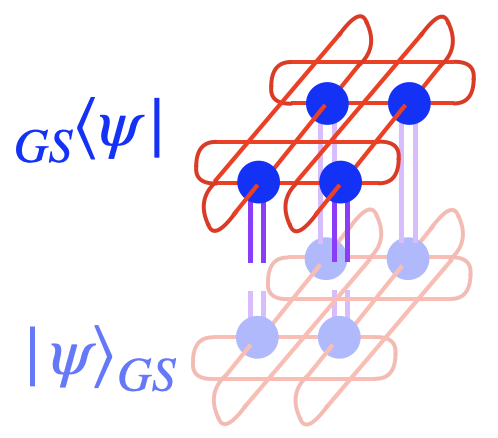}
\caption{(Left) Subsystems $A$, $B$, and $C$ used for the TEE. (Right) Evaluation of a component of the reduced density matrix.}\label{fig:eval-TEE}
\end{figure}

In this study, we use the subsystem arrangement shown in the left panel of Fig.~\ref{fig:eval-TEE}, where the subsystems $A$ and $B$ consist of one link each and $C$ of two links.
%%%%%%%%%%%%%%%%%%%%%%%%%%%%%%%%%%%%
%%%%%%%%%%%%%%%%%%%%%%%%%%%%%%%%%%%%
Since this arrangement fits within a $2\times 2$ lattice, we place it inside the local cell.

We first outline how to calculate the EE for each subsystem.
Writing the density matrix of the full system as $\rho=|\psi\rangle\langle\psi|/\langle\psi|\psi\rangle$, we define the reduced density matrix for the subsystem $A$ by
\beq
\rho_A={\rm Tr}_{\bar A}\rho.
\eeq
Using the product basis $|q\rangle_A$ of the links in the subsystem $A$, its matrix elements can be written as
\beq
(\rho_A)_{q'q}={}_A\langle q'|\rho_A|q\rangle_A
=\frac{\langle\psi|\big(|q\rangle\langle q'|\big)_A\otimes \mathbb{1}_{\bar A}|\psi\rangle}{\langle\psi|\psi\rangle}.
\eeq
Thus, the $(q',q)$ component of the reduced density matrix is obtained by the same procedure as for the local operators in Sec.~\ref{sec:local-O}.
Namely, we contract the physical legs in $\bar A$ between the ket and bra, and insert the local operator $|q\rangle\langle q'|$, a matrix whose only nonzero entry is $1$, on the physical legs in the subsystem $A$.
If the subsystem $A$ contains $n$ links, we measure all the components by varying $(q',q)$ over the $2^n\times2^n$ combinations.
We then calculate the eigenvalues $\mu_i$ of the resulting matrix $\rho_A$ and obtain the EE as
\beq
S_A=-\sum_i\mu_i\ln\mu_i.
\eeq

From the viewpoint of the TEE, the $\mathbb{Z}_N$ gauge theory has a gapped topological order with $S_{\rm topo}=-\ln N$, which diverges as $N \rightarrow \infty$.
This is consistent with the disappearance of the deconfined phase for $U(1)$.
Reference~\cite{Ghosh:2015iwa} provides a definition that allows a unified treatment of the constant term in the EE for discrete $\mathbb{Z}_N$ and continuous $U(1)$, and shows that the EE decomposes into a classical term originating from the edge modes/superselection sectors and a quantum term.
The constant term in the EE in the $(3+1)$-dimensional $U(1)$ deconfined phase was discussed in Ref.~\cite{Pretko:2015zva}.

%%%%%%%%%%%%%%%%%%%%%%%%%%%%%%%%%%%%%%%%%%%%%%%%%%%%%%%%%%
\section{Calculation methods: Grassmann tensor network and HOTRG}\label{sec:calc-methods}
%%%%%%%%%%%%%%%%%%%%%%%%%%%%%%%%%%%%%%%%%%%%%%%%%%%%%%%%%%
In the previous work~\cite{Emonts:2022yom}, the numerical calculations to obtain the ground state based on the GGFPEPS ansatz were performed using the covariance-matrix formalism for Gaussian states, with the gauge-field configurations $q$ sampled by the VMC.
In general, physical observables written in terms of fermionic operators are difficult to handle.
Indeed, an original motivation for imposing a Gaussian form on the virtual fermions in this ansatz was that such products of Gaussian states allow various physical observables to be calculated from covariance matrices.

In the present work, we take a different approach, in which the calculations are performed by mapping fermionic operators to classical Grassmann numbers.
This is the Grassmann TN approach~\cite{Gu:2010yh, Akiyama:2020sfo}, which makes direct use of the fact that the ansatz is built from virtual fermions.
Even in this approach, a Gaussian ansatz state has the advantage that the complicated minus signs arising from the interchange of fermionic operators are easier to handle.
Here, the sum over the gauge-field configurations $q$, which is sampled in the VMC, is carried out exactly as a part of the tensor contraction, and the whole network is contracted deterministically.
This allows us to examine precisely whether the ansatz remains valid over the whole range of the coupling $\lambda$, and in particular 
whether it accurately describes the system in the vicinity of the critical point, $\lambda_c$, obtained by continuous-time QMC~\cite{wu2012phase}.

For $L=2$ and $3$, the network can be contracted exactly.
For larger volumes, $L \geq 4$, we apply HOTRG~\cite{Xie:2012mjn} to the environment tensors, with the bond dimension truncated to $D_{\rm cut}$.
There is thus no statistical error from MC sampling.
Instead, a systematic error from the truncation to $D_{\rm cut}$ enters, which we control by examining the singular-value spectrum.

In this section, we outline the Grassmann TN in Sec.~\ref{sec:grassmann-TN} and HOTRG in Sec.~\ref{sec:hotrg}.
In Sec.~\ref{sec:sv-hotrg}, we also describe the properties of the singular values appearing in the SVD within the HOTRG process in the toric code ($\lambda=0$) case.

%%%%%%%%%%%%%%%%%%%%%%%%%%%%%%%%%%%%%%%%%%%%
\subsection{Grassmann tensor network}\label{sec:grassmann-TN}
%%%%%%%%%%%%%%%%%%%%%%%%%%%%%%%%%%%%%%%%%%%%
The Grassmann TN~\cite{Gu:2010yh, Akiyama:2020sfo} is an efficient way to extend the ordinary TN for bosonic systems to fermionic systems. 
The basic idea is to replace fermionic operators by classical Grassmann numbers. 
Fermionic operators obey the anticommutation relation
\beq
\{a_i^\dagger,a_j\}=\delta_{ij},\label{eq:def-anti-comm}
\eeq
while Grassmann numbers obey
\beq
\{\bar{\theta}_i,\theta_j\}=0.
\eeq
The constant on the right-hand side of Eq.~\eqref{eq:def-anti-comm} is
transferred to the measure of the Grassmann integral,
$\int d\bar\theta d\theta e^{-\bar\theta\theta}$, so that the contraction of fermionic operators, namely the inner product in Fock space, is written exactly as a Grassmann integral.

A tensor contraction is thus formally expressed as a Grassmann integral.
The calculation proceeds in two steps.
First, the fermionic operators defining each tensor are reordered into the form $\langle\Omega|(\text{annihilation operators})(\text{creation operators})|\Omega\rangle$.
The factors of $(-1)$ generated by this reordering are absorbed into the coefficients of the Grassmann tensor.
Second, the contractions between tensors are carried out as Grassmann integrals.
The signs associated with the order of contraction are supplied automatically by the anticommutativity of the Grassmann numbers.
Thus, we need not count the complicated signs by hand.

In the GGFPEPS ansatz, Eq.~\eqref{eq:Gauged-PEPS-ansatz}, $w^\dagger$ contains only annihilation operators, $A$ contains only creation operators, and $\mathcal U_q$ is a function of number operators.
All of them are written as exponentials of bilinear forms in the fermionic operators, and hence have even fermion parity.
The ansatz is therefore well suited to the Grassmann TN.

Our code is written in Julia from scratch, with the design of the Grassmann tensor operations following the Python package GrassmannTN~\cite{Yosprakob:2023flr}.

%%%%%%%%%%%%%%%%%%%%%%%%%%%%%%%%%%%%%%%%%%%%%%%%%%%%%%%%%%
\subsection{HOTRG and the singular-value spectrum}
\label{sec:hotrg}
%%%%%%%%%%%%%%%%%%%%%%%%%%%%%%%%%%%%%%%%%%%%%%%%%%%%%%%%%%
In this subsection, we explain the HOTRG applied to the environment tensors of size $2^k$.
In the HOTRG, the bond dimension is truncated based on the SVD.
We also discuss the structure of the singular values in the toric code ground state, where the meaning of the bond and the degeneracy of the singular values differ between the $F=1$ and $F=2$ ansatzes.
%%%%%%%%%%%%%%%%%%%%%%%%%%%%%%%%%%%%%%%%%%%
\subsubsection{Outline of HOTRG}
%%%%%%%%%%%%%%%%%%%%%%%%%%%%%%%%%%%%%%%%%%%
Let $S[r,u,l,d,q_x,q_y]$ denote the single-layer tensor shown in the middle panel of Fig.~\ref{fig:state-prep}, which has four types of virtual legs ($r$, $u$, $l$, and $d$) and two physical legs on each site.
Note that since each type of virtual leg carries $F$ virtual fermion modes, its dimension is $2^F$, namely $4$ for the $F=2$ ansatz.

HOTRG contracts the network of the norm $\langle\psi|\psi\rangle$ rather
than $|\psi\rangle$ itself. 
The tensor on each site, denoted \texttt{Tii}, is obtained by contracting the physical legs of $S$ (ket) with those of its complex conjugate $S^\ast$ (bra),
\begin{equation}
\texttt{Tii}[(r,r'),(u,u'),(l,l'),(d,d')]
=\sum_{q_x,q_y} S[r,u,l,d,q_x,q_y] S^\ast[r',u',l',d',q_x,q_y].
\label{eq:Tii}
\end{equation}
Thus, each leg of \texttt{Tii} is labeled by a pair (ket leg, bra leg),
e.g., $(r,r')$, and has dimension $2^F\times 2^F$.
The environment tensors \texttt{Tx}, \texttt{Ty}, and \texttt{Tc} in
Fig.~\ref{fig:eval-O-HOTRG} are obtained by coarse-graining \texttt{Tii} with HOTRG.

As an example, one coarse-graining step in the $y$ direction proceeds as follows.
Two neighboring tensors \texttt{Tii} are contracted along $y$, i.e., the $d$-leg of the upper tensor with the $u$-leg of the lower one.
The two right legs $(r,r')_{\rm top}$ and $(r,r')_{\rm bot}$ of the resulting two-site tensor are then combined into a single leg $R=\big((r,r')_{\rm top},(r,r')_{\rm bot}\big)$ of dimension $(2^F\times2^F)^2$, namely $256$ for the $F=2$ ansatz, and likewise the two left legs into $L=\big((l,l')_{\rm top},(l,l')_{\rm bot}\big)$.

To keep the bond dimension finite, $R$ is truncated to $D_{\rm cut}$.
For this purpose, the two-site tensor is reshaped into a matrix with row index $R$ and column index $(L,u,d)$,
\begin{eqnarray}
M_{R,(L,u,d)}&=&\sum_{(m,m')}
\texttt{Tii}_{\rm top}\big[(r,r')_{\rm top},(u,u'),(l,l')_{\rm top},(m,m')\big]
\nonumber\\
&&\times
\texttt{Tii}_{\rm bot}\big[(r,r')_{\rm bot},(m,m'),(l,l')_{\rm bot},(d,d')\big],
\label{eq:M}
\end{eqnarray}
where $(m,m')$ is the contracted intermediate leg, $(u,u')$ is the upper leg of the upper tensor, and $(d,d')$ is the lower leg of the lower tensor.
In the indices of $M$, we write $u$ and $d$ for $(u,u')$ and $(d,d')$.
We then diagonalize
\begin{equation}
(MM^\dagger)_{R,\tilde R}=\sum_{L,u,d} M_{R,(L,u,d)}M^\ast_{\tilde R,(L,u,d)},
\label{eq:MMdag}
\end{equation}
and denote the square roots of its eigenvalues in decreasing order by $s_1\ge s_2\ge\cdots$, which we call the singular values.
The eigenvectors corresponding to the largest $D_{\rm cut}$ eigenvalues define the isometry $U$ that projects $R$ onto a $D_{\rm cut}$-dimensional space.
The same procedure with $L$ in place of $R$ gives the isometry for the left leg, and the $x$-direction step is obtained by exchanging the roles of $x$ and $y$.

Note that the ket leg and the bra leg of \texttt{Tii} have the same fermion parity, so that the odd-parity block of each leg vanishes identically.
Indeed, $\mathcal U_q(\vb*{x},1)=(-1)^{qN_r(\vb*{x})}$ in Eq.~\eqref{eq:def-Uq1} involves only the gauge-field configuration $q$ of the link $(\vb*{x},1)$ and the $r$ modes at the starting site of that link, so that the $q_x$-dependent factor in the sum in Eq.~\eqref{eq:Tii} is
\beq
\sum_{q_x=0,1}(-1)^{q_xN_r}(-1)^{q_xN_{r'}}=1+(-1)^{N_r+N_{r'}},
\label{eq:parity-ket-bra}
\eeq
which equals 2 when $N_r+N_{r'}$ is even, and is 0 when it is odd.
Hence \texttt{Tii} can be nonzero only if the ket leg $r$ and the bra leg $r'$ have the same fermion parity.
The same holds for the other legs, and hence also for the $x$-direction step.
Consequently, the effective dimension of each leg of \texttt{Tii} is $2^{2F-1}$, namely $8$ for the $F=2$ ansatz, and the dimension of $R$ is $8\times8=64$.
Since every tensor has even parity, neither the contractions nor the construction of the HOTRG isometries mix the parity blocks.
We have also confirmed numerically that the odd-parity block of $MM^\dagger$ in Eq.~\eqref{eq:MMdag} vanishes, so that its nonzero part is a $64\times64$ matrix.
The isometry $U$ then projects this $64$-dimensional space onto $D_{\rm cut}$ dimensions, and the same truncation is repeated at each coarse-graining step.

%%%%%%%%%%%%%%%%%%%%%%%%%%%%%%%%%%%%%%%%%
\subsubsection{Degeneracy of the singular values in the toric code ground state}\label{sec:sv-hotrg}
%%%%%%%%%%%%%%%%%%%%%%%%%%%%%%%%%%%%%%%%%
We now consider the structure of the singular values at $\lambda=0$, where the state is the toric code ground state $|\psi_B\rangle$ in Eq.~\eqref{eq:def-psi-B}.
In the p-basis, it is an equal-weight superposition of all closed-loop configurations of the electric flux.
Consequently, when a bond is cut, every allowed configuration on the links crossing the cut appears with the same weight, and the corresponding singular values are all equal.
The leading singular values are therefore degenerate, and the degeneracy equals the number of independent configurations carried by the cut bond.

\begin{figure}[htb]
\centering
\includegraphics[scale=0.2]{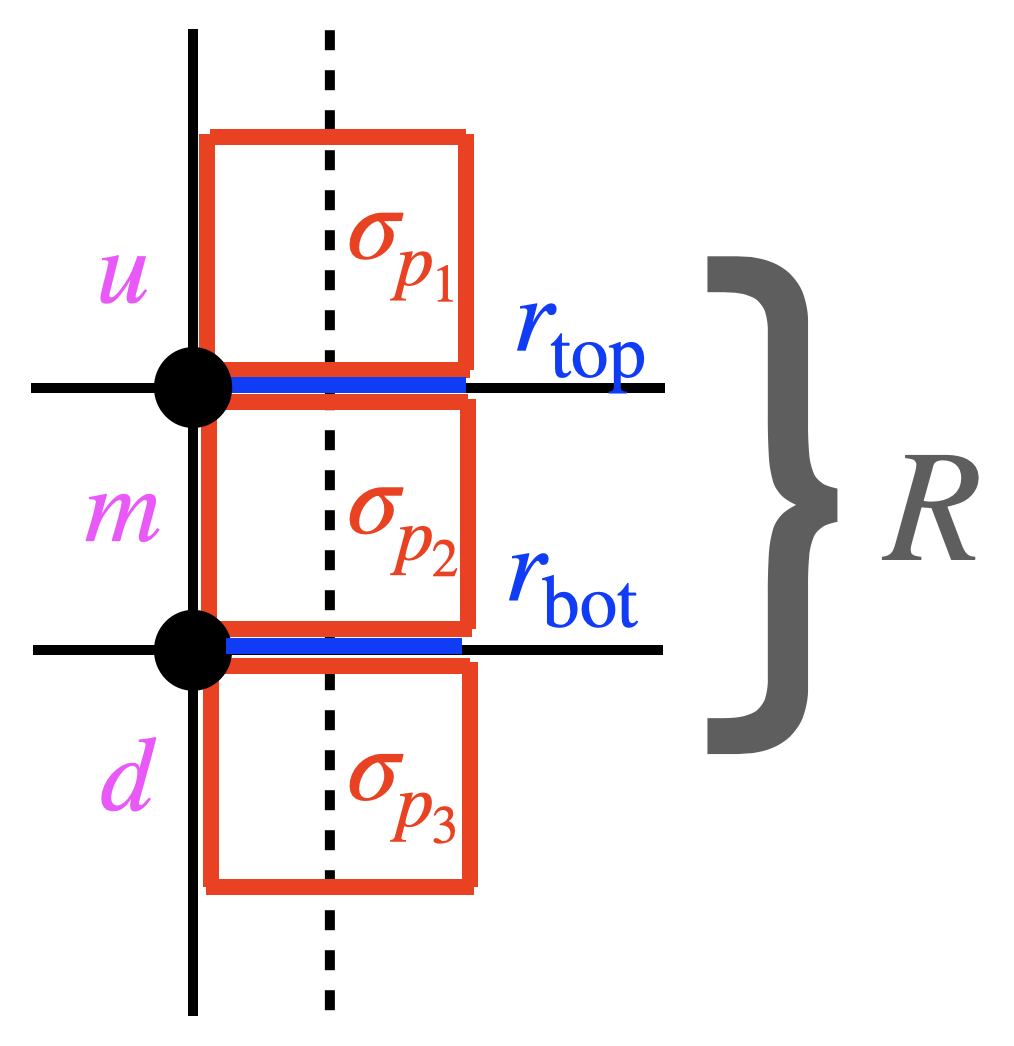}
\caption{The bond $R$ cut in the $y$-direction coarse-graining (dashed line). The two links crossing the cut carry the virtual legs $r_{\rm top}$ and $r_{\rm bot}$, and the three plaquettes $p_1$, $p_2$, and $p_3$ touch these links, with $p_2$ shared by the two. The link $m$ between the two sites is contracted.}\label{fig:image-SVD}
\end{figure}

As an example, let us consider the $y$-direction coarse-graining described above.
The bond being cut, $R$, corresponds to the two links shown in Fig.~\ref{fig:image-SVD}, namely, the right link of the upper site with the virtual leg $r_{\rm top}$, and that of the lower site with $r_{\rm bot}$.
%%%%%%%%%%%%%%%%%%%%%%%%%%%%%%%%%%%%
%%%%%%%%%%%%%%%%%%%%%%%%%%%%%%%%%%%%

For $F=1$, the occupation $r\in\{0,1\}$ of the virtual mode on a link equals the electric flux $p$ of that link.
Since the ket leg $r$ and the bra leg $r'$ have the same fermion parity, Eq.~\eqref{eq:parity-ket-bra}, which means $r=r'$ for $F=1$, the bond $R$ is effectively labeled by the pair $(r_{\rm top},r_{\rm bot})$, which takes $2^2=4$ values.
The closed-loop condition, namely, that an even number of links with $p=1$ meet at every site, gives for the two-site tensor the constraint
\begin{equation}
r_{\rm top}+r_{\rm bot}+u+d+l_{\rm top}+l_{\rm bot}\equiv 0 \pmod 2.
\label{eq:parity}
\end{equation}
A loop entering the two-site column through $r_{\rm top}$ need not leave through $r_{\rm bot}$.
It may leave through any of $u$, $d$, $l_{\rm top}$, or $l_{\rm bot}$, so the parity of $r_{\rm top}+r_{\rm bot}$ alone is not fixed.
However, Eq.~\eqref{eq:parity} depends on $(r_{\rm top},r_{\rm bot})$ only through $r_{\rm top}+r_{\rm bot}\bmod 2$, so the rows $(0,0)$ and $(1,1)$ of $M$ are identical up to a phase, as are the rows $(1,0)$ and $(0,1)$.
Hence $M$ has rank 2, and then the leading singular values are twofold degenerate.

For $F=2$, the two modes on the virtual leg of a link correspond to the two plaquettes sharing that link, as shown in Fig.~\ref{fig:def-corner}.
The electric flux of the link is given by $\sigma_{p_1}+\sigma_{p_2}\bmod 2$, where $\sigma_p\in\{0,1\}$ indicates whether the plaquette $p$ is ``on'' (i.e., $p\in S$ in Eq.~\eqref{eq:def-psi-B}) or ``off''.
The bond $R$ therefore carries $\sigma_p$ for the three plaquettes $p_1$, $p_2$, and $p_3$ in Fig.~\ref{fig:image-SVD}, the middle one $p_2$ being shared by the two links.
Since any choice of $\{\sigma_p\}$ automatically satisfies the closed-loop condition, no constraint analogous to Eq.~\eqref{eq:parity} arises.
However, $p_2$ touches only the two links in $R$ and the contracted link $m$, so that the column index $(L,u,d)$ does not depend on $\sigma_{p_2}$.
Flipping $\sigma_{p_2}$ thus multiplies a row of $M$ only by a constant, as in the case of $r_{\rm top}+r_{\rm bot}$ for $F=1$.
Then, the independent rows are labeled by $(\sigma_{p_1},\sigma_{p_3})$, which takes $2^2=4$ values.
In addition, since \texttt{Tii} is a double-layer tensor, the bra pattern need not coincide with the ket pattern.
The pattern with all $\sigma_p$ flipped gives the same flux and satisfies the parity condition, Eq.~\eqref{eq:parity-ket-bra}.
Moreover, the corresponding rows of $M$ have support on different columns.
This doubles the number of independent rows, and the leading singular values are eightfold degenerate for the $F=2$ ansatz at $\lambda=0$.

The difference in the degeneracy, 2 for $F=1$ and 8 for $F=2$, reflects the difference in the information carried by the virtual bond in the two ansatzes, not a difference in the physical state, which is the same $|\psi_B\rangle$ in both cases.
In the $F=2$ ansatz, the patterns $\sigma_p$ and $1-\sigma_p$ describe the same physical state, and the twofold degeneracy from the bra pattern is a manifestation of this redundancy.
As $\lambda$ increases and the state becomes different from $|\psi_B\rangle$, the degeneracy changes.
In the numerical calculations, we look at the singular values at $\lambda=0.1$, where a remnant of this degeneracy is seen, in Sec.~\ref{sec:feasibility}.

%%%%%%%%%%%%%%%%%%%%%%%%%%%%%%%%%%%%%%%%%%%%%%%%%%%%%%%%%%
%%%%%%%%%%%%%%%%%%%%%%%%%%%%%%%%%%%%%%%%%%%%%%%%%%%%%%%%%%
\section{Results for finite volume calculations}\label{sec:results-finite-vol}
%%%%%%%%%%%%%%%%%%%%%%%%%%%%%%%%%%%%%%%%%%%%%%%%%%%%%%%%%%
%%%%%%%%%%%%%%%%%%%%%%%%%%%%%%%%%%%%%%%%%%%%%%%%%%%%%%%%%%
In this section we present our numerical results at finite volume, and the results of infinite volume simulations are given in the next section.

\subsection{Simulation details}\label{sec:sim-detail}
Here, we first compare the results for $L=2$, $3$, and $4$ with those of ED in Sec.~\ref{sec:feasibility}.
For $L=2$ and $3$ the network is contracted exactly, which serves as a test of the reformulation of the ansatz as a Grassmann TN.
For $L=4$, one coarse-graining step is applied in each of the $x$ and $y$ directions, so that we can check that the calculation with HOTRG still reproduces the ED results.
We also examine the behavior of the singular values and the role of the two cutoffs.

In Sec.~\ref{sec:large-volume}, we apply the coarse-graining repeatedly and carry out calculations from $L=6$ to $34$, namely $L=2+2^k$ with $k=2,\dots,5$.
After the optimization of variational parameters, we evaluate the energy, the winding 't Hooft loop, and the TEE, and study their volume dependence.
In particular, we compare the results with the strong and weak coupling expansions of Sec.~\ref{sec:pert}.
In Sec.~\ref{sec:suscept}, we compare with the value of the critical coupling obtained by continuous-time QMC for the toric code Hamiltonian in an external field, Eq.~\eqref{eq:toric-code-H}~\cite{wu2012phase}, and discuss whether the present $F=2$ ansatz can capture the critical point or not.
In addition, in Sec.~\ref{sec:comp-plaq-prod-state}, we find that near the critical point the optimized state goes outside the plaquette product family of Sec.~\ref{sec:E-fam}, which is also suggested by the fact that variational parameters other than $b$ take nonzero values.

In the calculations, we mainly use ${\rm svcut}=10^{-10}$ and $D_{\rm cut}=32$.
Here ${\rm svcut}$ is the cutoff on the singular values in the HOTRG truncation, i.e., singular values with $s_i/s_1<{\rm svcut}$ are discarded, while $D_{\rm cut}$ is the maximum number of singular values kept.
These are two independent cutoffs, and which of them is the effective one depends on the coupling region, as discussed in Sec.~\ref{sec:feasibility}.
The variational parameters are optimized by the Nelder--Mead method, with the energy density $E_{tot}$, Eq.~\eqref{eq:def-Etot}, as the cost function.
For the $F=2$ ansatz, there are eight complex parameters, i.e., 16 real parameters, in Eq.~\eqref{eq:def-T-matrix-F2}.
The tolerance for the change of $E_{tot}$ is set to $10^{-14}$, and the optimization is continued until convergence.
The number of iterations required for the convergence depends on $\lambda$ and ranges from $\mathcal{O}(10^2)$ to $\mathcal{O}(10^5)$.

The initial values of the parameters are taken to be those of the toric code ground state plus small random numbers.
Namely, we take $b=(1-i)/2\sqrt{2}$, one of the four solutions of $b^4=-1/16$ in Eq.~\eqref{eq:f-b-relation}, with all other parameters set to zero, and add uniform random numbers in $[-0.05,0.05)$ to each real parameter.
To examine the dependence on the initial values, we use 10 random seeds for $L\le4$ and 5 random seeds for $L=6$ to $34$.

%%%%%%%%%%%%%%%%%%%%%%%%%%%%%%%%%%%%%%%%%%%%%%%%%%%%%%%%%%
\subsection{Feasibility tests in small lattices}\label{sec:feasibility}
%%%%%%%%%%%%%%%%%%%%%%%%%%%%%%%%%%%%%%%%%%%%%%%%%%%%%%%%%%
Here we show the results for the small lattices, $L=2$, $3$, and $4$, where ED is feasible and can be compared with.
As mentioned above, $L=2$ and $3$ serve as a test of the reformulation as a Grassmann TN, while $L=4$ serves as a test of HOTRG.

We first look at the singular values in HOTRG for $L=4$.
The coarse-graining is applied once in each of the $x$ and $y$ directions.
We show the singular values in the second step, i.e., the $y$ direction, since the hierarchy among them is larger and the seed dependence is more visible than in the first step, as seen for the $F=1$ ansatz in Appendix~\ref{sec:F1-numerical}.

%%%%%%%%%%%%%%%%%%%%%%%%%%%%%%
\begin{figure}[htb]
\centering
\includegraphics[scale=0.45]{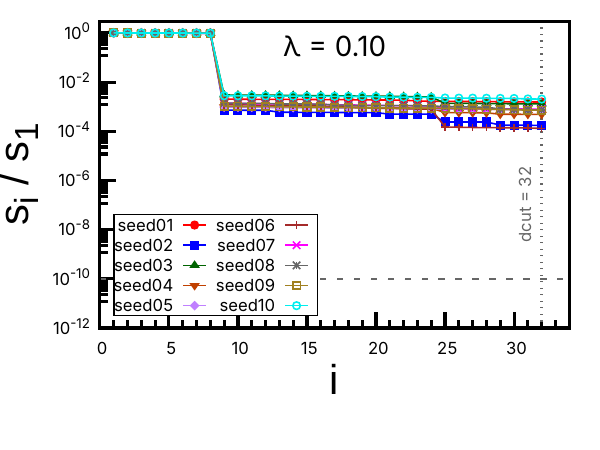}
\includegraphics[scale=0.45]{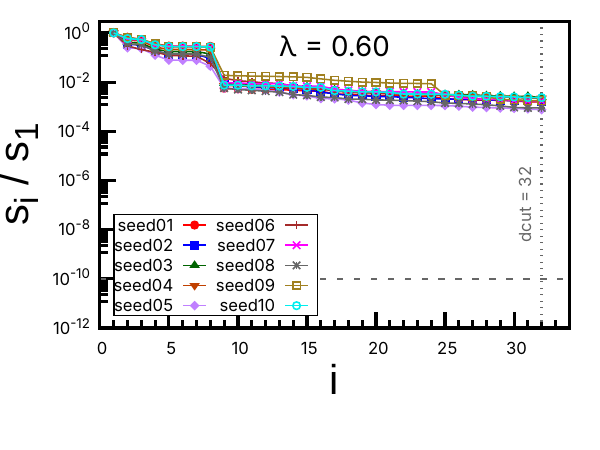}  
\includegraphics[scale=0.45]{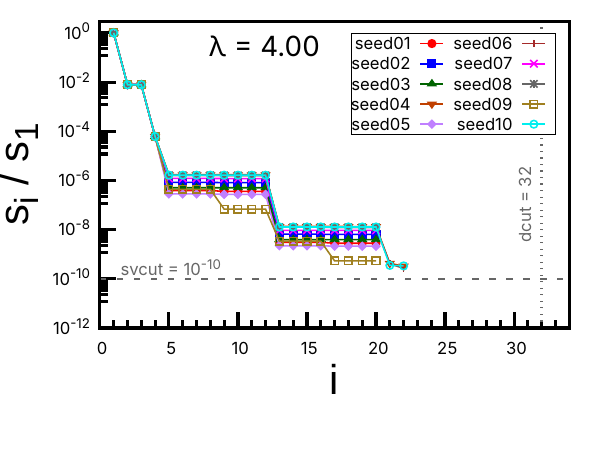}
\caption{Singular values $s_i/s_1$ in the $y$-direction coarse-graining step for $L=4$ with the $F=2$ ansatz at $\lambda=0.1$ (left), $0.6$ (middle), and $4.0$ (right). The results for the 10 random seeds are overlaid.}\label{fig:sv-scaling-lambda40}
\end{figure}
%%%%%%%%%%%%%%%%%%%%%%%%%%%%%%
Figure~\ref{fig:sv-scaling-lambda40} shows the singular values defining the HOTRG isometry for the state after the optimization, with $D_{\rm cut}=32$ and ${\rm svcut}=10^{-10}$.
The vertical axis is $s_i/s_1$, normalized by the largest singular value $s_1$.
In the left panel, for the nearly toric code ground state at $\lambda=0.1$, the leading eight singular values are nearly degenerate, as discussed in Sec.~\ref{sec:sv-hotrg}.
The middle panel is for $\lambda=0.6$, near the critical region.
The eightfold degeneracy seen at $\lambda=0.1$ is partially lifted, although a remnant is still visible in the leading eight singular values.
We also see that the hierarchy among $s_i$ is small near $\lambda_c$.
In both cases all the $32$ singular values survive, so that the effective cutoff is $D_{\rm cut}$.
In particular, at $\lambda=0.6$ the truncation error is expected to be larger since the discarded singular values are not sufficiently small compared with the kept ones.
The seed dependence of the singular values is also larger than at the other two couplings.
The right panel is for $\lambda=4.0$ in the strong coupling region.
Here the effective cutoff becomes ${\rm svcut}=10^{-10}$, and the effective bond dimension is $20$--$22$.
Since the singular values are kept down to a sufficiently small magnitude, the truncation error in the physical quantities evaluated with the obtained state is expected to be negligible.

%%%%%%%%%%%%%%%%%%%%%%%%%%%%%%
\begin{figure}[h]
\centering
\includegraphics[scale=0.6]{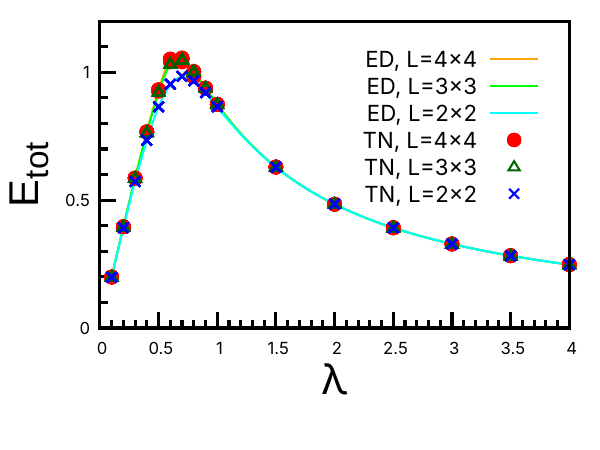}  
\includegraphics[scale=0.6]{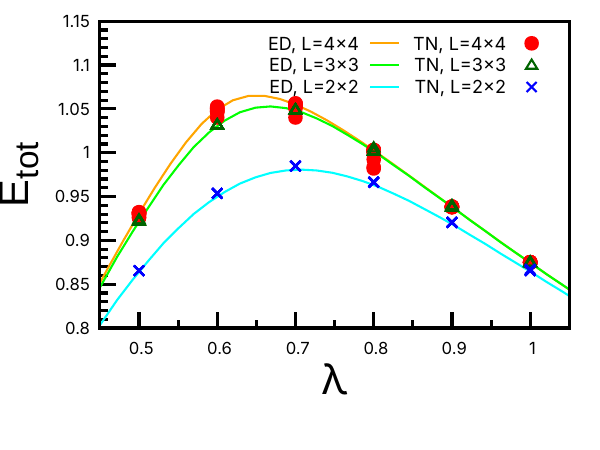}
\caption{Energy density $E_{tot}$ for $L=2$, $3$, and $4$ with the $F=2$ ansatz, compared with ED (solid curves). The right panel is an enlarged plot of the region near $\lambda_c$.}\label{fig:data-Etot-small-L}
\end{figure}
%%%%%%%%%%%%%%%%%%%%%%%%%%%%%%
Figure~\ref{fig:data-Etot-small-L} shows the energy density $E_{tot}$ for $L=2$, $3$, and $4$.
Here we depict the scatter plots for ten data points for each $(\lambda, L)$. 
The solid curves are the ED results for each lattice size.
Here, the ED results are obtained with the QuSpin package~\cite{Weinberg:2017,Weinberg:2019}.
Our numerical results agree well with ED in the whole range of $\lambda$.
Note that $E_{tot}$ has a maximum around $\lambda\simeq0.7$. 
This region is close to the critical point $\lambda_c=0.573$, Eq.~\eqref{eq:MC-lambda_c}, obtained by the continuous-time QMC.
The right panel of Fig.~\ref{fig:data-Etot-small-L} is an enlarged plot of the region near $\lambda_c$.
The agreement with ED is good also in this region.
For $L=2$ and $3$ the network is contracted exactly, so that the deviation from ED comes only from the limitation of the variational ansatz, while for $L=4$ the truncation error in HOTRG may be added.

We also find that the results for $L=2$ and $3$ differ from each other, while those for $L=3$ and $4$ are close to each other.
It indicates that the volume dependence becomes small for $L \gtrsim 4$.
A similar observation was reported in Ref.~\cite{Emonts:2022yom}, where the energies for $L=4$ and $6$ agree well.

\begin{figure}[h]
\centering
\includegraphics[scale=0.6]{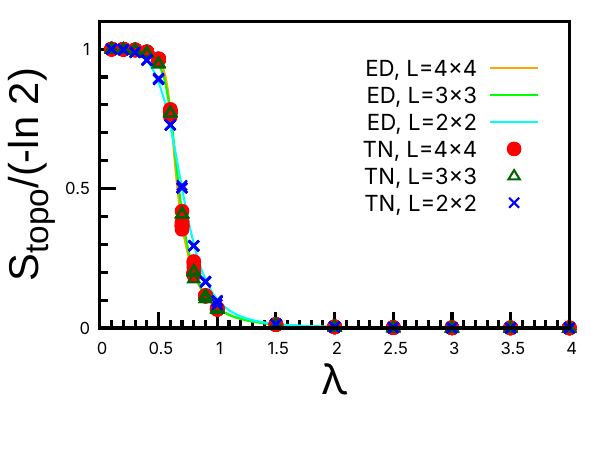}  
\includegraphics[scale=0.6]{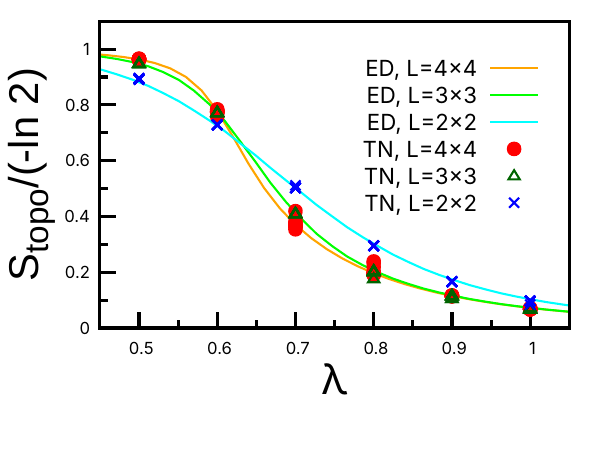}
\caption{TEE normalized by $-\ln2$ for $L=2$, $3$, and $4$ with the $F=2$ ansatz, compared with ED (solid curves). The right panel is an enlarged plot of the region near $\lambda_c$.}\label{fig:data-TEE-small-L}
\end{figure}

The results for the TEE on these small lattices are shown in Fig.~\ref{fig:data-TEE-small-L}.
Here the vertical axis is normalized by the value in the toric code ground state, $S_{\rm topo}=-\ln2$.
%%%%%%%%%%%%%%%%%%%%%%%%%%%%%%
%%%%%%%%%%%%%%%%%%%%%%%%%%%%%%
The results show a nearly flat plateau, $S_{\rm topo}/(-\ln2)\simeq1$, in the small $\lambda$ region, a rapid decrease around the maximum of $E_{tot}$, and a value consistent with zero in the strong coupling region.
The right panel is again an enlarged plot near $\lambda_c$.
The TEE also agrees well with ED in this region.
The volume dependence between $L=3$ and $4$ is still visible but relatively small compared with the seed dependence.

\begin{figure}[h]
\centering
\includegraphics[scale=0.6]{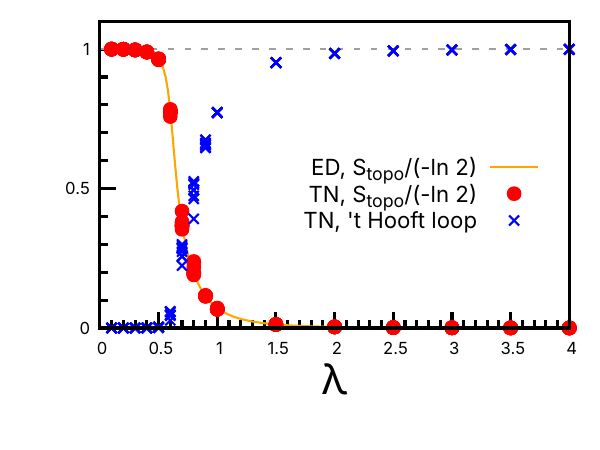}  
\includegraphics[scale=0.6]{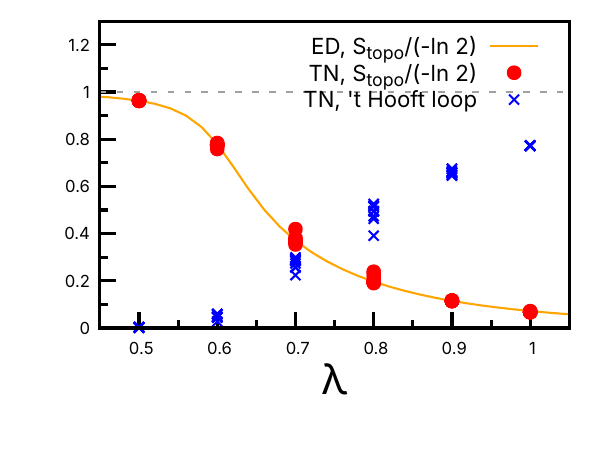}
\caption{TEE normalized by $-\ln2$ and the winding 't Hooft loop $\langle W\rangle$ for $L=4$ with the $F=2$ ansatz. The solid curve is the ED result for the TEE. The right panel is an enlarged plot of the region near $\lambda_c$.}\label{fig:data-tHooft-small-L}
\end{figure}

Finally, Fig.~\ref{fig:data-tHooft-small-L} shows the TEE and the winding 't Hooft loop, Eq.~\eqref{eq:def-W}, for $L=4$.
The winding 't Hooft loop increases from $0$ to $1$ in the region where the TEE changes.
%%%%%%%%%%%%%%%%%%%%%%%%%%%%%%
%%%%%%%%%%%%%%%%%%%%%%%%%%%%%%
Both data show smooth changes and there is no signal of discontinuity.
It indicates that the transition is not of first order but of second order or a crossover.

%%%%%%%%%%%%%%%%%%%%%%%%%%%%%%%%%%%%%%%%%%%%%%%%%%%%%%%%%%
\subsection{Large volume calculations: $L=6$ to $34$ lattices}\label{sec:large-volume}
%%%%%%%%%%%%%%%%%%%%%%%%%%%%%%%%%%%%%%%%%%%%%%%%%%%%%%%%%%
%%%%%%%%%%%%%%%%%%%%%%%%%%%%%%%%%%%%%%%%%%%%%%%%%%%%%%%%%%
\subsubsection{Seed dependence and truncation error}\label{sec:seed-dcut-deps}
%%%%%%%%%%%%%%%%%%%%%%%%%%%%%%%%%%%%%%%%%%%%%%%%%%%%%%%%%%
Let us see what kind of state is obtained by the optimization of the $F=2$ GGFPEPS in large volume calculations.

\begin{figure}[htb]
\centering
\includegraphics[scale=0.48]{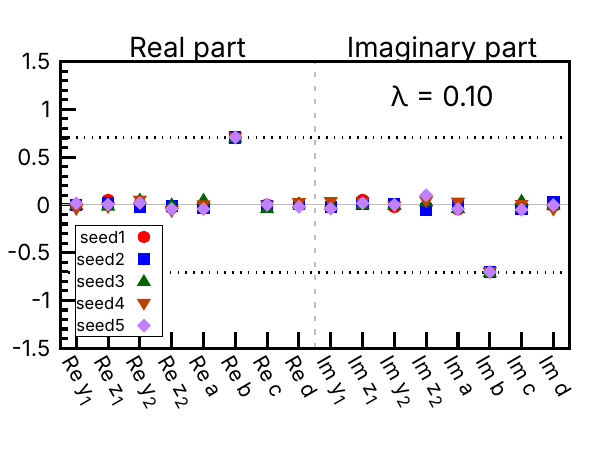}  
\includegraphics[scale=0.48]{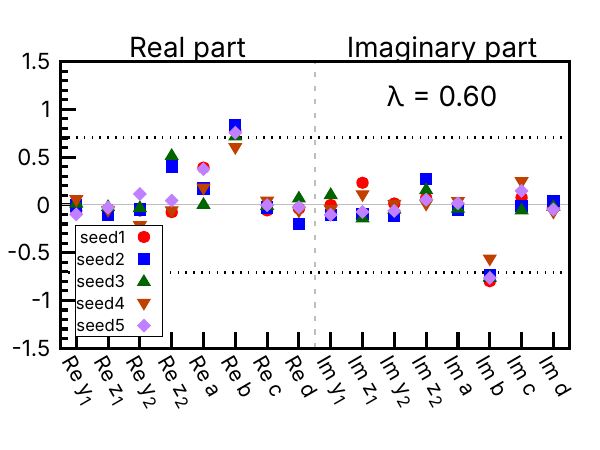}  
\includegraphics[scale=0.48]{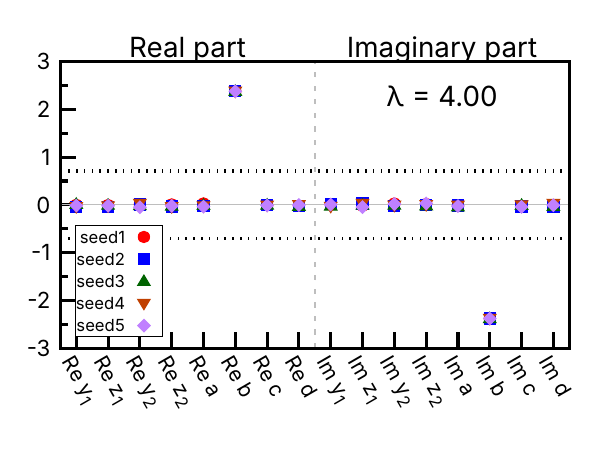}  
\caption{Optimized parameters for $L=34$ at $\lambda=0.1$ (left), $0.6$ (middle), and $4.0$ (right) with the $F=2$ ansatz. The results for the five seeds are overlaid.}\label{fig:opt-params}
\end{figure}

Figure~\ref{fig:opt-params} shows the optimized parameters for $L=34$ at $\lambda=0.1$ (left), $0.6$ (middle), and $4.0$ (right), where the results for the five seeds are presented.
%%%%%%%%%%%%%%%%%%%%%%%%%%%%%%
%%%%%%%%%%%%%%%%%%%%%%%%%%%%%%

At $\lambda=0.1$ (left), the parameters are almost those of the toric code ground state, Eq.~\eqref{eq:f-b-relation}, namely $b=(1-i)/2\sqrt{2}$ with all other parameters zero, which corresponds to $t=1$ in the plaquette product family of Sec.~\ref{sec:E-fam}.
At $\lambda=4.0$ (right), the other parameters are again almost zero, while $b$ takes a large value for all the five seeds.
In Sec.~\ref{sec:E-fam}, the strong coupling limit was described by $t\rightarrow0$, but here the optimization goes to the opposite side, $t\simeq8\lambda^2$, i.e., toward $t\rightarrow\infty$.

On the other hand, at $\lambda=0.6$ (middle), near the critical point, the parameters other than $b$ take nonzero values for all the seeds in different ways.
This suggests that the state near $\lambda_c$ is not in the plaquette product family, which will be further examined in Sec.~\ref{sec:comp-plaq-prod-state}.
In addition, the seed dependence of the parameters is large.
Table~\ref{tab:seed-comparison-L34-g0.6} shows the physical quantities for the five seeds at $L=34$ and $\lambda=0.6$.
%%%%%%%%%%%%%%%%%%%%%%%%%%%%%%
\begin{table}[htbp]
  \centering
  \begin{tabular}{crrrcc}
    \hline
    seed & $E_{tot}$ & $E_e$ & $E_m$ & $\langle W\rangle$ & $S_{\rm topo}/(-\ln 2)$ \\
    \hline
    1 & 1.05967 & 0.67781 & 0.38186 & $1.09\times10^{-10}$ & 0.7300 \\
    2 & 1.06669 & 0.55971 & 0.50699 & $2.98\times10^{-9}$  & 0.6042 \\
    3 & 1.05947 & 0.65527 & 0.40419 & $8.85\times10^{-12}$ & 0.7167 \\
    4 & 1.05918 & 0.68817 & 0.37101 & $2.39\times10^{-10}$ & 0.7407 \\
    5 & 1.05964 & 0.68144 & 0.37820 & $6.89\times10^{-10}$ & 0.7341 \\
    \hline
  \end{tabular}
 \caption{Physical quantities for the five seeds at $L=34$ and $\lambda=0.6$ with the $F=2$ ansatz. $S_{\rm topo}$ is normalized by $-\ln 2$.}
  \label{tab:seed-comparison-L34-g0.6}  
\end{table}
%%%%%%%%%%%%%%%%%%%%%%%%%%%%%%
The physical quantities themselves also vary among the seeds, which indicates that each run falls into a different physical state rather than that the parameters are redundant.
Such a behavior is not seen in the small or large $\lambda$ regime.
This suggests that the optimization landscape becomes complicated near $\lambda_c$, possibly containing multiple local minima with similar energies.

Next, we look at the $D_{\rm cut}$ dependence.
For $L=34$ and $\lambda=0.1$, $0.6$, $2.0$, and $4.0$, we take the variational parameters of seed 1 optimized with $D_{\rm cut}=32$ and evaluate $E_{tot}$ with $D_{\rm cut}=16$, $32$, $48$, and $64$.
The results are shown in Table~\ref{tab:dcut-dependence-L34}.
%%%%%%%%%%%%%%%%%%%%%%%%%%%%%%
\begin{table}[htbp]
  \centering
  \begin{tabular}{crrrr}
    \hline
    $D_{\rm cut}$ & $\lambda=0.1$ & $\lambda=0.6$ & $\lambda=2.0$ & $\lambda=4.0$ \\
    \hline
    16 & 0.19950006 & 1.05865105 & 0.48437498 & 0.24804687 \\
    32 & 0.19950010 & 1.05966892 & 0.48437498 & 0.24804687 \\
    48 & 0.19950007 & 1.06012384 & 0.48437498 & 0.24804687 \\
    64 & 0.19950013 & 1.06016481 & 0.48437498 & 0.24804687 \\
    \hline
  \end{tabular}
  \caption{$D_{\rm cut}$ dependence of $E_{tot}$ for $L=34$ with the $F=2$ ansatz. The variational parameters of seed 1 optimized with $D_{\rm cut}=32$ are used for all $D_{\rm cut}$.}
  \label{tab:dcut-dependence-L34}
\end{table}
%%%%%%%%%%%%%%%%%%%%%%%%%%%%%%

Except for $\lambda=0.6$ near the critical point, the differences among $D_{\rm cut}$ are less than $10^{-7}$.
At $\lambda=4.0$, the number of singular values remaining after ${\rm svcut}$ is about $20$, so that no truncation occurs for $D_{\rm cut}\ge32$ and the results naturally coincide.
At $\lambda=2.0$, about $40$ singular values remain after ${\rm svcut}$, so that $D_{\rm cut}=32$ actually truncates the spectrum, but its effect on $E_{tot}$ is of order $10^{-11}$.
At $\lambda=0.1$, all the $64$ singular values remain even for $D_{\rm cut}=64$, but the hierarchy of the singular values is strong as seen in Sec.~\ref{sec:feasibility}, and the contribution of the discarded ones is less than $10^{-7}$.

At $\lambda=0.6$, on the other hand, all the $64$ singular values also remain for $D_{\rm cut}=64$, and the singular values decay slowly near the critical point, so that $E_{tot}$ changes by $5\times10^{-4}$ from $D_{\rm cut}=32$ to $64$.
This is smaller than the seed spread in Table~\ref{tab:seed-comparison-L34-g0.6}.
Note that the origins of the two are different. 
The seed spread comes from the flatness of the cost function, while the $D_{\rm cut}$ dependence comes from the slow decay of the singular values.
We also comment on the fact that $E_{tot}$ at $\lambda=0.6$ increases with $D_{\rm cut}$.
A variational energy is in general an upper bound on the true ground state energy, but this shows that the truncation can make $E_{tot}$ smaller than its true value for the given variational parameters.
Table~\ref{tab:dcut-dependence-L34} shows that this effect is small and under control.
From these results, we use $D_{\rm cut}=32$ in this paper.

In the following, the large volume results are shown with error bars, which denote the spread of the results among the five seeds for simplicity of presentation.
We regard it as an estimate of the uncertainty of the optimization.
Note that it is not a statistical error in the MC sense.

%%%%%%%%%%%%%%%%%%%%%%%%%%%%%%%%%%%%%%%%%%%%%%%%%%%%%%%%%%
\subsubsection{Energy density}
\begin{figure}[h]
\centering
\includegraphics[scale=0.6]{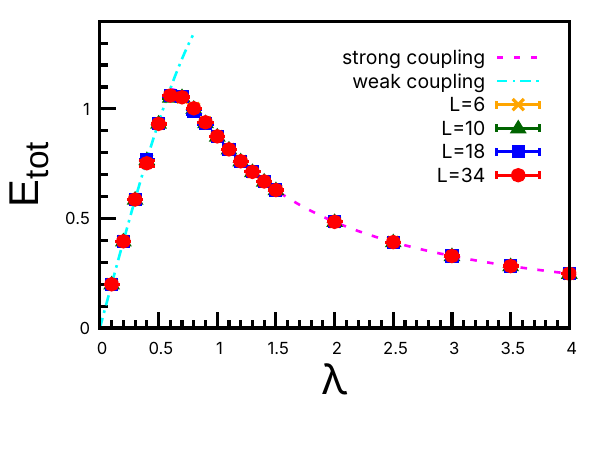}  
\includegraphics[scale=0.6]{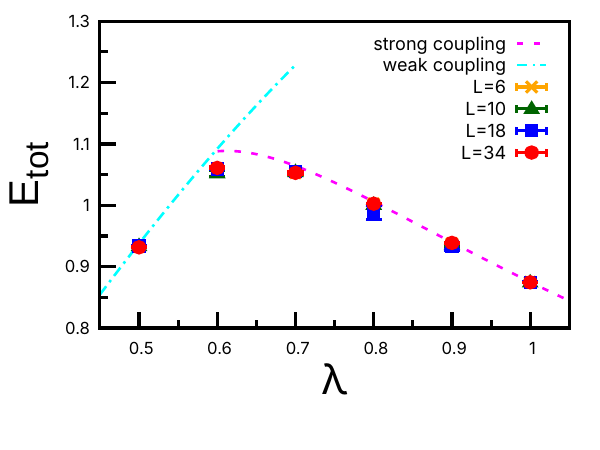}
\caption{Energy density $E_{tot}$ for $L=6$--$34$ with the $F=2$ ansatz, compared with the strong (magenta dotted curve) and weak coupling expansions (cyan dashed curve). The right panel is an enlarged plot of the region near $\lambda_c$. The size of errors is comparable to the symbol size.}\label{fig:large-vol-Etot}
\end{figure}

%%%%%%%%%%%%%%%%%%%%%%%%%%%%%%%%%%%%%%%%%%%%%%%%%%%%%%%%%%
From here on we show the results for $L=6$, $10$, $18$, and $34$.
Since ED is no longer available for these lattice sizes, we first compare the energy density with the strong coupling expansion, Eq.~\eqref{eq:Etot-strong-coupling}, and the weak coupling expansion, Eq.~\eqref{eq:wc-energy-result}, in Fig.~\ref{fig:large-vol-Etot}.
In particular, the right panel, an enlarged plot near $\lambda_c$, shows that the results agree well with the weak coupling expansion for $\lambda\lesssim 0.5$ and with the strong coupling expansion for $\lambda\gtrsim 0.9$.
%%%%%%%%%%%%%%%%%%%%%%%%%%%%%%
%%%%%%%%%%%%%%%%%%%%%%%%%%%%%%
We also find that no volume dependence is visible at these volumes, even near $\lambda_c$.

%%%%%%%%%%%%%%%%%%%%%%%%%%%%%%%%%%%%%%%%%%%%%%%%%%%%%%%%%%
\subsubsection{Winding 't Hooft loop}\label{sec:thooft-large}
\begin{figure}[htb]
\centering
\includegraphics[scale=0.6]{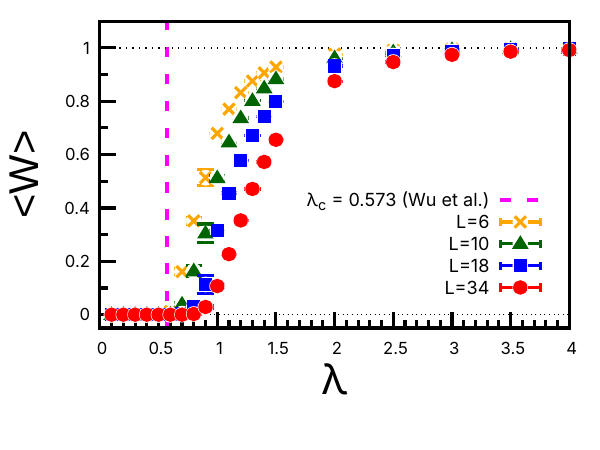}  
\caption{Winding 't Hooft loop $\langle W\rangle$ for $L=6$--$34$ with the $F=2$ ansatz. The magenta dotted line indicates the critical point $\lambda_c=0.573$ obtained by the continuous-time QMC~\cite{wu2012phase}.}\label{fig:large-vol-tHooft}
\end{figure}

%%%%%%%%%%%%%%%%%%%%%%%%%%%%%%%%%%%%%%%%%%%%%%%%%%%%%%%%%%
Next, we show the results for the winding 't Hooft loop, Eq.~\eqref{eq:def-W}.
Figure~\ref{fig:large-vol-tHooft} depicts the results for $L=6$ to $34$.
As a reference, the critical point obtained by the continuous-time QMC, $\lambda_c=0.573$ in Eq.~\eqref{eq:MC-lambda_c}, is shown by the magenta dotted line.
%%%%%%%%%%%%%%%%%%%%%%%%%%%%%%
%%%%%%%%%%%%%%%%%%%%%%%%%%%%%%
In contrast to the energy density, a clear volume dependence is seen here, as expected.
The value of $\lambda$ at which $\langle W\rangle$ becomes nonzero, which we call the onset, moves to larger $\lambda$ as $L$ increases.
The onset is therefore a finite volume effect and does not indicate the transition point.

In fact, $\langle W\rangle$ vanishes for any $\lambda$ in the infinite volume limit.
In the strong coupling region, $\langle W\rangle=e^{-\alpha L}$ from Eq.~\eqref{eq:def-alpha} with a finite $\alpha$ as in Eq.~\eqref{eq:alpha-sc}, so that it goes to zero as $L\rightarrow\infty$ at fixed $\lambda$.
On the other hand, in the weak coupling region, it vanishes already at the leading order of the weak coupling expansion for $L>2$, as shown in Sec.~\ref{sec:WCE}.
For $L=4$ in Sec.~\ref{sec:feasibility}, $\langle W\rangle$ increased in the region where the TEE changes, but this is because $e^{-\alpha L}$ is still large for such a small $L$.
Note that $\langle W\rangle$ is exactly zero in our calculations below the onset because of the property of the isometry described in Sec.~\ref{sec:def-winding-loop}.
In perturbation theory, however, $\langle W\rangle$ in this region would scale as $\mathcal{O}(\lambda^{2L})$, which is practically indistinguishable from zero.

Therefore, the physical information at finite volume is carried not by $\langle W\rangle$ itself but by $\alpha(\lambda)$ defined in Eq.~\eqref{eq:def-alpha}.
%%%%%%%%%%%%%%%%%%%%%%%%%%%%%%
\begin{figure}[htb]
\centering
\includegraphics[scale=0.6]{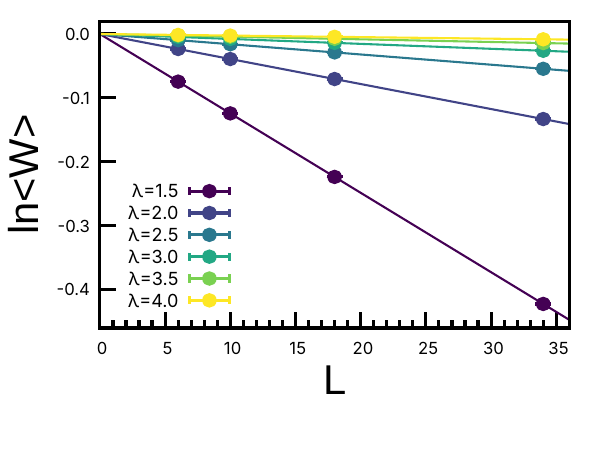}  
\includegraphics[scale=0.6]{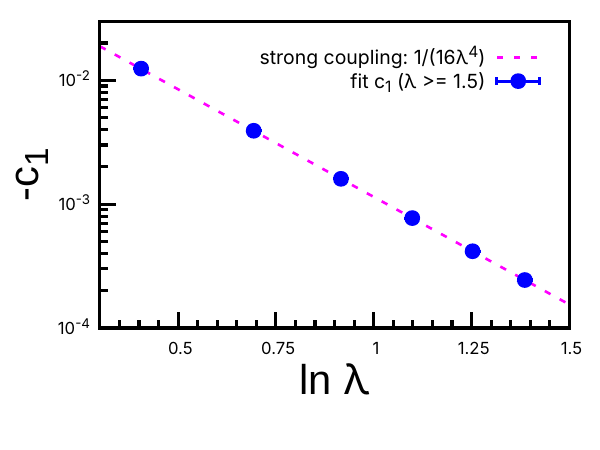}
\caption{(Left) $L$ dependence of $\ln\langle W\rangle$ for $\lambda\ge1.5$. The lines are the fits $\ln\langle W\rangle=c_0+c_1L$. (Right) The slope $-c_1$ as a function of $\lambda$, compared with the strong coupling expansion $\alpha^{(SC)}=1/16\lambda^4$ (magenta dotted line).}\label{fig:fit-tHooft}
\end{figure}
%%%%%%%%%%%%%%%%%%%%%%%%%%%%%%
The left panel of Fig.~\ref{fig:fit-tHooft} shows the $L$ dependence of $\ln\langle W\rangle$ in the strong coupling region.
The lines present the fit functions, $\ln\langle W\rangle=c_0+c_1L$, for each $\lambda\ge 1.5$.
In this range of $\lambda$, our data can be well fitted by a linear function of $L$ and $\langle W\rangle$ converges to zero as $L\rightarrow\infty$.
The linearity in $L$ is consistent with the cluster decomposition $\langle W\rangle\simeq\langle P\rangle^L$.

The right panel of Fig.~\ref{fig:fit-tHooft} shows the obtained $(-c_1)$, which corresponds to $\alpha(\lambda)$, for each $\lambda$.
It agrees well with the prediction of the strong coupling expansion, $\alpha^{(SC)}=1/16\lambda^4$ in Eq.~\eqref{eq:alpha-sc}.
Together with the results for the energy density, we conclude that the region $\lambda\ge1.5$ is well described by the strong coupling expansion.

%%%%%%%%%%%%%%%%%%%%%%%%%%%%%%%%%%%%%%%%%%%%%%%%%%%%%%%%%%
\subsubsection{Topological entanglement entropy}
\begin{figure}[htb]
\centering
\includegraphics[scale=0.6]{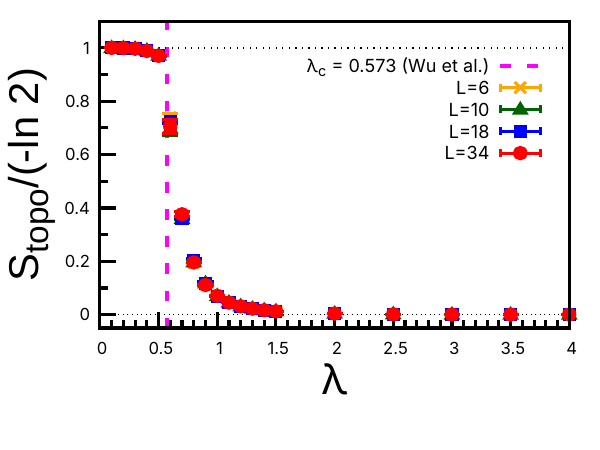}  
\includegraphics[scale=0.6]{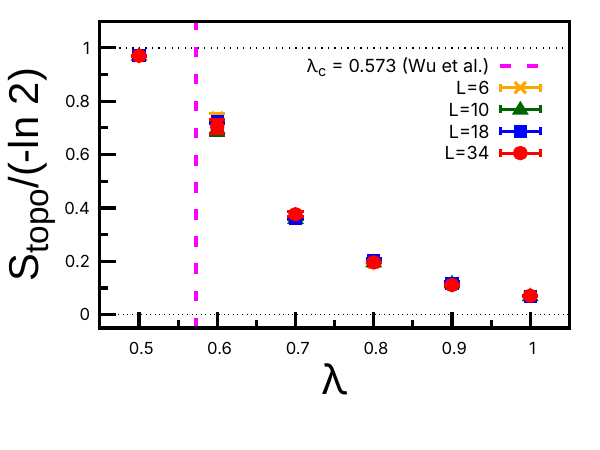}
\caption{TEE normalized by $-\ln2$ for $L=6$--$34$ with the $F=2$ ansatz. The magenta dotted line indicates the critical point $\lambda_c=0.573$ obtained by the continuous-time QMC~\cite{wu2012phase}. The right panel is an enlarged plot of the region near $\lambda_c$.}\label{fig:large-vol-TEE}
\end{figure}

%%%%%%%%%%%%%%%%%%%%%%%%%%%%%%%%%%%%%%%%%%%%%%%%%%%%%%%%%%
Finally, we look at the results for the TEE at large volumes.
Figure~\ref{fig:large-vol-TEE} shows the results of TEE for $L=6$ to $34$, normalized by $(-\ln 2)$ as in Sec.~\ref{sec:feasibility}.
The critical point obtained by the continuous-time QMC, $\lambda_c=0.573$, is again shown by the magenta dotted line.
%%%%%%%%%%%%%%%%%%%%%%%%%%%%%%
%%%%%%%%%%%%%%%%%%%%%%%%%%%%%%

The left panel shows that the point where the TEE starts to decrease rapidly is located almost at $\lambda_c$.
The right panel, an enlarged plot near $\lambda_c$, shows that the TEE has no volume dependence even near $\lambda_c$ for these sizes, in contrast to the small volumes $L=2$, $3$, and $4$ in Sec.~\ref{sec:feasibility}.
The change of the TEE is relatively smooth and no discontinuity is seen, which is consistent with the second-order transition expected from the 3D Ising universality class.
In addition, note that the error bars at $\lambda=0.6$, which is close to $\lambda_c$, are slightly larger than those at the other couplings.

Here, the sizes of the subsystems are fixed as in Fig.~\ref{fig:eval-TEE}, and only the total volume $L$ is varied.
If both the volume and the subsystems were infinitely large, the TEE would be $(-\ln2)$ in the deconfined phase and $0$ in the confined phase, with a sharp change at $\lambda_c$.
The smooth change observed here is therefore not due to the finite volume, which has no visible effect, but due to the finite size of the subsystems.
We expect that the TEE approaches the step function as the subsystem size increases.
In the present method, however, enlarging the subsystems is costly, since the number of components of the reduced density matrix to be measured is $2^n\times2^n$ for a subsystem with $n$ links.
We leave this for future work.
In Sec.~\ref{sec:ctmrg}, we examine the same point in the infinite volume limit by CTMRG with the same subsystems.

%%%%%%%%%%%%%%%%%%%%%%%%%%%%%%%%%%%%%%%%%%%%%%%%%%%%%%%%%%
\subsubsection{Susceptibilities and the critical coupling estimation}\label{sec:suscept}
%%%%%%%%%%%%%%%%%%%%%%%%%%%%%%%%%%%%%%%%%%%%%%%%%%%%%%%%%%
We now examine more precisely whether the changes of $E_{tot}$ and the TEE correspond to the topological phase transition found by the continuous-time QMC.
For this purpose, we evaluate their susceptibilities and compare the peak positions with $\lambda_c$.
We added data points around $\lambda_c$, namely $0.50\le\lambda\le0.80$ with the interval $\Delta\lambda=0.03$ for $L=18$ and $34$, with five seeds for each point.

\begin{figure}[htb]
\centering
\includegraphics[scale=0.5]{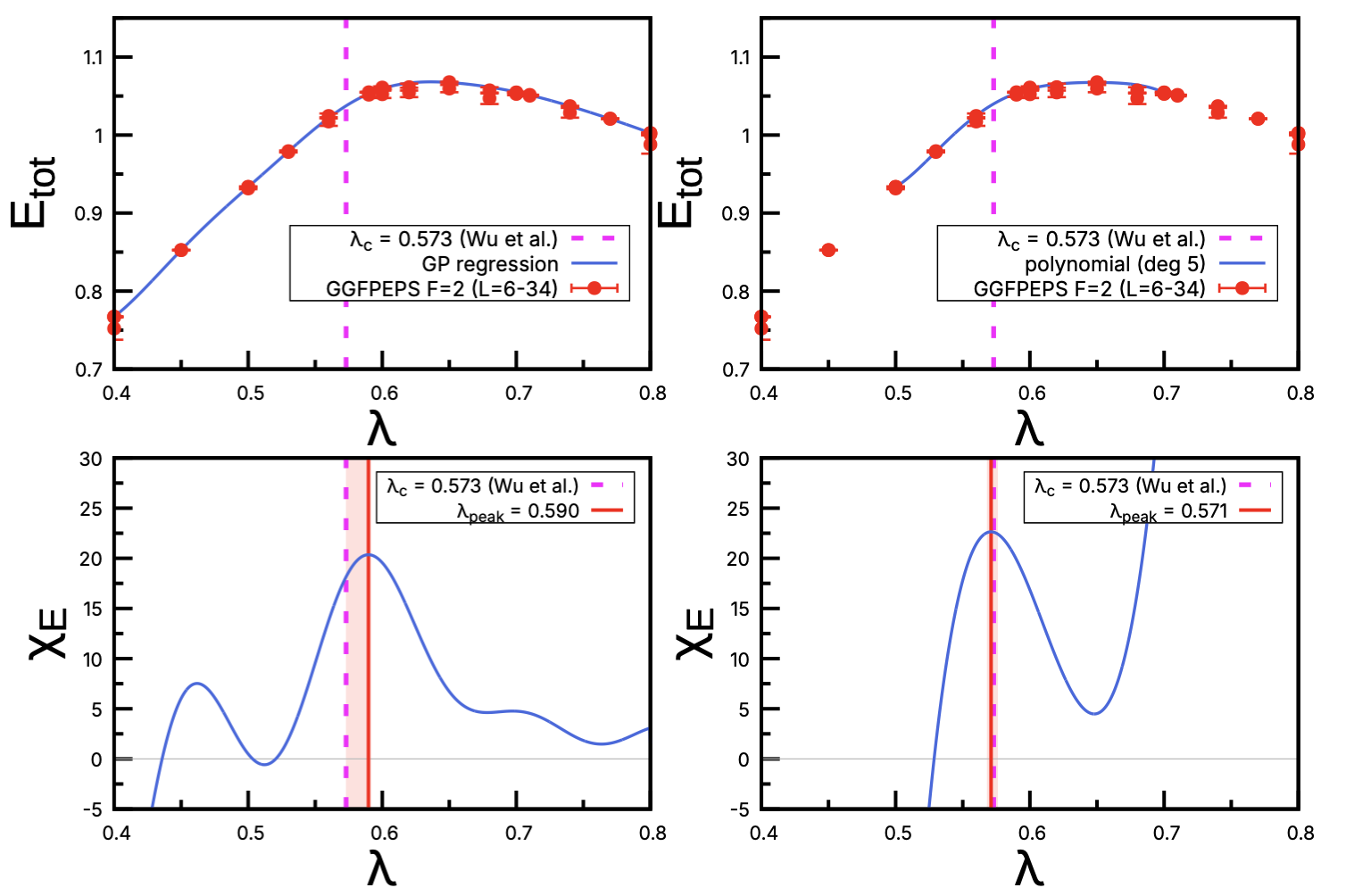}  
\caption{Energy density $E_{tot}$ (top) and its susceptibility $\chi_E$ (bottom) near $\lambda_c$, obtained by the Gaussian process regression (left) and the polynomial fit (right). In the bottom panels, the red solid line indicates the peak position and the shaded region the 68\% confidence interval. The magenta dotted line indicates $\lambda_c=0.573$ obtained by the continuous-time QMC~\cite{wu2012phase}.}\label{fig:Etot-suscpt}
\end{figure}

Figure~\ref{fig:Etot-suscpt} shows the results for $E_{tot}$.
We fit the data of $E_{tot}$ by a smooth function and evaluate the susceptibility, defined as the second derivative $\chi_E(\lambda)=-\partial^2E_{tot}/\partial\lambda^2$, from the fit function.
Since no volume dependence is seen in the data, all the data for $L=6$ to $34$ are used in the fit, where each data point $(L,\lambda)$ is weighted by its error.
%%%%%%%%%%%%%%%%%%%%%%%%%%%%%%
%%%%%%%%%%%%%%%%%%%%%%%%%%%%%%
To evaluate the systematic uncertainty of the fitting function, 
here we perform two types of fit, the Gaussian process (GP) regression in the left panel and the polynomial fit in the right panel.
The top panels show the fits to the data and the bottom panels the susceptibilities.
In the bottom panels, the peak position is shown by the red solid line and the 68\% confidence interval by the shaded region.

For the GP regression~\cite{Rasmussen2006}, we use \texttt{GaussianProcessRegressor} in scikit-learn~\cite{scikit-learn} with the kernel $k(\lambda,\lambda')=A\exp[-(\lambda-\lambda')^2/2\ell^2]$.
The fit range is $0.40\le\lambda\le0.80$.
The hyperparameters $A$ and $\ell$ are determined by maximizing the log marginal likelihood with 8 restarts and $\ell\in[0.05,1.0]$, which gives $\ell=0.085$.
The residual is $\chi^2/N=1.45$.
The error is estimated by the parametric bootstrap with 200 samples.
We obtain $\lambda_{\rm peak}=0.590$ with the 68\% confidence interval $[0.573,0.592]$.

For the polynomial fit, we use $E_{tot}(\lambda)=\sum_{n=0}^{5}c_n\lambda^n$ and minimize $\chi^2$ with the errors $\sigma$ of the data points.
The fits fail to describe the sharp peak around $\lambda\simeq0.6$ if we set the same fit range, $[0.40,0.80]$.
We therefore use the window $[0.50,0.70]$ (24 data points), where the fifth-order polynomial gives $\chi^2/{\rm dof}=2.05$.
Here $\chi_E$ is obtained by the analytic second derivative, the peak is searched for in $[0.53,0.67]$, and the same bootstrap analysis gives $\lambda_{\rm peak}=0.571$ with the 68\% confidence interval $[0.568,0.576]$.
In both methods, the continuous-time QMC result $\lambda_c=0.573$ is consistent with the shaded region.

\begin{figure}[htb]
\centering
\includegraphics[scale=0.5]{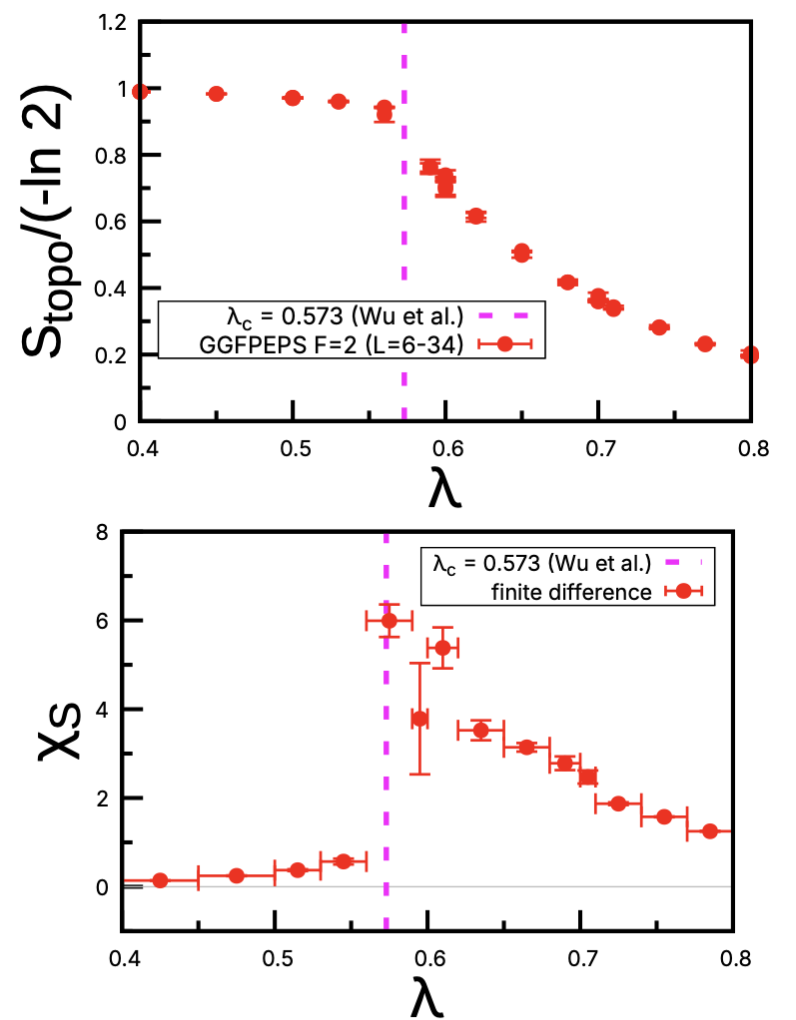}  
\caption{TEE for $L=18$ and $34$ with the additional data points near $\lambda_c$ (top) and its susceptibility $\chi_S$ evaluated by the finite difference (bottom). The horizontal error bars indicate the half width of the interval, and the vertical ones are propagated from the errors of the TEE. The magenta dotted line indicates $\lambda_c=0.573$ obtained by the continuous-time QMC~\cite{wu2012phase}.}\label{fig:TEE-suscpt}
\end{figure}

We perform a similar analysis for the TEE.
The top panel of Fig.~\ref{fig:TEE-suscpt} shows the TEE for $L=18$ and $34$ with the additional data points.
%%%%%%%%%%%%%%%%%%%%%%%%%%%%%%
%%%%%%%%%%%%%%%%%%%%%%%%%%%%%%
Since the TEE is almost constant at small $\lambda$ and drops rapidly around $\lambda_c$, it is difficult to fit it by a smooth function.
We therefore define the susceptibility as the first derivative,
\beq
\chi_S(\lambda)= \frac{1}{\ln 2}\frac{\partial S_{\rm topo}}{\partial\lambda},
\eeq
and evaluate it by the finite difference.
Namely, for the neighboring data points $\lambda_i$ and $\lambda_{i+1}$, we plot $\chi_S(\lambda_{\rm mid})=-(S_{i+1}-S_i)/(\lambda_{i+1}-\lambda_i)/(-\ln 2)$ at the midpoint $\lambda_{\rm mid}=(\lambda_i+\lambda_{i+1})/2$, with the half width of the interval as the horizontal error bar and the error propagated from those of the TEE as the vertical one.
Here, the bottom panel shows the results for the data of $L=18$ and $34$ together.
The highest peak appears at $\lambda_{\rm mid}=0.575$, which again agrees with the continuous-time QMC result $\lambda_c=0.573$ within the width of the interval.

The three estimates of the peak position, $\lambda_{\rm peak}=0.590$ and $0.571$ from $E_{tot}$ and $\lambda_{\rm mid}=0.575$ from the TEE, capture the continuous-time QMC value $\lambda_c=0.573$ within their errors, although the sizes of the errors differ among them.
From these results, we conclude that the GGFPEPS with the $F=2$ ansatz, contracted by the Grassmann TN, is consistent with the continuous-time QMC result even around the critical point, where the state is expected to be the most difficult to represent.
The number of variational parameters is only eight complex parameters, so that the ansatz is a highly efficient one.

%%%%%%%%%%%%%%%%%%%%%%%%%%%%%%%%%%%%%%%%%%%%%%%%%%%%%%%%%%
\subsubsection{Comparison with the plaquette product family}\label{sec:comp-plaq-prod-state}
%%%%%%%%%%%%%%%%%%%%%%%%%%%%%%%%%%%%%%%%%%%%%%%%%%%%%%%%%%
Finally, let us look at the obtained states in more detail.
As explained in Sec.~\ref{sec:more-F2}, our ansatz is constructed so that the plaquette product family of Sec.~\ref{sec:E-fam}, which contains the toric code ground state, is easily reached.
As seen in Sec.~\ref{sec:seed-dcut-deps}, this is indeed the case in the weak and strong coupling regimes, where the optimized states correspond to $t=1$ and $t\simeq8\lambda^2$, respectively.
However, are the states obtained near the critical point also in this family?
In other words, is the $F=2$ ansatz useful for representing states beyond this family?

\begin{figure}[htb]
\centering
\includegraphics[scale=0.6]{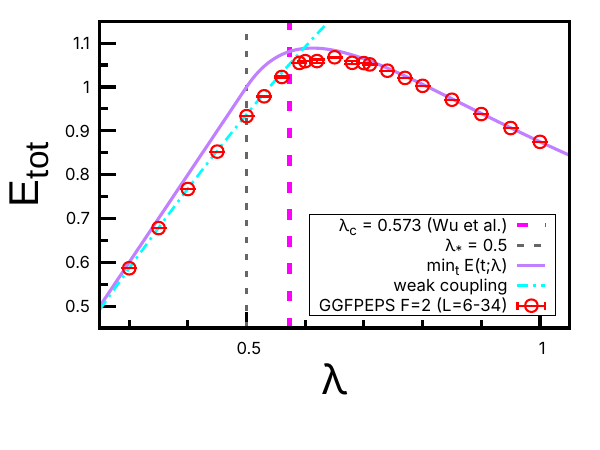}  
\includegraphics[scale=0.6]{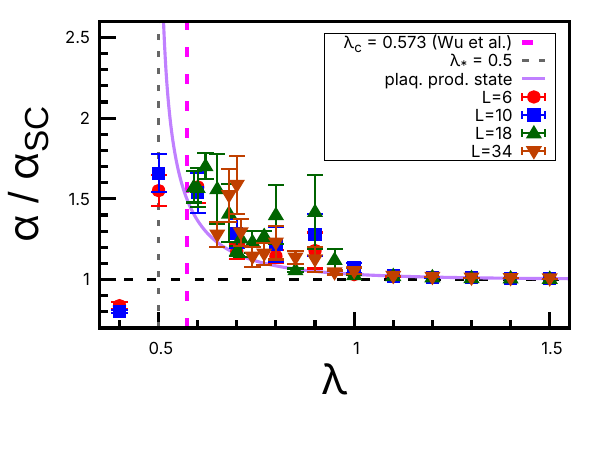} 
\caption{Comparison with the plaquette product family. (Left) $E_{tot}$ for $L=6$--$34$ compared with the lower bound within the family, $\min_tE_{\rm fam}(t;\lambda)$ (purple solid curve), and the weak coupling expansion (cyan dash-dotted curve). (Right) $\alpha(\lambda)$ normalized by $\alpha^{(SC)}=1/16\lambda^4$ for $L=6$--$34$, compared with the prediction of the family (purple solid curve), which diverges at $\lambda=1/2$. The magenta dotted line indicates $\lambda_c=0.573$ obtained by the continuous-time QMC~\cite{wu2012phase}, and the gray dotted line $\lambda=1/2$.}\label{fig:comp-plaq-prod}
\end{figure}

To answer this, we compare $E_{tot}$ with the lower bound within the family, $\min_tE_{\rm fam}(t;\lambda)$ in Eq.~\eqref{eq:E-fam-min}, in the left panel of Fig.~\ref{fig:comp-plaq-prod}.
%%%%%%%%%%%%%%%%%%%%%%%%%%%%%%
%%%%%%%%%%%%%%%%%%%%%%%%%%%%%%
Our numerical results are lower than $\min_tE_{\rm fam}$ near $\lambda_c$.
Since a variational energy within the family cannot be lower than $\min_tE_{\rm fam}$, this shows that the optimized state of the $F=2$ GGFPEPS is not in the family.\footnote{Strictly speaking, the variational inequality is not guaranteed here because of the truncation by $D_{\rm cut}$ and ${\rm svcut}$ in HOTRG. However, as shown in Table~\ref{tab:dcut-dependence-L34}, the $D_{\rm cut}$ dependence of $E_{tot}$ at $\lambda=0.6$ is of order $10^{-4}$, which is much smaller than the difference from $\min_tE_{\rm fam}$.}
This is expected from the weak coupling expansion, as noted in Sec.~\ref{sec:E-fam}.
The first-order correction in Eq.~\eqref{eq:leading-state-weak}, generated by $P_l$, is outside the plaquette product family.
Indeed, $E_{tot}=0.1995$ at $\lambda=0.1$ in Table~\ref{tab:dcut-dependence-L34} agrees with $2\lambda-\lambda^3/2$ of the weak coupling expansion rather than $2\lambda$ of the family.

Similarly, the right panel of Fig.~\ref{fig:comp-plaq-prod} shows $\alpha(\lambda)$ in Eq.~\eqref{eq:def-alpha} obtained from the winding 't Hooft loop, normalized by $\alpha^{(SC)}=1/16\lambda^4$.
As shown in Sec.~\ref{sec:E-fam}, $\alpha$ within the plaquette product family diverges at $\lambda=1/2$, and $\langle W\rangle=0$ for $\lambda<1/2$.
Our results, on the other hand, remain finite at $\lambda=1/2$, and for $\lambda<1/2$ the ratio $\alpha/\alpha^{(SC)}$ is even smaller than one.
This again shows that the $F=2$ GGFPEPS represents a state more nontrivial than those in the one-parameter plaquette product family.

%%%%%%%%%%%%%%%%%%%%%%%%%%%%%%%%%%%%%%%%%%%%%%%%%%%%%%%%%%
%%%%%%%%%%%%%%%%%%%%%%%%%%%%%%%%%%%%%%%%%%%%%%%%%%%%%%%%%%
\section{Results for infinite volume calculations using CTMRG}\label{sec:ctmrg}
%%%%%%%%%%%%%%%%%%%%%%%%%%%%%%%%%%%%%%%%%%%%%%%%%%%%%%%%%%
%%%%%%%%%%%%%%%%%%%%%%%%%%%%%%%%%%%%%%%%%%%%%%%%%%%%%%%%%%
In this section, we present the results of the infinite volume calculations with the $F=2$ GGFPEPS ansatz using the CTMRG method~\cite{Baxter1968, NishinoO1996, OrusV2009, CorbozWVT2011}.
In the finite volume calculations, we imposed the periodic boundary conditions and used HOTRG to contract the Grassmann TN.
For the infinite volume, we consider an open boundary condition and apply the CTMRG method to contract the Grassmann TN.

\subsection{Simulation details}
\begin{figure}[h]
\centering
\includegraphics[scale=1]{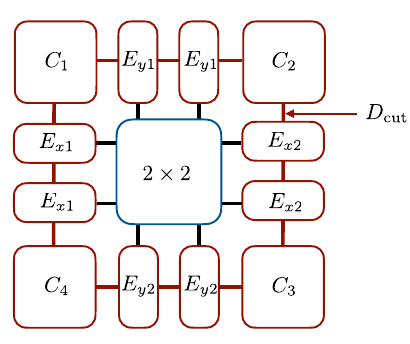}  
\caption{Diagram of the CTMRG environment for calculating the expectation value $\langle \hat{O} \rangle$ of a local observable. We approximate the infinitely large environment by a network consisting of corner ($C_i$) and edge ($E_{xi}, E_{yi}$) tensors. The accuracy of the approximation is controlled by the maximum bond dimension $D_{\mathrm{cut}}$ of the corner and edge tensors.}
\label{fig:CTMRG_method}
\end{figure}

We perform the CTMRG calculation with the $F=2$ GGFPEPS ansatz to obtain the energy density and the TEE in the infinite volume limit.
The CTMRG method is an iterative method to obtain the environment tensors corresponding to the infinite volume starting from an initial guess, as shown in Fig.~\ref{fig:CTMRG_method}.
Since the GGFPEPS ansatz is translationally invariant, we consider only a single-site unit cell also for the CTMRG environment.
We do not explicitly impose the lattice rotational symmetry.
Therefore, we consider four different corner and edge tensors as shown in Fig.~\ref{fig:CTMRG_method}.
Due to the topological nature of the weak coupling phase, the fixed point of the CTMRG iteration could be sensitive to the initial conditions.
To avoid this issue, we prepared the initial corner and edge tensors from the bulk site tensor by imposing the condition that there are no virtual fermions at the boundary.
Once we obtain the converged corner and edge tensors, we can evaluate the energy density and the TEE in the infinite volume limit, similarly to the finite volume case, using local cells of up to $2\times 2$ sites, as shown in Fig.~\ref{fig:CTMRG_method}.
%%%%%%%%%%%%%%%%%%%%%%%%%%%%%%
%%%%%%%%%%%%%%%%%%%%%%%%%%%%%%

The accuracy of the CTMRG method is controlled by the bond dimension $D_{\mathrm{cut}}$ of the corner and edge tensors.\footnote{Note that this $D_{\mathrm{cut}}$ is different from the one in HOTRG in Sec.~\ref{sec:hotrg}.}
In the CTMRG iteration, we repeatedly absorb the bulk tensors into the edge and corner tensors so that the corner and edge tensors represent the environment consisting of many bulk tensors.
To avoid increasing the size of the corner and edge tensors, we construct a projector based on the SVD and keep up to $D_{\mathrm{cut}}$ singular values~\cite{CorbozWVT2011}.
The iteration is repeated until the change of the energy density $E_{tot}$ is smaller than the tolerance $10^{-5}$.
As shown in Appendix~\ref{sec:Dcut_CTMRG}, we confirmed that the energy density and the TEE are converged within the accuracy of $10^{-5}$ at $D_{\mathrm{cut}}=80$.
Therefore, in the following, we show the results obtained with $D_{\mathrm{cut}}=80$.

In the optimization of the variational parameters, we used the same procedure as in the finite volume case, except that the tolerance for the change of $E_{tot}$ was set to $10^{-5}$.
We prepared five different seeds for each $\lambda$ and performed the optimization independently.
The errors and the derivatives of the energy density and the TEE are based on these five independent samples.

We calculated the derivatives of $E_{tot}$ and the TEE with respect to $\lambda$ by the finite difference method using three neighboring data points.
The uncertainties in the derivatives arising from the seed dependence of the optimization were estimated by propagating the standard errors of the means over the five seeds, treating the data at different $\lambda$ as independent.
Error bars indicate one standard error.

\subsection{Results}
\begin{figure}[h]
\centering
\includegraphics[scale=0.6]{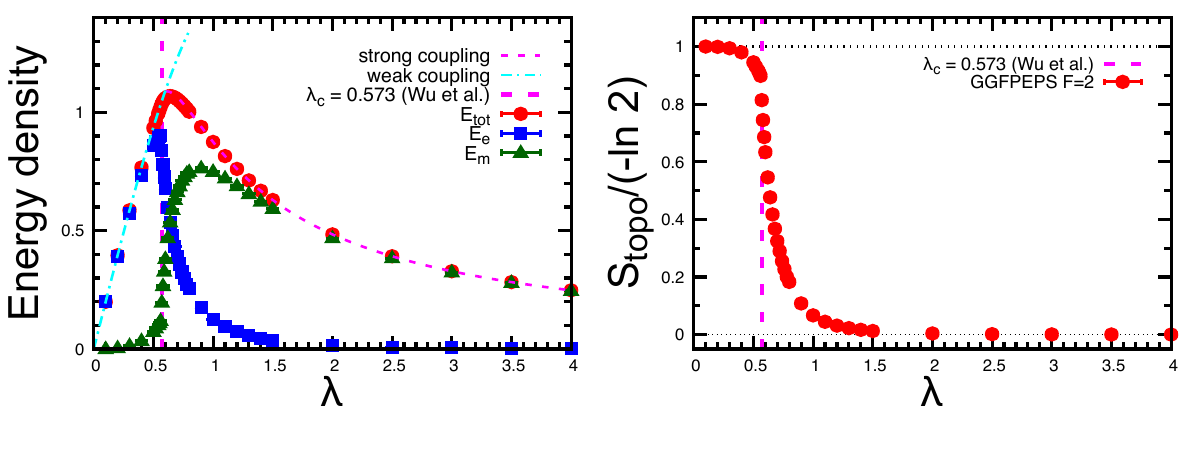}  
\caption{(Left) Energy density $E_{tot}$ and its electric and magnetic contributions, $E_e$ and $E_m$, in the infinite volume limit using the $F=2$ ansatz, compared with the strong coupling expansion (magenta dotted curve) and the weak coupling expansion (cyan dashed curve). (Right) The TEE normalized by $(-\ln 2)$. The magenta vertical dotted line indicates the critical point $\lambda_c=0.573$ obtained by the continuous-time QMC~\cite{wu2012phase}.}
\label{fig:CTMRG-global}
\end{figure}

\begin{figure}[h]
\centering
\includegraphics[scale=0.6]{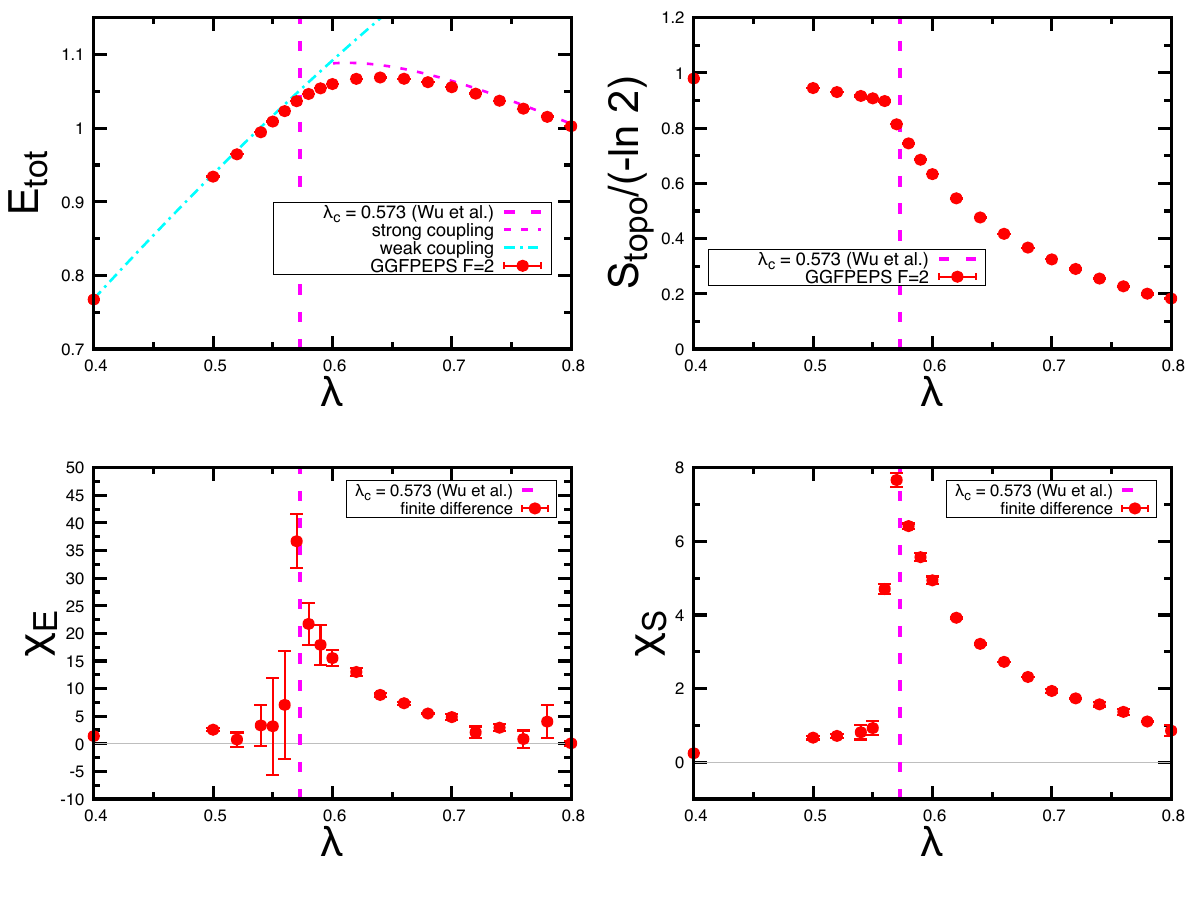}  
\caption{(Left) Energy density $E_{tot}$ (top) and its susceptibility $\chi_E$ (bottom) near $\lambda_c$ in the infinite volume limit using the $F=2$ ansatz. $E_{tot}$ is compared with the strong coupling expansion (magenta dotted curve) and the weak coupling expansion (cyan dashed curve). (Right) The TEE normalized by $(-\ln 2)$ (top) and its susceptibility $\chi_S$ (bottom) near $\lambda_c$. The susceptibilities are obtained by the finite difference method. The magenta vertical dotted line indicates the critical point $\lambda_c=0.573$ obtained by the continuous-time QMC~\cite{wu2012phase}.}
\label{fig:CTMRG-local}
\end{figure}

In the left panel of Fig.~\ref{fig:CTMRG-global}, we show the energy density $E_{tot}$ and its electric and magnetic contributions, $E_e$ and $E_m$, in the infinite volume limit using the $F=2$ ansatz, compared with the strong coupling and weak coupling expansions.
As expected from the large finite volume results, the energy density in the infinite volume is also consistent with the weak coupling expansion for $\lambda\lesssim 0.5$ and with the strong coupling expansion for $\lambda\gtrsim 0.9$.
See also the top-left panel of Fig.~\ref{fig:CTMRG-local}.
The results also agree quantitatively with those for $L=34$.
At $\lambda=0.1$, $2.0$, and $4.0$, $E_{tot}$ in Table~\ref{tab:dcut-dependence-Linfty-energy} agrees with that for $L=34$ in Table~\ref{tab:dcut-dependence-L34} within $10^{-5}$.
At $\lambda=0.6$, the difference is of order $10^{-4}$, which is within the $D_{\mathrm{cut}}$ dependence and the seed spread of the $L=34$ results.
Thus, we conclude that the finite volume effect at $L=34$ is not visible within these uncertainties.
Both $E_e$ and $E_m$ exhibit rapid changes near $\lambda_c=0.573$ obtained by the continuous-time QMC~\cite{wu2012phase}, although these quantities are not discontinuous at a continuous phase transition.
%%%%%%%%%%%%%%%%%%%%%%%%%%%%%%
%%%%%%%%%%%%%%%%%%%%%%%%%%%%%%

The right panel of Fig.~\ref{fig:CTMRG-global} shows the TEE normalized by $(-\ln 2)$.
In addition to exhibiting a rapid change near $\lambda_c=0.573$, the normalized TEE deviates from the expected value of $S_{\mathrm{topo}}/(-\ln 2)=1$ for $\lambda < \lambda_c$.
This deviation is likely due to the finite size of the subsystems used to evaluate the TEE.
Although we consider the infinite volume limit, the TEE is evaluated with the same finite subsystems as in Fig.~\ref{fig:eval-TEE}.
We expect that the TEE approaches $S_{\mathrm{topo}}=-\ln 2$ as the subsystem size increases.
Similarly, the nonzero values of $S_{\mathrm{topo}}$ for $\lambda > \lambda_c$ are likely attributable to the finite size of the subsystems.

To further investigate the nature of the phase transition, we examine the susceptibilities $\chi_E$ of $E_{tot}$ and $\chi_S$ of the TEE by the finite difference method, as shown in Fig.~\ref{fig:CTMRG-local}.
Both susceptibilities exhibit a peak at $\lambda \simeq 0.570 \pm 0.005$, where the uncertainty corresponds to half the spacing between adjacent data points.
This is consistent with $\lambda_c = 0.573$ obtained by the continuous-time QMC~\cite{wu2012phase}.
In a previous bosonic infinite PEPS (iPEPS) calculation of the toric code model in a magnetic field~\cite{Crone:2019ybv}, the critical point was estimated to be $\lambda_c=0.5771(9)$ in our notation.
It is slightly higher than the continuous-time QMC result due to the finite bond dimension $D=3$.
Our $F=2$ GGFPEPS, which corresponds to a bond dimension of $D=4$, yields a critical point slightly closer to the continuous-time QMC result.
However, since the uncertainty in our estimate is based on the spacing between discrete data points, the numerical resolution is insufficient to establish whether our result is more accurate than the previous iPEPS result.
Nevertheless, our GGFPEPS result reproduces the critical point of the continuous-time QMC with an accuracy comparable to that of the previous $D=3$ iPEPS result.
This suggests that the Gaussian ansatz provides a good approximation for the pure $\mathbb{Z}_2$ gauge theory in $(2+1)$ dimensions.
%%%%%%%%%%%%%%%%%%%%%%%%%%%%%%
%%%%%%%%%%%%%%%%%%%%%%%%%%%%%%

%%%%%%%%%%%%%%%%%%%%%%%%%%%%%%%%%%%%%%%%%%%%%%%%%%%%%%%%%%
%%%%%%%%%%%%%%%%%%%%%%%%%%%%%%%%%%%%%%%%%%
\section{Conclusions and Outlook}\label{sec:summary}
%%%%%%%%%%%%%%%%%%%%%%%%%%%%%%%%%%%%%%%%%%
%%%%%%%%%%%%%%%%%%%%%%%%%%%%%%%%%%%%%%%%%%%%%%%%%%%%%%%%%%
In this paper, we have examined the expressive power of the GGFPEPS ansatz for the pure $\mathbb{Z}_2$ gauge theory in $(2+1)$ dimensions.
In the GGFPEPS, the gauge field on each link is represented by coupling its configuration $q$ to the fermion number of the virtual fermions at the two ends of the link, in the same way as a link variable couples to a charge.
The ansatz is manifestly gauge invariant and well suited to the TN methods.
We have reformulated the calculation as the Grassmann TN, in which the whole network including the sum over $q$ is contracted deterministically.
The optimization of the variational parameters is therefore free from statistical errors.
The only systematic error comes from the truncation of the bond dimension to $D_{\rm cut}$, which we have confirmed to be under control.
This allowed us to test the ansatz systematically.
We have introduced the HOTRG to perform large volume calculations up to $L=34$, and the CTMRG to evaluate the infinite volume limit.

We have analyzed the expressive power of the ansatz from two aspects.
First, we have constructed the $F=2$ ansatz so that the one-parameter plaquette product family, which interpolates between the strong and weak coupling ground states, is easily realized.
At $\lambda=0$, we have examined the degeneracy of the singular values in the HOTRG.
The degeneracy is twofold for $F=1$ and eightfold for $F=2$, which reflects the difference in the information carried by the virtual bond in the two ansatzes, while the physical state is the same toric code ground state.
Second, we have compared the optimized state with the plaquette product family.
Around the critical point, the optimized state has a lower energy than any state in the family.
This shows that the state goes outside this one-parameter family.
We find that the $F=2$ ansatz can represent such a nontrivial state near the critical point.

As physical observables, we have calculated the energy density, the TEE, and the winding 't Hooft loop.
In the small volumes, the results agree with those of the ED.
In the large volumes, the results reproduce the perturbative behaviors in both the strong and weak coupling limits.
From the susceptibilities of the energy density and the TEE, we have estimated the critical coupling of the topological phase transition.
The estimates capture the continuous-time QMC result within their errors.
The CTMRG calculation in the infinite volume limit also shows the peaks of the susceptibilities at the coupling consistent with the continuous-time QMC result.

There are several directions for future work.
The first one is to include dynamical fermions.
The GGFPEPS ansatz with dynamical fermions has already been formulated and studied by the VMC~\cite{Kelman:2024exo}.
The Grassmann TN is naturally applicable also to this case, where the fermionic degrees of freedom are placed on the site tensors.
It enables a more precise determination of the variational parameters.
Then, it is an important task to examine the expressive power of the ansatz in the same way as in this paper.

The second one is to extend the gauge group.
The extension of the method to other discrete gauge groups is straightforward.
In the previous study of the $\mathbb{Z}_3$ gauge theory~\cite{Emonts:2020drm}, the weak coupling region was reported to be challenging for the ansatz with a small number of virtual fermions.
It is worth reconsidering the construction of the ansatz based on the analysis in this paper, as we did for the $F=2$ ansatz in Sec.~\ref{sec:more-F2}.
For continuous gauge groups, the physical leg $q$ takes continuous values.
Then, the contraction of the physical legs by the TN requires an additional treatment of $q$.
In this respect, the VMC has an advantage, since it samples the continuous variable directly.

The third one is to combine the Grassmann TN with the MC sampling of the gauge-field configurations.
Our method has two limitations.
The volume is restricted to the values reached by the HOTRG steps, $L=2+2^k$.
The measurement of extended observables is costly, as we have seen for the TEE.
The MC sampling of $q$ in Refs.~\cite{Emonts:2022yom, Kelman:2024gjt} has neither limitation, while the optimization suffers from the statistical error.
We therefore propose the following hybrid method.
The variational parameters are determined by the Grassmann TN without statistical errors.
This may provide a more accurately optimized variational state than the one obtained by the VMC.
Then, the expectation values are evaluated on this state by the MC sampling of $q$.
In this way, the calculation can be done in any volume, and extended observables can be measured, such as the spatial Wilson loop and the R\'enyi version of the TEE with larger subsystems.
We leave these extensions for future work.

\acknowledgments
We would like to thank P.~Emonts, A.~Kelman, and E.~Zohar for providing us with the details of Ref.~\cite{Emonts:2022yom} and their code~\cite{Kelman:2026goj}.
We thank J.~Vidal for bringing the interesting related works~\cite{Dusuel:2015sta, Vanderstraeten:2017noh} to our attention.
We also thank A.~Yosprakob for his contribution at the early stage of this work.
E.~I. and T.~O. are supported by the Center of Innovations for Sustainable Quantum AI (JST Grant No.~JPMJPF2221). % SQAI
E.~I. is also supported by JSPS KAKENHI Grant Numbers
JP23H05439 % Kiban-S
and JP25K01001, % Kiban-B
JST Grant Number
JPMJCR24I3, % CREST
the Program for Promoting Researches on the Supercomputer ``Fugaku'' (Simulation for basic science: from fundamental laws of particles to creation of nuclei, and Simulation for basic science: approaching the new quantum era), and the Joint Institute for Computational Fundamental Science (JICFuS), Grant Number JPMXP1020230411. % Fugaku
T.~O. is also supported by MEXT Next-Generation Computational Science Global Reach \& Innovation Program Japan (Grant No.~JPMXP1920260301).
The numerical calculations were carried out on the computer cluster at RIKEN iTHEMS, and on a workstation and the computer Heian at Yukawa Institute for Theoretical Physics.
This work is also supported by the Center for Gravitational Physics and Quantum Information (CGPQI) at Yukawa Institute for Theoretical Physics.
The English of the manuscript was improved with the assistance of Claude (Anthropic) and ChatGPT Astra (OpenAI).
The authors take full responsibility for the content.

\appendix
%%%%%%%%%%%%%%%%%%%%%%%%%%%%%%%%%%%%%%%%%%%%%%%%%%%%%%%%%%
\section{$F=1$ ansatz}\label{sec:F1-ansatz}
%%%%%%%%%%%%%%%%%%%%%%%%%%%%%%%%%%%%%%%%%%%%%%%%%%%%%%%%%%
\subsection{Perturbative constraint for the toric code ground state}\label{sec:F1-pert}
%%%%%%%%%%%%%%%%%%%%%%%%%%%%%
In the previous work, Ref.~\cite{Emonts:2022yom}, the authors argued that the $F=1$ ansatz cannot represent the toric code ground state $|\psi_B\rangle$ by matching the amplitudes of low-order plaquette excitations.
The amplitude of two plaquettes sharing a site leads to a contradiction, as shown in Appendix A.2 of Ref.~\cite{Emonts:2022yom}.~\footnote{See also Ref.~\cite{Vanderstraeten:2017noh} for a bosonic tensor network representation of the toric code ground state.}
In this subsection, we repeat the same perturbative analysis and show that there is a consistent solution for the toric code ground state even in the $F=1$ ansatz.
We also confirm numerically that the solution gives the exact toric code ground state.

For the $F=1$ ansatz, the matrix $T_{\alpha\beta}$ in the site tensor $A(\vb*{x})$, with the legs ordered as $(r,u,l,d)$, is given by
\beq
T=
\begin{pmatrix}
0 & -z & -iy & -iz \\
z & 0 & -iz & y \\
iy & iz & 0 & z \\
iz & -y & -z & 0
\end{pmatrix}.\label{eq:def-T-matrix-F1}
\eeq
It contains two complex parameters $y$ and $z$~\cite{Emonts:2022yom}.
The matrices in $w(\vb*{x},i)$ are $W^1=1$ and $W^2=\eta^2$ with $\eta=e^{i\pi/4}$ as in Sec.~\ref{sec:more-F2}.
Expanding the $A$ tensor in this $F=1$ ansatz, we obtain
\beq
A &=&1-2zr^\dag u^\dag -2i y r^\dag l^\dag -2i zr^\dag d^\dag -2i z u^\dag l^\dag +2y u^\dag d^\dag+2z l^\dag d^\dag + 4 (iy^2 -2z^2)r^\dag u^\dag l^\dag d^\dag \nonumber\\
&=& 1 - 2z\,\legbox{r,u} - 2iy\,\legbox{r,l} - 2iz\,\legbox{r,d}
    - 2iz\,\legbox{u,l} + 2y\,\legbox{u,d} + 2z\,\legbox{l,d}
    + 4\left(iy^2-2z^2\right)\legbox{r,u,l,d}. \nonumber\\
    \label{eq:A-tensor-expand-F1}
\eeq
Here the coefficient of the four-leg term $r^\dag u^\dag l^\dag d^\dag$ receives contributions from the three ways of pairing the four legs, $(ru)(ld)$, $(rl)(ud)$, and $(rd)(ul)$, with the signs from the reordering of the fermionic operators.
From this expression, the parameters can be interpreted as follows: $z$ is associated with corners where the links turn, while $y$ is associated with straight segments connecting links.

Using $\mathcal{O}$ defined by writing the GGFPEPS state as in Eq.~\eqref{eq:def-O}, we write
\beq
\mathcal{O}| \psi_E \rangle = \sum_{ \{p \} }c_{\{p \} } \left( \prod_{p \in \{p \} } B_p \right) | \psi_E \rangle.
\eeq
Here, $\{ p \}$ runs over all sets of excited plaquettes, and we refer to the coefficients $c_{ \{ p \} }$ as amplitudes.
In the toric code ground state, all the amplitudes should be equal.
Here, we normalize them to $1$.
We now examine perturbatively whether there exist variational parameters $y$ and $z$ satisfying this condition.

\begin{enumerate}
\item A single excited plaquette:\\
From the expansion of the $A$ tensor, a single plaquette is formed by the product of four corners,
\beq
\left(-2z\,\legbox{r,u}\right)\times\left(-2iz\,\legbox{r,d}\right)
\times\left(2z\,\legbox{l,d}\right)\times\left(-2iz\,\legbox{u,l}\right)
\;=\; 16z^4\,\plqbox .
\eeq
Furthermore, this diagram contains two vertical links, connected by $w(\vb*{x},2)$, which gives a factor of $\eta^4=-1$.
Reordering the virtual fermions in $w(\vb*{x},1)$ and $w(\vb*{x},2)$ before the contraction may also produce a sign, but in this case it is $+1$.
We therefore obtain
\beq
z^4 = -1/16 \label{eq:z4-cond-F1}
\eeq
from the equal amplitude condition.

\item Two separated excited plaquettes:\\
If the two plaquettes share neither links nor sites, the amplitude is simply the product,
\beq
\left( 16z^4\,\plqbox \right) \times \left( 16z^4\,\plqbox \right) = 256z^8 \left( \plqbox \,\plqbox \right).
\eeq
The condition $256z^8=1$ gives nothing new beyond Eq.~\eqref{eq:z4-cond-F1}.

\item Two plaquettes sharing a link:\\
Let us consider two excited plaquettes aligned horizontally, which form a rectangle.
We obtain
\beq
&&\left( - 2iz\,\legbox{r,d} \right) \times \left( - 2iy\,\legbox{r,l} \right) \times \left( 2z\,\legbox{l,d} \right) \times \left( - 2iz\,\legbox{u,l} \right) \times \left( - 2iy\,\legbox{r,l} \right) \times \left( - 2z\,\legbox{r,u} \right) \nonumber\\
&=& -64z^4 y^2 \rectboxH .
\eeq
In this case, $w(\vb*{x},2)$ appears twice in the contractions of the vertical links, which gives $(\eta^2)^2=-1$.
The sign from the reordering is $+1$.
We therefore obtain the condition
\beq
64z^4 y^2  =1 .
\eeq
Substituting Eq.~\eqref{eq:z4-cond-F1}, we find
\beq
y^2=-1/4 \label{eq:y-cond-F1}.
\eeq

Similarly, for two plaquettes aligned vertically, the coefficient is
\beq
&&\left(-2z\,\legbox{r,u}\right)\times\left(2y\,\legbox{u,d}\right)
\times\left(-2iz\,\legbox{r,d}\right)\times\left(2z\,\legbox{l,d}\right)
\times\left(2y\,\legbox{u,d}\right)\times\left(-2iz\,\legbox{u,l}\right)\nonumber\\
&=& 64z^4y^2\,\rectboxV.
\eeq
In this case, $w(\vb*{x},2)$ appears four times, which gives $(\eta^2)^4 = 1$, and the reordering produces no sign either.
We thus obtain
\beq
64z^4 y^2  =1,
\eeq
which is the same condition on the variational parameters.

\item Two plaquettes sharing a site:\\
This configuration is obtained by
\beq
&& \left(-2z\,\legbox{r,u}\right)\times\left(-2iz\,\legbox{u,l}\right)
\times\left(-2iz\,\legbox{r,d}\right)
\times\Big(4\big(iy^2-2z^2\big)\,\legbox{r,u,l,d}\Big) \nonumber\\
&&\times\left(-2iz\,\legbox{u,l}\right)\times\left(-2iz\,\legbox{r,d}\right)
\times\left(2z\,\legbox{l,d}\right)\nonumber\\
&=&\big(512z^8-256i\,z^6y^2\big)\,\diagbox .
\eeq
Since there are four vertical links, the phase from $w(\vb*{x},2)$ is $\eta^8=1$.
The sign from the reordering is $+1$, so we obtain the condition
\beq
512z^8-256i\,z^6y^2 = 1 
\eeq
for the variational parameters.
Substituting $y^2=-1/4$, we find that $z^2 = -i/4$ gives a solution consistent with all the conditions obtained so far.
Note that the corresponding coefficient in Ref.~\cite{Emonts:2022yom} is $256(z^8-iz^6y^2)$, with which the conditions become inconsistent.
The difference comes from the coefficient of the four-leg term in Eq.~\eqref{eq:A-tensor-expand-F1}, which receives contributions from all the three pairings of the legs as mentioned above.
\end{enumerate}

From the above analysis, we find that the $F=1$ ansatz can represent the toric code ground state if we choose
\beq
y^2=-1/4, \quad z^2 = -i/4.\label{eq:F1-init-param}
\eeq
There are apparently four solutions, $y=\pm i/2$ and $z=\pm (1-i)/2\sqrt{2}$.
To distinguish the resulting states, we consider the amplitude of each configuration, 
$c_{\{p\}}\propto (\pm z)^{n_z} (\pm y)^{n_y}$, where $n_z$ and $n_y$ are the numbers of corners and straight segments, respectively.
Since the number of corners $n_z$ is always even for closed loops, changing the sign as $z \rightarrow -z$ does not change the state.
On the other hand, changing the sign of $y$ as $y \rightarrow -y$ can change the expectation value of the winding 't Hooft loop, Eq.~\eqref{eq:def-W}.
For even lattice size $L$, $n_y$ is even also for the loops winding around the torus, so that the state is independent of the sign of $y$.
For odd $L$, however, a straight winding loop has $n_y=L$, and the sign of its amplitude changes. 
Then we can create a state orthogonal to the original one by changing the sign of the parameter $y$.
These two states correspond to two of the four degenerate toric code ground states.

In addition, we have evaluated the state with the parameters in Eq.~\eqref{eq:F1-init-param} without optimization for $L=3$, $4$, and $6$, and confirmed that $\langle B_p\rangle=1$ and $S_{\rm topo}=-\ln2$ hold to machine precision.
Since $B_p$ has eigenvalues $\pm1$, this means that the state is an exact eigenstate of all $B_p$ with eigenvalue $+1$, and hence a toric code ground state.

\subsection{Numerical results}\label{sec:F1-numerical}
We perform numerical calculations with the $F=1$ ansatz for $L=4$.
The initial values of the variational parameters are taken to be the solution in Eq.~\eqref{eq:F1-init-param} with small random numbers added.
Here, we use five independent seeds to see the initial parameter dependence.
Since $L=4$ is even, the sign of $y$ does not matter as discussed above.
The other settings of the optimization are the same as those for the $F=2$ ansatz in Sec.~\ref{sec:sim-detail}.

\begin{figure}[htb]
\centering
\includegraphics[scale=0.65]{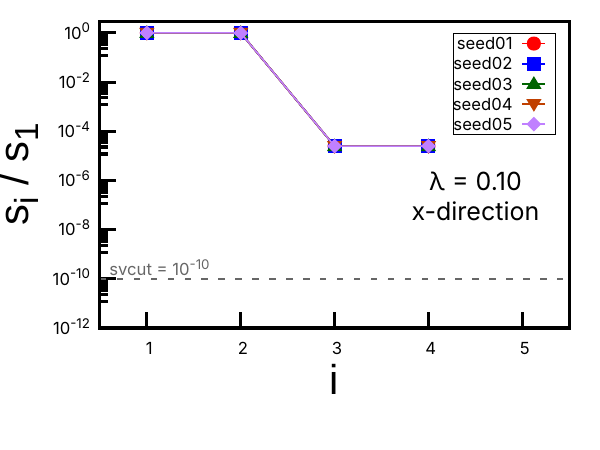}
\includegraphics[scale=0.65]{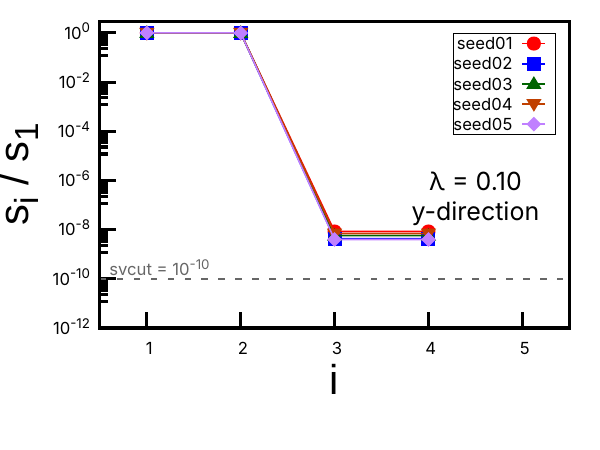}
\caption{Singular values $s_i/s_1$ in the $x$-direction (left) and $y$-direction (right) coarse-graining steps for $L=4$ with the $F=1$ ansatz at $\lambda=0.1$. The results for the five seeds are overlaid.}\label{fig:sv-scaling-F1}
\end{figure}

We first look at the singular values at $\lambda=0.1$.
For $L=4$, HOTRG is applied once in the $x$ direction and then once in the $y$ direction.
Figure~\ref{fig:sv-scaling-F1} shows the singular values defining the isometries in the two steps, normalized by the largest one $s_1$.
%%%%%%%%%%%%%%%%%%%%%%%%%%%%%%
%%%%%%%%%%%%%%%%%%%%%%%%%%%%%%

For the $F=1$ ansatz, the bond dimension of $R$ is only $4$, since each leg of \texttt{Tii} has the effective dimension $2^{2F-1}=2$.
As discussed in Sec.~\ref{sec:sv-hotrg}, the leading singular values are expected to be twofold degenerate for $F=1$, in contrast to the eightfold degeneracy for $F=2$.
This is indeed seen in Fig.~\ref{fig:sv-scaling-F1}.
The seed dependence is larger in the second ($y$-direction) step than in the first ($x$-direction) step.

\begin{figure}[htb]
\centering
\includegraphics[scale=0.6]{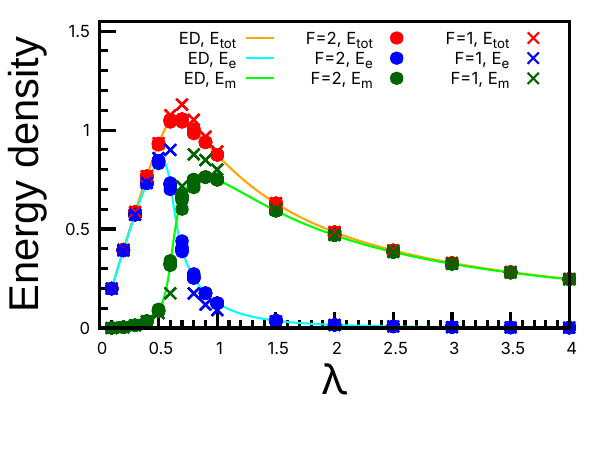}  
\includegraphics[scale=0.6]{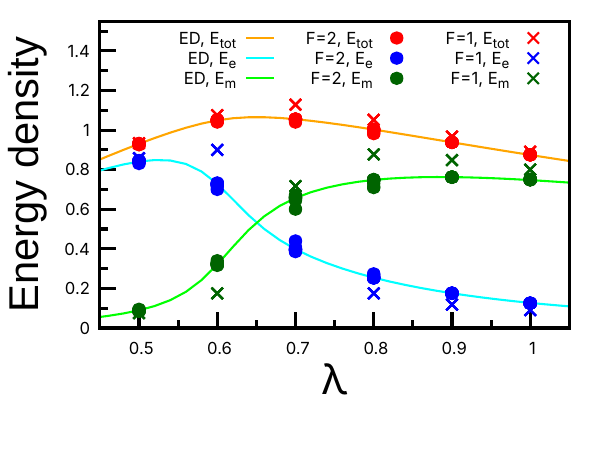}
\caption{Energy density $E_{tot}$ and its electric and magnetic parts, $E_e$ and $E_m$, for $L=4$ with the $F=1$ and $F=2$ ansatzes, compared with ED (solid curves). The right panel is an enlarged plot of the region near $\lambda_c$.}\label{fig:comp-E-F1-F2}
\end{figure}

Figure~\ref{fig:comp-E-F1-F2} shows the results for a wide range of $\lambda$ from the strong to weak coupling regions.
Here, the energy density $E_{tot}$ and its electric and magnetic parts, $E_e$ and $E_m$, are plotted for the $F=1$ and $F=2$ ansatzes and compared with ED (solid curves).
%%%%%%%%%%%%%%%%%%%%%%%%%%%%%%
%%%%%%%%%%%%%%%%%%%%%%%%%%%%%%
For $\lambda\lesssim0.3$ and $\lambda\gtrsim1.5$, there is almost no difference between $F=1$ and $2$, and both agree well with ED.
However, the enlarged plot in the right panel shows that the $F=1$ ansatz changes abruptly from a state with an excess of $E_e$ to a state with an excess of $E_m$ compared with ED.
Note that the variational principle holds only for $E_{tot}$, and $E_e$ and $E_m$ can individually be lower than the ED values.
This indicates that the two variational parameters of the $F=1$ ansatz cannot describe the intermediate state in the transition region, where the two terms compete.
A similar behavior was reported in Ref.~\cite{Emonts:2022yom} for the $2\times2$ lattice.
On the other hand, we can see that the $F=2$ ansatz reproduces not only $E_{tot}$ but also $E_e$ and $E_m$ well in this region.

\section{$D_{\mathrm{cut}}$ dependence in the CTMRG calculation}\label{sec:Dcut_CTMRG}
In this appendix, we examine the $D_{\mathrm{cut}}$ dependence of the energy density and the TEE in the infinite volume limit using the $F=2$ ansatz.
We pick a single set of optimized parameters for each $\lambda$ and perform the CTMRG calculation with $D_{\mathrm{cut}}=16$, $24$, $32$, $48$, $64$, and $80$ to evaluate them.

Tables~\ref{tab:dcut-dependence-Linfty-energy} and~\ref{tab:dcut-dependence-Linfty-entropy} show the $D_{\mathrm{cut}}$ dependence of $E_{tot}$ and the TEE, respectively, for $\lambda=0.1$, $0.6$, $2.0$, and $4.0$.
Except for $\lambda=0.6$, which is near $\lambda_c = 0.573$, the results are almost converged even at the smallest $D_{\mathrm{cut}}=16$.
For $\lambda=0.6$, the results are slightly different between $D_{\mathrm{cut}}=16$ and $D_{\mathrm{cut}}=24$, but they are eventually converged for $D_{\mathrm{cut}}\ge 32$, at least within the accuracy of $10^{-5}$.
These results indicate that the $D_{\mathrm{cut}}$ dependence is well under control for $D_{\mathrm{cut}}=80$, which is used in the main text.

\begin{table}[htbp]
  \centering
  \begin{tabular}{crrrr}
    \hline
    $D_{\rm cut}$ & $\lambda=0.1$ & $\lambda=0.6$ & $\lambda=2.0$ & $\lambda=4.0$ \\
    \hline
    16 & 0.199503471960 & 1.060085828991 & 0.484374990655 & 0.248046875415 \\
    24 & 0.199503471952 & 1.060081875059 & 0.484374990655 & 0.248046875415 \\
    32 & 0.199503471949 & 1.060081892047 & 0.484374990655 & 0.248046875415 \\
    48 & 0.199503471931 & 1.060081891931 & 0.484374990655 & 0.248046875415 \\
    64 & 0.199503471940 & 1.060081891410 & 0.484374990655 & 0.248046875415 \\
    80 & 0.199503471940 & 1.060081891390 & 0.484374990655 & 0.248046875415 \\
    \hline
  \end{tabular}
    \caption{$D_{\rm cut}$ dependence of $E_{tot}$ for $L=\infty$ with the $F=2$ ansatz.}
  \label{tab:dcut-dependence-Linfty-energy}
\end{table}

\begin{table}[htbp]
  \centering
  \begin{tabular}{crrrr}
    \hline
    $D_{\rm cut}$ & $\lambda=0.1$ & $\lambda=0.6$ & $\lambda=2.0$ & $\lambda=4.0$ \\
    \hline
    16 & 0.999924234794 & 0.631977651444 & 0.003937180423 & 0.000244519408 \\
    24 & 0.999924234795 & 0.631975718104 & 0.003937180423 & 0.000244519408 \\
    32 & 0.999924234795 & 0.631975785451 & 0.003937180423 & 0.000244519408 \\
    48 & 0.999924234795 & 0.631975782945 & 0.003937180423 & 0.000244519408 \\
    64 & 0.999924234795 & 0.631975782848 & 0.003937180423 & 0.000244519408 \\
    80 & 0.999924234795 & 0.631975782844 & 0.003937180423 & 0.000244519408 \\
    \hline
  \end{tabular}
  \caption{$D_{\rm cut}$ dependence of $S_{\mathrm{topo}}/(-\ln 2)$ for $L=\infty$ with the $F=2$ ansatz.}
  \label{tab:dcut-dependence-Linfty-entropy}
\end{table}

\clearpage
% Bibliography

%% [A] Recommended: using JHEP.bst file
\bibliographystyle{JHEP}
\bibliography{PEPS}

@article{Dusuel:2015sta,
    author = "Dusuel, S. and Vidal, J.",
    title = "{Mean-field ansatz for topological phases with string tension}",
    eprint = "1506.03259",
    archivePrefix = "arXiv",
    primaryClass = "cond-mat.str-el",
    doi = "10.1103/PhysRevB.92.125150",
    journal = "Phys. Rev. B",
    volume = "92",
    number = "12",
    pages = "125150",
    year = "2015"
}

@article{Vanderstraeten:2017noh,
    author = {Vanderstraeten, Laurens and Mari{\"e}n, Micha{\"e}l and Haegeman, Jutho and Schuch, Norbert and Vidal, Julien and Verstraete, Frank},
    title = "{Bridging Perturbative Expansions with Tensor Networks}",
    eprint = "1703.04112",
    archivePrefix = "arXiv",
    primaryClass = "cond-mat.str-el",
    doi = "10.1103/PhysRevLett.119.070401",
    journal = "Phys. Rev. Lett.",
    volume = "119",
    pages = "070401",
    year = "2017"
}

@article{OrusV2009,
  title = {Simulation of two-dimensional quantum systems on an infinite lattice revisited: Corner transfer matrix for tensor contraction},
  author = {Or{\'u}s, Rom{\'a}n and Vidal, Guifr{\'e}},
  journal = {Phys. Rev. B},
  volume = {80},
  issue = {9},
  pages = {094403},
  numpages = {4},
  year = {2009},
  month = {Sep},
  publisher = {American Physical Society},
  doi = {10.1103/PhysRevB.80.094403},
  url = {https://link.aps.org/doi/10.1103/PhysRevB.80.094403}
}

@article{CorbozWVT2011,
  title = {Stripes in the two-dimensional $t$-$J$ model with infinite projected entangled-pair states},
  author = {Corboz, Philippe and White, Steven R. and Vidal, Guifr{\'e} and Troyer, Matthias},
  journal = {Phys. Rev. B},
  volume = {84},
  issue = {4},
  pages = {041108(R)},
  numpages = {5},
  year = {2011},
  month = {Jul},
  publisher = {American Physical Society},
  doi = {10.1103/PhysRevB.84.041108},
  url = {https://link.aps.org/doi/10.1103/PhysRevB.84.041108}
}

@article{NishinoO1996,
  author  = {Nishino, Tomotoshi and Okunishi, Kouichi},
  title   = {Corner Transfer Matrix Renormalization Group Method},
  journal = {J. Phys. Soc. Jpn.},
  volume  = {65},
  number  = {4},
  pages   = {891--894},
  year    = {1996},
  doi     = {10.1143/JPSJ.65.891}
}

@article{Baxter1968,
  author  = {Baxter, R. J.},
  title   = {Dimers on a Rectangular Lattice},
  journal = {J. Math. Phys.},
  volume  = {9},
  number  = {4},
  pages   = {650--654},
  year    = {1968},
  doi     = {10.1063/1.1664623}
}

@article{Gomelski:2025enu,
    author = "Gomelski, Itay and Elyovich, Jonathan and Kelman, Ariel and Zohar, Erez and Emonts, Patrick",
    title = "{Algorithmic aspects of gauged Gaussian fermionic projected entangled pair states: Gauge fixing and translation invariance}",
    eprint = "2512.13812",
    archivePrefix = "arXiv",
    primaryClass = "hep-lat",
    doi = "10.1103/xjvk-hks4",
    journal = "Phys. Rev. Res.",
    volume = "8",
    number = "2",
    pages = "023240",
    year = "2026"
}

@article{Kelman:2026goj,
    author = "Kelman, Ariel and Gomelski, Itay and Elyovich, Jonathan and Zohar, Erez and Emonts, Patrick",
    title = "{ggpeps: A package for Gauged Gaussian PEPS}",
    eprint = "2609.06486",
    archivePrefix = "arXiv",
    primaryClass = "hep-lat",
    month = "9",
    year = "2026"
}

@article{Kelman:2024exo,
    author = "Kelman, Ariel and Borla, Umberto and Emonts, Patrick and Zohar, Erez",
    title = "{Projected entangled pair states for lattice gauge theories with dynamical fermions}",
    eprint = "2412.16951",
    archivePrefix = "arXiv",
    primaryClass = "hep-lat",
    doi = "10.1038/s42005-025-02482-7",
    journal = "Commun. Phys.",
    volume = "9",
    number = "1",
    pages = "50",
    year = "2026"
}

@article{Emonts:2020drm,
    author = "Emonts, Patrick and Ba{\~n}uls, Mari Carmen and Cirac, Ignacio and Zohar, Erez",
    title = "{Variational Monte Carlo simulation with tensor networks of a pure $\mathbb{Z}_3$ gauge theory in (2+1)d}",
    eprint = "2008.00882",
    archivePrefix = "arXiv",
    primaryClass = "quant-ph",
    doi = "10.1103/PhysRevD.102.074501",
    journal = "Phys. Rev. D",
    volume = "102",
    number = "7",
    pages = "074501",
    year = "2020"
}

@article{Zohar:2017yxl,
    author = "Zohar, Erez and Cirac, J. Ignacio",
    title = "{Combining tensor networks with Monte Carlo methods for lattice gauge theories}",
    eprint = "1710.11013",
    archivePrefix = "arXiv",
    primaryClass = "quant-ph",
    doi = "10.1103/PhysRevD.97.034510",
    journal = "Phys. Rev. D",
    volume = "97",
    number = "3",
    pages = "034510",
    year = "2018"
}

@article{Zohar:2015eda,
    author = "Zohar, Erez and Burrello, Michele and Wahl, Thorsten B. and Cirac, J. Ignacio",
    title = "{Fermionic Projected Entangled Pair States and Local U(1) Gauge Theories}",
    eprint = "1507.08837",
    archivePrefix = "arXiv",
    primaryClass = "quant-ph",
    doi = "10.1016/j.aop.2015.10.009",
    journal = "Annals Phys.",
    volume = "363",
    pages = "385--439",
    year = "2015"
}

@article{Kadam:2022ipf,
    author = "Kadam, Saurabh V. and Raychowdhury, Indrakshi and Stryker, Jesse R.",
    title = "{Loop-string-hadron formulation of an SU(3) gauge theory with dynamical quarks}",
    eprint = "2212.04490",
    archivePrefix = "arXiv",
    primaryClass = "hep-lat",
    reportNumber = "UMD-PP-022-11",
    doi = "10.1103/PhysRevD.107.094513",
    journal = "Phys. Rev. D",
    volume = "107",
    number = "9",
    pages = "094513",
    year = "2023"
}

@article{Chandrasekharan:1996ih,
    author = "Chandrasekharan, S. and Wiese, U. J.",
    title = "{Quantum link models: A Discrete approach to gauge theories}",
    eprint = "hep-lat/9609042",
    archivePrefix = "arXiv",
    reportNumber = "MIT-CTP-2573, CTP-2573",
    doi = "10.1016/S0550-3213(97)00006-0",
    journal = "Nucl. Phys. B",
    volume = "492",
    pages = "455--474",
    year = "1997"
}

@article{Berenstein:2022wlm,
    author = "Berenstein, David and Kawai, Hiroki and Brower, Richard",
    title = "{U(1) fields from qubits: An approach via D-theory algebra}",
    eprint = "2201.02412",
    archivePrefix = "arXiv",
    primaryClass = "hep-th",
    doi = "10.1103/PhysRevD.110.014506",
    journal = "Phys. Rev. D",
    volume = "110",
    number = "1",
    pages = "014506",
    year = "2024"
}

@article{Bauer:2021gek,
    author = "Bauer, Christian W. and Grabowska, Dorota M.",
    title = "{Efficient representation for simulating U(1) gauge theories on digital quantum computers at all values of the coupling}",
    eprint = "2111.08015",
    archivePrefix = "arXiv",
    primaryClass = "hep-ph",
    reportNumber = "CERN-TH-2021-188",
    doi = "10.1103/PhysRevD.107.L031503",
    journal = "Phys. Rev. D",
    volume = "107",
    number = "3",
    pages = "L031503",
    year = "2023"
}

@article{Ciavarella:2025bsg,
    author = "Ciavarella, Anthony N. and Burbano, Ivan M. and Bauer, Christian W.",
    title = "{Efficient truncations of SU(Nc) lattice gauge theory for quantum simulation}",
    eprint = "2503.11888",
    archivePrefix = "arXiv",
    primaryClass = "hep-lat",
    doi = "10.1103/ylqb-phv5",
    journal = "Phys. Rev. D",
    volume = "112",
    number = "5",
    pages = "054514",
    year = "2025"
}

@article{Zache:2023dko,
    author = "Zache, Torsten V. and Gonz{\'a}lez-Cuadra, Daniel and Zoller, Peter",
    title = "{Quantum and Classical Spin-Network Algorithms for q-Deformed Kogut-Susskind Gauge Theories}",
    eprint = "2304.02527",
    archivePrefix = "arXiv",
    primaryClass = "quant-ph",
    doi = "10.1103/PhysRevLett.131.171902",
    journal = "Phys. Rev. Lett.",
    volume = "131",
    number = "17",
    pages = "171902",
    year = "2023"
}

@article{Halimeh:2024bth,
    author = {Halimeh, Jad C. and Hanada, Masanori and Matsuura, Shunji and Nori, Franco and Rinaldi, Enrico and Sch{\"a}fer, Andreas},
    title = "{A universal framework for the quantum simulation of Yang{\textendash}Mills theory}",
    eprint = "2411.13161",
    archivePrefix = "arXiv",
    primaryClass = "quant-ph",
    reportNumber = "RIKEN-iTHEMS-Report-24",
    doi = "10.1038/s42005-025-02421-6",
    journal = "Commun. Phys.",
    volume = "9",
    number = "1",
    pages = "67",
    year = "2026"
}

@article{Hayata:2023bgh,
    author = "Hayata, Tomoya and Hidaka, Yoshimasa",
    title = "{q deformed formulation of Hamiltonian SU(3) Yang-Mills theory}",
    eprint = "2306.12324",
    archivePrefix = "arXiv",
    primaryClass = "hep-lat",
    reportNumber = "KEK-TH-2533, J-PARC-TH-0291, RIKEN-iTHEMS-Report-23",
    doi = "10.1007/JHEP09(2023)123",
    journal = "JHEP",
    volume = "09",
    pages = "123",
    year = "2023"
}

@article{Kaplan:2018vnj,
    author = "Kaplan, David B. and Stryker, Jesse R.",
    title = "{Gauss{\textquoteright}s law, duality, and the Hamiltonian formulation of U(1) lattice gauge theory}",
    eprint = "1806.08797",
    archivePrefix = "arXiv",
    primaryClass = "hep-lat",
    reportNumber = "INT-PUB-18-034",
    doi = "10.1103/PhysRevD.102.094515",
    journal = "Phys. Rev. D",
    volume = "102",
    number = "9",
    pages = "094515",
    year = "2020"
}

@article{Bergner:2024qjl,
    author = "Bergner, Georg and Hanada, Masanori and Rinaldi, Enrico and Schafer, Andreas",
    title = "{Toward QCD on quantum computer: orbifold lattice approach}",
    eprint = "2401.12045",
    archivePrefix = "arXiv",
    primaryClass = "hep-th",
    reportNumber = "RIKEN-iTHEMS-Report-24",
    doi = "10.1007/JHEP05(2024)234",
    journal = "JHEP",
    volume = "05",
    pages = "234",
    year = "2024"
}

@article{Brower:1997ha,
    author = "Brower, R. and Chandrasekharan, S. and Wiese, U. J.",
    title = "{QCD as a quantum link model}",
    eprint = "hep-th/9704106",
    archivePrefix = "arXiv",
    reportNumber = "MIT-CTP-2623",
    doi = "10.1103/PhysRevD.60.094502",
    journal = "Phys. Rev. D",
    volume = "60",
    pages = "094502",
    year = "1999"
}

@article{Raychowdhury:2019iki,
    author = "Raychowdhury, Indrakshi and Stryker, Jesse R.",
    title = "{Loop, string, and hadron dynamics in SU(2) Hamiltonian lattice gauge theories}",
    eprint = "1912.06133",
    archivePrefix = "arXiv",
    primaryClass = "hep-lat",
    reportNumber = "INT-PUB-19-059, UMD-PP-019-08",
    doi = "10.1103/PhysRevD.101.114502",
    journal = "Phys. Rev. D",
    volume = "101",
    number = "11",
    pages = "114502",
    year = "2020"
}

@article{Zache:2021ggw,
    author = "Zache, Torsten V. and Van Damme, Maarten and Halimeh, Jad C. and Hauke, Philipp and Banerjee, Debasish",
    title = "{Toward the continuum limit of a 1+1D quantum link Schwinger model}",
    eprint = "2104.00025",
    archivePrefix = "arXiv",
    primaryClass = "hep-lat",
    doi = "10.1103/PhysRevD.106.L091502",
    journal = "Phys. Rev. D",
    volume = "106",
    number = "9",
    pages = "L091502",
    year = "2022"
}

@article{Paulson:2020zjd,
    author = "Paulson, Danny and others",
    title = "{Simulating 2D Effects in Lattice Gauge Theories on a Quantum Computer}",
    eprint = "2008.09252",
    archivePrefix = "arXiv",
    primaryClass = "quant-ph",
    doi = "10.1103/PRXQuantum.2.030334",
    journal = "PRX Quantum",
    volume = "2",
    number = "3",
    pages = "030334",
    year = "2021"
}

@article{Zohar:2014qma,
    author = "Zohar, Erez and Burrello, Michele",
    title = "{Formulation of lattice gauge theories for quantum simulations}",
    eprint = "1409.3085",
    archivePrefix = "arXiv",
    primaryClass = "quant-ph",
    doi = "10.1103/PhysRevD.91.054506",
    journal = "Phys. Rev. D",
    volume = "91",
    number = "5",
    pages = "054506",
    year = "2015"
}

@article{Haase:2020kaj,
    author = "Haase, Jan F. and Dellantonio, Luca and Celi, Alessio and Paulson, Danny and Kan, Angus and Jansen, Karl and Muschik, Christine A.",
    title = "{A resource efficient approach for quantum and classical simulations of gauge theories in particle physics}",
    eprint = "2006.14160",
    archivePrefix = "arXiv",
    primaryClass = "quant-ph",
    reportNumber = "DESY-20-146",
    doi = "10.22331/q-2021-02-04-393",
    journal = "Quantum",
    volume = "5",
    pages = "393",
    year = "2021"
}

@article{Itou:2024psm,
    author = "Itou, Etsuko and Matsumoto, Akira and Tanizaki, Yuya",
    title = "{DMRG study of the theta-dependent mass spectrum in the 2-flavor Schwinger model}",
    eprint = "2407.11391",
    archivePrefix = "arXiv",
    primaryClass = "hep-lat",
    reportNumber = "YITP-24-79, RIKEN-iTHEMS-Report-24",
    doi = "10.1007/JHEP09(2024)155",
    journal = "JHEP",
    volume = "09",
    pages = "155",
    year = "2024"
}

@article{Okuda:2022hsq,
    author = "Okuda, Takuya",
    title = "{Schwinger model on an interval: Analytic results and DMRG}",
    eprint = "2210.00297",
    archivePrefix = "arXiv",
    primaryClass = "hep-lat",
    reportNumber = "UT-Komaba/22-4",
    doi = "10.1103/PhysRevD.107.054506",
    journal = "Phys. Rev. D",
    volume = "107",
    number = "5",
    pages = "054506",
    year = "2023"
}

@article{Kanno:2024elz,
    author = "Kanno, Hayato and Akiyama, Shinichiro and Murakami, Kotaro and Takeda, Shinji",
    title = "{Grassmann tensor renormalization group for the massive Schwinger model with a {\ensuremath{\theta}} term using staggered fermions}",
    eprint = "2412.08959",
    archivePrefix = "arXiv",
    primaryClass = "hep-lat",
    reportNumber = "UTCCS-P-156, KANAZAWA 24-07, RIKEN-iTHEMS-Report-24",
    doi = "10.1007/JHEP11(2025)036",
    journal = "JHEP",
    volume = "11",
    pages = "036",
    year = "2025"
}

@article{Fujii:2024reh,
    author = "Fujii, Hirotsugu and Fujikura, Kohei and Kikukawa, Yoshio and Okuda, Takuya and Pedersen, Juan W.",
    title = "{Critical behavior of the Schwinger model via gauge-invariant variational uniform matrix product states}",
    eprint = "2412.03569",
    archivePrefix = "arXiv",
    primaryClass = "hep-lat",
    reportNumber = "UT-Komaba/24-10",
    doi = "10.1103/PhysRevD.111.094505",
    journal = "Phys. Rev. D",
    volume = "111",
    number = "9",
    pages = "094505",
    year = "2025"
}

@article{ArguelloCruz:2024xzi,
    author = "Arguello Cruz, Erick and Tarnopolsky, Grigory and Xin, Yuan",
    title = "{Precision study of the massive Schwinger model near quantum criticality}",
    eprint = "2412.01902",
    archivePrefix = "arXiv",
    primaryClass = "hep-th",
    doi = "10.1103/vf7m-kzk2",
    journal = "Phys. Rev. D",
    volume = "112",
    number = "3",
    pages = "034023",
    year = "2025"
}

@article{Bazavov:2024spg,
    author = "Bazavov, Alexei and Henke, Brandon and Hostetler, Leon and Lee, Dean and Lin, Huey-Wen and Pederiva, Giovanni and Shindler, Andrea",
    title = "{Efficient state preparation for the Schwinger model with a theta term}",
    eprint = "2411.00243",
    archivePrefix = "arXiv",
    primaryClass = "hep-lat",
    reportNumber = "TTK-24-44",
    doi = "10.1103/PhysRevD.111.074515",
    journal = "Phys. Rev. D",
    volume = "111",
    number = "7",
    pages = "074515",
    year = "2025"
}

@article{Mueller:2022xbg,
    author = {Mueller, Niklas and Carolan, Joseph A. and Connelly, Andrew and Davoudi, Zohreh and Dumitrescu, Eugene F. and Yeter-Aydeniz, K{\"u}bra},
    title = "{Quantum Computation of Dynamical Quantum Phase Transitions and Entanglement Tomography in a Lattice Gauge Theory}",
    eprint = "2210.03089",
    archivePrefix = "arXiv",
    primaryClass = "quant-ph",
    doi = "10.1103/PRXQuantum.4.030323",
    journal = "PRX Quantum",
    volume = "4",
    number = "3",
    pages = "030323",
    year = "2023"
}

@article{Lee:2023urk,
    author = "Lee, Kyle and Mulligan, James and Ringer, Felix and Yao, Xiaojun",
    title = "{Liouvillian dynamics of the open Schwinger model: String breaking and kinetic dissipation in a thermal medium}",
    eprint = "2308.03878",
    archivePrefix = "arXiv",
    primaryClass = "quant-ph",
    reportNumber = "JLAB-THY-23-3894, MIT-CTP-5592, YITP-SB-2023-23, IQuS@UW-21-061",
    doi = "10.1103/PhysRevD.108.094518",
    journal = "Phys. Rev. D",
    volume = "108",
    number = "9",
    pages = "094518",
    year = "2023"
}

@article{Farrell:2024fit,
    author = "Farrell, Roland C. and Illa, Marc and Ciavarella, Anthony N. and Savage, Martin J.",
    title = "{Quantum simulations of hadron dynamics in the Schwinger model using 112 qubits}",
    eprint = "2401.08044",
    archivePrefix = "arXiv",
    primaryClass = "quant-ph",
    reportNumber = "IQuS@UW-21-069, NT@UW-24-1",
    doi = "10.1103/PhysRevD.109.114510",
    journal = "Phys. Rev. D",
    volume = "109",
    number = "11",
    pages = "114510",
    year = "2024"
}

@article{Mildenberger:2022jqr,
    author = "Mildenberger, Julius and Mruczkiewicz, Wojciech and Halimeh, Jad C. and Jiang, Zhang and Hauke, Philipp",
    title = "{Confinement in a ${{\mathbb{Z}}}_{2}$ lattice gauge theory on a quantum computer}",
    eprint = "2203.08905",
    archivePrefix = "arXiv",
    primaryClass = "quant-ph",
    doi = "10.1038/s41567-024-02723-6",
    journal = "Nature Phys.",
    volume = "21",
    number = "2",
    pages = "312--317",
    year = "2025"
}

@article{Chen:2026sul,
    author = "Chen, Haobin and Cheng, Lin and Guo, Xingyu",
    title = "{Digital Quantum Simulation of Nonequilibrium Dynamics in the Schwinger Model under a Strong External Electric Field}",
    eprint = "2607.02894",
    archivePrefix = "arXiv",
    primaryClass = "hep-lat",
    month = "7",
    year = "2026"
}

@article{Lamm:2026cfq,
    author = "Lamm, Henry",
    title = "{MORE Thermal Gauge Theories at Finite $\theta$ and $\mu$ from Real-Time Quantum Simulation}",
    eprint = "2609.38569",
    archivePrefix = "arXiv",
    primaryClass = "hep-lat",
    reportNumber = "FERMILAB-PUB-26-0658-T",
    month = "9",
    year = "2026"
}

@article{Honda:2021aum,
    author = "Honda, Masazumi and Itou, Etsuko and Kikuchi, Yuta and Nagano, Lento and Okuda, Takuya",
    title = "{Classically emulated digital quantum simulation for screening and confinement in the Schwinger model with a topological term}",
    eprint = "2105.03276",
    archivePrefix = "arXiv",
    primaryClass = "hep-lat",
    reportNumber = "YITP-21-12, UT-Komaba-21-4",
    doi = "10.1103/PhysRevD.105.014504",
    journal = "Phys. Rev. D",
    volume = "105",
    number = "1",
    pages = "014504",
    year = "2022"
}

@article{Honda:2021ovk,
    author = "Honda, Masazumi and Itou, Etsuko and Kikuchi, Yuta and Tanizaki, Yuya",
    title = "{Negative string tension of a higher-charge Schwinger model via digital quantum simulation}",
    eprint = "2110.14105",
    archivePrefix = "arXiv",
    primaryClass = "hep-th",
    reportNumber = "YITP-21-111, RIKEN-iTHEMS-Report-21",
    doi = "10.1093/ptep/ptac007",
    journal = "PTEP",
    volume = "2022",
    number = "3",
    pages = "033B01",
    year = "2022"
}

@article{Honda:2022edn,
    author = "Honda, Masazumi and Itou, Etsuko and Tanizaki, Yuya",
    title = "{DMRG study of the higher-charge Schwinger model and its \textquoteright{}t Hooft anomaly}",
    eprint = "2210.04237",
    archivePrefix = "arXiv",
    primaryClass = "hep-lat",
    reportNumber = "YITP-22-100, RIKEN-iTHEMS-Report-22",
    doi = "10.1007/JHEP11(2022)141",
    journal = "JHEP",
    volume = "11",
    pages = "141",
    year = "2022"
}

@article{Itou:2023img,
    author = "Itou, Etsuko and Matsumoto, Akira and Tanizaki, Yuya",
    title = "{Calculating composite-particle spectra in Hamiltonian formalism and demonstration in 2-flavor QED$_{1+1d}$}",
    eprint = "2307.16655",
    archivePrefix = "arXiv",
    primaryClass = "hep-lat",
    reportNumber = "YITP-23-98, RIKEN-iTHEMS-Report-23",
    doi = "10.1007/JHEP11(2023)231",
    journal = "JHEP",
    volume = "11",
    pages = "231",
    year = "2023"
}

@article{Dempsey:2022nys,
	archiveprefix = {arXiv},
	author = {Dempsey, Ross and Klebanov, Igor R. and Pufu, Silviu S. and Zan, Bernardo},
	doi = {10.1103/PhysRevResearch.4.043133},
	eprint = {2206.05308},
	journal = {Phys. Rev. Res.},
	number = {4},
	pages = {043133},
	primaryclass = {hep-th},
	reportnumber = {PUPT-2634},
	title = {{Discrete chiral symmetry and mass shift in the lattice Hamiltonian approach to the Schwinger model}},
	volume = {4},
	year = {2022}}

@article{Angelides:2023noe,
    author = {Angelides, Takis and Naredi, Pranay and Crippa, Arianna and Jansen, Karl and K\"uhn, Stefan and Tavernelli, Ivano and Wang, Derek S.},
    title = "{First-Order Phase Transition of the Schwinger Model with a Quantum Computer}",
    eprint = "2312.12831",
    archivePrefix = "arXiv",
    primaryClass = "hep-lat",
    month = "12",
    year = "2023"
}

@article{Kaikov:2024acw,
    author = "Kaikov, Oleg and Saporiti, Theo and Sazonov, Vasily and Tamaazousti, Mohamed",
    title = "{Phase Diagram of the Schwinger Model by Adiabatic Preparation of States on a Quantum Simulator}",
    eprint = "2407.09224",
    archivePrefix = "arXiv",
    primaryClass = "hep-lat",
    month = "7",
    year = "2024"
}

@article{Nagano:2023uaq,
    author = "Nagano, Lento and Bapat, Aniruddha and Bauer, Christian W.",
    title = "{Quench dynamics of the Schwinger model via variational quantum algorithms}",
    eprint = "2302.10933",
    archivePrefix = "arXiv",
    primaryClass = "hep-ph",
    doi = "10.1103/PhysRevD.108.034501",
    journal = "Phys. Rev. D",
    volume = "108",
    number = "3",
    pages = "034501",
    year = "2023"
}

@article{Magnifico:2024eiy,
    author = "Magnifico, Giuseppe and Cataldi, Giovanni and Rigobello, Marco and Majcen, Peter and Jaschke, Daniel and Silvi, Pietro and Montangero, Simone",
    title = "{Tensor networks for lattice gauge theories beyond one dimension}",
    eprint = "2407.03058",
    archivePrefix = "arXiv",
    primaryClass = "hep-lat",
    doi = "10.1038/s42005-025-02125-x",
    journal = "Commun. Phys.",
    volume = "8",
    number = "1",
    pages = "322",
    year = "2025"
}

@article{Dalmonte:2016alw,
    author = "Dalmonte, M. and Montangero, S.",
    title = "{Lattice gauge theory simulations in the quantum information era}",
    eprint = "1602.03776",
    archivePrefix = "arXiv",
    primaryClass = "cond-mat.quant-gas",
    reportNumber = "CONT.-PHYS.-(2016), 1-25",
    doi = "10.1080/00107514.2016.1151199",
    journal = "Contemp. Phys.",
    volume = "57",
    pages = "388",
    year = "2016"
}

@article{DiMeglio:2023nsa,
    author = "Di Meglio, Alberto and others",
    title = "{Quantum Computing for High-Energy Physics: State of the Art and Challenges}",
    eprint = "2307.03236",
    archivePrefix = "arXiv",
    primaryClass = "quant-ph",
    reportNumber = "FERMILAB-PUB-23-468-ETD",
    doi = "10.1103/PRXQuantum.5.037001",
    journal = "PRX Quantum",
    volume = "5",
    number = "3",
    pages = "037001",
    year = "2024"
}

@article{Bauer:2022hpo,
    author = "Bauer, Christian W. and others",
    title = "{Quantum Simulation for High-Energy Physics}",
    eprint = "2204.03381",
    archivePrefix = "arXiv",
    primaryClass = "quant-ph",
    reportNumber = "UMD-PP-022-04, LA-UR-22-22100, RIKEN-iTHEMS-Report-22, RIKEN-iTHEMS-Report-22,
  FERMILAB-PUB-22-249-SQMS-T, FERMILAB-PUB-22-249-SQMS-T, IQuS@UW-21-027, MITRE-21-03848-2, FERMILAB-PUB-22-249-SQMS-T",
    doi = "10.1103/PRXQuantum.4.027001",
    journal = "PRX Quantum",
    volume = "4",
    number = "2",
    pages = "027001",
    year = "2023"
}

@article{Funcke:2023jbq,
    author = {Funcke, Lena and Hartung, Tobias and Jansen, Karl and K{\"u}hn, Stefan},
    title = "{Review on Quantum Computing for Lattice Field Theory}",
    eprint = "2302.00467",
    archivePrefix = "arXiv",
    primaryClass = "hep-lat",
    reportNumber = "MIT-CTP/5482",
    doi = "10.22323/1.430.0228",
    journal = "PoS",
    volume = "LATTICE2022",
    pages = "228",
    year = "2023"
}

@article{Banuls:2019bmf,
    author = "Ba{\~n}uls, M. C. and others",
    title = "{Simulating Lattice Gauge Theories within Quantum Technologies}",
    eprint = "1911.00003",
    archivePrefix = "arXiv",
    primaryClass = "quant-ph",
    doi = "10.1140/epjd/e2020-100571-8",
    journal = "Eur. Phys. J. D",
    volume = "74",
    number = "8",
    pages = "165",
    year = "2020"
}

@article{Banuls:2019rao,
    author = "Ba{\~n}uls, Mari Carmen and Cichy, Krzysztof",
    title = "{Review on Novel Methods for Lattice Gauge Theories}",
    eprint = "1910.00257",
    archivePrefix = "arXiv",
    primaryClass = "hep-lat",
    doi = "10.1088/1361-6633/ab6311",
    journal = "Rept. Prog. Phys.",
    volume = "83",
    number = "2",
    pages = "024401",
    year = "2020"
}

@article{Weinberg:2017,
  title={QuSpin: a Python package for dynamics and exact diagonalisation of quantum many body systems part I: spin chains},
  author={Weinberg, Phillip and Bukov, Marin},
  journal={arXiv preprint arXiv:1610.03042},
  year={2016}
}

@article{Weinberg:2019,
  title={QuSpin: a Python package for dynamics and exact diagonalisation of quantum many body systems. Part II: bosons, fermions and higher spins},
  author={Weinberg, Phillip and Bukov, Marin},
  journal={arXiv preprint arXiv:1804.06782},
  year={2018}
}

@book{Rasmussen2006,
  author    = {Rasmussen, Carl Edward and Williams, Christopher K. I.},
  title     = {Gaussian Processes for Machine Learning},
  publisher = {MIT Press},
  address   = {Cambridge, MA},
  year      = {2006}
}

@article{scikit-learn,
  author  = {Pedregosa, F. and Varoquaux, G. and Gramfort, A. and Michel, V. and Thirion, B. and Grisel, O. and Blondel, M. and Prettenhofer, P. and Weiss, R. and Dubourg, V. and Vanderplas, J. and Passos, A. and Cournapeau, D. and Brucher, M. and Perrot, M. and Duchesnay, E.},
  title   = {Scikit-learn: Machine Learning in {P}ython},
  journal = {Journal of Machine Learning Research},
  volume  = {12},
  pages   = {2825--2830},
  year    = {2011}
}

@article{levin2006detecting,
  title={Detecting topological order in a ground state wave function},
  author={Levin, Michael and Wen, Xiao-Gang},
  journal={Physical review letters},
  volume={96},
  number={11},
  pages={110405},
  year={2006},
  publisher={APS}
}

@article{Kitaev:2005dm,
    author = "Kitaev, Alexei and Preskill, John",
    title = "{Topological entanglement entropy}",
    eprint = "hep-th/0510092",
    archivePrefix = "arXiv",
    reportNumber = "CALT-68-2578, CALT-68-2578",
    doi = "10.1103/PhysRevLett.96.110404",
    journal = "Phys. Rev. Lett.",
    volume = "96",
    pages = "110404",
    year = "2006"
}

@article{Xie:2012mjn,
    author = "Xie, Z. Y. and Chen, J. and Qin, M. P. and Zhu, J. W. and Yang, L. P. and Xiang, T.",
    title = "{Coarse-graining renormalization by higher-order singular value decomposition}",
    eprint = "1201.1144",
    archivePrefix = "arXiv",
    primaryClass = "cond-mat.stat-mech",
    doi = "10.1103/PhysRevB.86.045139",
    journal = "Phys. Rev. B",
    volume = "86",
    number = "4",
    pages = "045139",
    year = "2012"
}

@article{Emonts:2022yom,
    author        = "Emonts, Patrick and Kelman, Ariel and Borla, Umberto and Moroz, Sergej and Gazit, Snir and Zohar, Erez",
    title         = "{Finding the ground state of a lattice gauge theory with fermionic tensor networks: A $2+1d$ $\mathbb{Z}_2$ demonstration}",
    eprint        = "2211.00023",
    archivePrefix = "arXiv",
    primaryClass  = "quant-ph",
    doi           = "10.1103/PhysRevD.107.014505",
    journal       = "Phys. Rev. D",
    volume        = "107",
    number        = "1",
    pages         = "014505",
    year          = "2023"
}

@article{Wilson:1974sk,
    author        = "Wilson, Kenneth G.",
    title         = "{Confinement of Quarks}",
    doi           = "10.1103/PhysRevD.10.2445",
    journal       = "Phys. Rev. D",
    volume        = "10",
    pages         = "2445--2459",
    year          = "1974"
}

@article{Satzinger:2021eqy,
    author        = "Satzinger, K. J. and others",
    title         = "{Realizing topologically ordered states on a quantum processor}",
    eprint        = "2104.01180",
    archivePrefix = "arXiv",
    primaryClass  = "quant-ph",
    doi           = "10.1126/science.abi8378",
    journal       = "Science",
    volume        = "374",
    number        = "6572",
    pages         = "1237--1241",
    year          = "2021"
}

@article{Sun:2022sux,
    author        = "Sun, Rong-Yang and Shirakawa, Tomonori and Yunoki, Seiji",
    title         = "{Parametrized quantum circuit for weight-adjustable quantum loop gas}",
    eprint        = "2210.14662",
    archivePrefix = "arXiv",
    primaryClass  = "quant-ph",
    doi           = "10.1103/PhysRevB.107.L041109",
    journal       = "Phys. Rev. B",
    volume        = "107",
    number        = "4",
    pages         = "L041109",
    year          = "2023"
}

@article{Crone:2019ybv,
    author        = "Crone, S. P. G. and Corboz, P.",
    title         = "{Detecting a $Z_2$ topologically ordered phase from unbiased infinite projected entangled-pair state simulations}",
    eprint        = "1912.00908",
    archivePrefix = "arXiv",
    primaryClass  = "cond-mat.str-el",
    doi           = "10.1103/PhysRevB.101.115143",
    journal       = "Phys. Rev. B",
    volume        = "101",
    number        = "11",
    pages         = "115143",
    year          = "2020"
}

@article{Yosprakob:2023flr,
    author = "Yosprakob, Atis",
    title = "{GrassmannTN: a Python package for Grassmann tensor network computations}",
    eprint = "2309.07557",
    archivePrefix = "arXiv",
    primaryClass = "hep-lat",
    doi = "10.21468/SciPostPhysCodeb.20",
    month = "9",
    year = "2023"
}

@article{Akiyama:2020sfo,
    author = "Akiyama, Shinichiro and Kadoh, Daisuke",
    title = "{More about the Grassmann tensor renormalization group}",
    eprint = "2005.07570",
    archivePrefix = "arXiv",
    primaryClass = "hep-lat",
    reportNumber = "UTHEP-751",
    doi = "10.1007/JHEP10(2021)188",
    journal = "JHEP",
    volume = "10",
    pages = "188",
    year = "2021"
}

@article{Gu:2010yh,
    author = "Gu, Zheng-Cheng and Verstraete, Frank and Wen, Xiao-Gang",
    title = "{Grassmann tensor network states and its renormalization for strongly correlated fermionic and bosonic states}",
    eprint = "1004.2563",
    archivePrefix = "arXiv",
    primaryClass = "cond-mat.str-el",
    month = "4",
    year = "2010"
}

@article{wu2012phase,
  title={Phase diagram of the toric code model in a parallel magnetic field},
  author={Wu, Fengcheng and Deng, Youjin and Prokof'ev, Nikolay},
  journal={Physical Review B—Condensed Matter and Materials Physics},
  volume={85},
  number={19},
  pages={195104},
  year={2012},
  publisher={APS}
}

@article{Ghosh:2015iwa,
    author        = "Ghosh, Sudip and Soni, Ronak M. and Trivedi, Sandip P.",
    title         = "{On The Entanglement Entropy For Gauge Theories}",
    eprint        = "1501.02593",
    archivePrefix = "arXiv",
    primaryClass  = "hep-th",
    doi           = "10.1007/JHEP09(2015)069",
    journal       = "JHEP",
    volume        = "09",
    pages         = "069",
    year          = "2015"
}

@article{Pretko:2015zva,
    author        = "Pretko, Michael and Senthil, T.",
    title         = "{Entanglement Entropy of U(1) Quantum Spin Liquids}",
    eprint        = "1510.03863",
    archivePrefix = "arXiv",
    primaryClass  = "cond-mat.str-el",
    doi           = "10.1103/PhysRevB.94.125112",
    journal       = "Phys. Rev. B",
    volume        = "94",
    pages         = "125112",
    year          = "2016"
}

@article{Polyakov:1976fu,
    author = "Polyakov, Alexander M.",
    title = "{Quark Confinement and Topology of Gauge Groups}",
    reportNumber = "NORDITA-76/33",
    doi = "10.1016/0550-3213(77)90086-4",
    journal = "Nucl. Phys. B",
    volume = "120",
    pages = "429--458",
    year = "1977"
}

@article{Wegner:1971app,
    author = "Wegner, F. J.",
    title = "{Duality in Generalized Ising Models and Phase Transitions Without Local Order Parameters}",
    doi = "10.1063/1.1665530",
    journal = "J. Math. Phys.",
    volume = "12",
    pages = "2259--2272",
    year = "1971"
}

@article{Kelman:2024gjt,
    author = "Kelman, Ariel and Borla, Umberto and Gomelski, Itay and Elyovich, Jonathan and Roose, Gertian and Emonts, Patrick and Zohar, Erez",
    title = "{Gauged Gaussian PEPS -- A High Dimensional Tensor Network Formulation for Lattice Gauge Theories}",
    eprint = "2404.13123",
    archivePrefix = "arXiv",
    primaryClass = "hep-lat",
    month = "4",
    year = "2024"
}

@article{Zohar:2021wqy,
    author = "Zohar, Erez",
    title = "{Wilson loops and area laws in lattice gauge theory tensor networks}",
    eprint = "2101.05289",
    archivePrefix = "arXiv",
    primaryClass = "quant-ph",
    doi = "10.1103/PhysRevResearch.3.033179",
    journal = "Phys. Rev. Res.",
    volume = "3",
    number = "3",
    pages = "033179",
    year = "2021"
}

\end{document}